\documentclass[runningheads, orivec]{llncs}
\usepackage[top=2cm, bottom=2cm, left=3cm, right=3cm]{geometry}
\usepackage{bm}  
\usepackage{amsmath}
\usepackage{amsfonts}

\usepackage{multicol}
\usepackage{color}
\usepackage{float}

\usepackage{braket}
\usepackage[T1]{fontenc}
\usepackage{graphicx}
\usepackage{booktabs}
\usepackage{siunitx}
\usepackage{makecell}
\usepackage{multirow}
\usepackage{array}
\usepackage{enumitem}
\usepackage{longtable}
\usepackage{tablefootnote}
\usepackage[hyphens]{url}
\usepackage[colorlinks=true, urlcolor=blue, pdfborder={0 0 0}]{hyperref}
\hypersetup{
  colorlinks   = true, 
  urlcolor     = blue, 
  linkcolor    = blue, 
  citecolor   = blue 
}

\newcommand{\dagfootnote}[1]{%
  \begingroup
    \renewcommand{\thefootnote}{\textdagger}%
    \footnote{#1}%
    \addtocounter{footnote}{-1}%
  \endgroup
}

\usepackage{caption}
\usepackage{array}
\newcolumntype{C}[1]{>{\centering\arraybackslash}p{#1}}
\newcommand{\indentitem}{\hspace*{0.5em}}

\newcommand{\rottext}[1]{\rotatebox[origin=c]{90}{#1}}
\begin{document}
\title{Dynamic Liquidity Provision in Decentralized Markets: Strategy Optimization and Performance Evaluation in Concentrated Liquidity AMMs}
\titlerunning{Abbreviated paper title}

\author{Andrey~Urusov\inst{1,3}\thanks{Corresponding author.} \and 
Rostislav~Berezovskiy\inst{1,2}\and 
Anatoly~Krestenko\inst{1,4} \and \\
Andrei~Kornilov\inst{1} \and
Yury~Yanovich\inst{5}} 

\authorrunning{A. Urusov et al.}
\titlerunning{Dynamic Liquidity Provision in Decentralized Markets}

\institute{Vega Institute Foundation, Moscow, Russia \\ 
\and 
International Laboratory of Stochastic Analysis and its Applications, \\ HSE University, Moscow, Russia \\ 
\and 
Chuvash State University, Cheboksary, Russia \\ 
\and 
Moscow Institute of Physics and Technology, Moscow, Russia \\ 
\and
Skolkovo Institute of Science and Technology, Moscow, Russia
}

\maketitle
{
\renewcommand{\thefootnote}{}
\footnotetext{%
  \textit{E‑mail addresses:} tapwi93@gmail.com$^*$ (A.~Urusov),
  rostislavberezovskiy@gmail.com (R.~Berezovskiy),
  acryptokrestenko@gmail.com (A.~Krestenko), 
  kornilov.ag94@gmail.com (A.~Kornilov),
  y.yanovich@skoltech.ru (Y.~Yanovich).
}
}
%
\begin{abstract}
Concentrated Liquidity Market Makers (CLMMs) represent a fundamental innovation in market microstructure, transforming liquidity provision from passive portfolio allocation to active risk management. This evolution creates significant challenges for performance evaluation and strategy optimization, particularly due to the absence of comprehensive historical liquidity data. We address these challenges through a novel methodological framework that reconstructs historical liquidity states from swap transaction data, enabling rigorous backtesting of dynamic liquidity provision strategies. Our parametric reconstruction method achieves high accuracy (approximation errors averaging around 2\%) without relying on historical liquidity snapshots, addressing a critical data gap in decentralized finance research. We apply this framework to evaluate tau-reset strategies--dynamic liquidity reallocation approaches that respond to market movements--across multiple Uniswap v3 pools. Using machine learning to optimize strategy parameters based on market conditions, we identify consistent outperformance (13--23\% higher fees) compared to uniform allocation benchmarks. Our analysis reveals important insights into the risk-return tradeoffs in automated market making, including the critical role of impermanent loss as a dominant risk factor and the effectiveness of asymmetric strategy modifications for capital preservation. These findings contribute to the broader understanding of market microstructure in decentralized exchanges, providing both methodological innovations for performance evaluation and practical insights for liquidity providers navigating this evolving financial landscape.

\end{abstract}

\section{Introduction}
Decentralized Finance (DeFi) has become one of the fastest-growing areas of the blockchain industry, offering an open, programmable architecture to create, execute, and automate financial services through smart contracts, which are self‑executing programs that run on-chain without intermediaries~\cite{ref_1}. The DeFi ecosystem has evolved rapidly to include decentralized exchanges, lending protocols, and derivative markets~\cite{ref_37}. A central component of this ecosystem is the Decentralized Exchange (DEX), which is based on the Automated Market Maker (AMM) mechanism. In an AMM, liquidity is pooled and trading prices are set algorithmically rather than through a traditional order book. Since the launch of Uniswap V1/V2~\cite{ref_2}, such protocols have provided a simple and accessible trading model in which pricing is governed by a predefined mathematical rule, most notably the constant product function~\cite{ref_3}. This represents a departure from traditional order-book exchanges and has stimulated significant research into decentralized exchange mechanisms~\cite{ref_38}.

Despite their elegance and novelty, early AMM protocols suffered from capital inefficiency: liquidity was distributed uniformly across the entire price axis, although most trading activity occurred within a relatively narrow range. Uniswap V3 addressed this shortcoming by introducing the Concentrated Liquidity Market Maker (CLMM) concept~\cite{ref_4}, which allows liquidity providers (LP) to allocate capital to selected price intervals. Under CLMM, an LP earns fees only when the pool price is within the chosen range, turning liquidity provision from a passive activity into an active management problem that requires continuous monitoring of market conditions~\cite{ref_5}.

Consequently, optimal liquidity provision has become a critical research objective. However, developing and evaluating such strategies presents significant challenges. A key difficulty is the absence of comprehensive historical liquidity data needed for accurate backtesting. While swap transaction histories are typically available, the precise distribution of liquidity across price ranges at historical moments is often inaccessible or incomplete. This data gap hinders the development of data-driven approaches to liquidity optimization. Active liquidity management requires LPs to make decisions that weigh current and expected market conditions, risk exposure, and competition within pools. Therefore, designing effective liquidity management strategies requires a deep analysis of pool dynamics and the development of specialized tools to test research hypotheses.

To address these challenges, this paper introduces a general methodology for developing models that support optimal liquidity provision decisions on Concentrated Liquidity Market Makers (CLMM) platforms. The core innovation lies in reconstructing historical liquidity states from readily available swap data, enabling rigorous strategy development without dependency on historical liquidity snapshots. This paper makes three primary contributions to the financial analysis literature:

\begin{enumerate}
    \item \textbf{Methodological Innovation:} We develop and validate a parametric methodology for reconstructing historical liquidity profiles solely from swap transaction data, eliminating dependency on historical liquidity snapshots. This approach enables accurate performance evaluation and strategy backtesting even when complete historical data is unavailable.
    
    \item \textbf{Empirical Strategy Analysis:} We implement a comprehensive framework for developing and evaluating dynamic $\tau$-reset strategies across multiple asset pairs and market conditions. Our analysis identifies consistent patterns in optimal strategy parameters and performance characteristics, providing actionable insights for liquidity providers.
    
    \item \textbf{Financial Market Insights:} Beyond methodological contributions, our findings offer important insights into market microstructure in decentralized exchanges. We demonstrate how liquidity provision strategies affect risk-return tradeoffs, how market conditions influence optimal strategy parameters, and how modified strategy designs can mitigate specific risks inherent in CLMM participation.
\end{enumerate}

The remainder of this paper is organized as follows. Section \ref{section:background} provides the necessary background on CLMM mechanics, $\tau$-reset strategies, Section \ref{section:relwk} describes the related work. Section \ref{section:problem_solution} formalizes the problem and describes our proposed methodology, including liquidity approximation, reward modeling, and machine learning architecture. Section \ref{section:Experiments} presents an empirical study of historical Uniswap V3 pool data, covering liquidity reconstruction, model training, performance evaluation, and comparison with alternative tools. Section \ref{section:discussion} discusses theoretical and practical implications, and Section \ref{section:Conclusions} summarizes the findings and provides directions for future research.

\section{Background}
\label{section:background}
This section provides the necessary background for understanding our proposed methodology. We first review key concepts and terminology related to concentrated liquidity markets, then discuss the $\tau$-reset strategy family that forms the basis of our approach, and finally survey related work in the field.

\subsection{Key Concepts and External Data}
\label{section:KCED}

\subsubsection{Prices.} After choosing the trading pair (e.g. USDC/ETH\dagfootnote{In this research, ETH denotes Wrapped ETH (WETH) — the ERC‑20 tokenized version of Ether that is used for interacting with AMM protocols.}) and the specific DEX pool with a given fee tier~$\Gamma$ for which the optimal liquidity model will be constructed, we select a historical time interval—the \emph{modeling period}—over which the simulation is carried out.  
Next, we extract the set of swap transactions executed in the selected pool during this period and reconstruct the in-pool prices from those swaps.  
The resulting set of historical prices is denoted
$\bm{P}= \{p_0,\, p_1,\, \dots,\, p_H\}$.
Price intervals that contain elements of~$\bm{P}$ are referred to as \emph{active} (or \emph{working}) ranges.

The first task to address before looking for optimal LP strategies is to model the historical fee rewards of the pool within the active ranges.  
In our framework, relying on aggregated minute-, hour-, or day-level price data is inadequate: precise modeling requires the full sequence of in-pool price updates generated by every swap during the modeling period, together with the corresponding trade volumes. With the assumed fixed liquidity across the pool ranges, any price movement implies a change in the token reserves held in the active ranges, driven by traders who swap one pool token for the other. Given the volumes required to move the price in the pool along the set \(\bm{P}\) under the fee tier~\(\Gamma\), we compute the fees paid by the pool for the liquidity supplied by LPs within the active ranges. Because trades within the pool are bidirectional, using aggregated data may omit part of the flow that actually passes through the pool and, at a structural level, misstate the modeled fee income.

\subsubsection{Volumes and TVL.}
Each swap is associated with a trade volume, which we express in the chosen numéraire and use as a historical benchmark to approximate the historical liquidity of the pool. We refer to these as \emph{historical volumes} and denote them by $\bm{V}= \{v_0,\, v_1,\, \dots,\, v_H\}$. The average total value locked in the pool (TVL) during the modeling period is denoted by \(\Sigma\); this quantity is required to recover the level of liquidity within the active ranges.

It is worth highlighting that the quantity \(\Sigma\) does not require an overly rigid estimation procedure. If, for any reason, the average TVL during the modeling period cannot be reliably assessed, one can simply fix \(\Sigma\) to the TVL of the pool observed at the beginning of the study. The historical reward modeling technique proposed in~\cite{ref_6} approximates the liquidity profile by benchmarking against the historical fee income of the pool. In this case, the modeled trade volumes that determine the modeled reward levels of the pool pass through the pool's \emph{local} price zone bounded by active historical ranges. This implies that the overall approximated form of the pool's historical liquidity is less important to us than the specific configuration of the local zone containing active ranges. Consequently, using an approximate level of TVL for \(\Sigma\) is admissible, since the algorithm dynamically approximates the liquidity levels within the working ranges.

\subsection{Baseline Strategy Type: $\tau$-Reset Strategies}
\label{sec:bst}
The baseline strategy considered in this study is the dynamic \emph{$\tau$-reset} strategy with symmetric shape. Building on the exposition in \cite{ref_5}, the mechanism is described in detail in \cite{ref_6}; we summarize it here for completeness.

\subsubsection{Bucket partition}
Before the strategy is applied, the price axis is partitioned into \(N\) contiguous buckets of equal width, creating a set $\beta=\{B_1,\ldots,B_N\}$. Bucket \(i\) is delimited by a lower and an upper price, \(p_{a_i}\) and \(p_{b_i}\), such that \(p_{a_i}<p_{b_i}\) for \( i \in [1,N] \). Adjacency is enforced at the boundaries of the buckets: \(p_{a_i}=p_{b_{i-1}}\) for \(i>1\) and \(p_{b_i}=p_{a_{i+1}}\) for \(i<N\). Each bucket has the same fixed width $d = (p_{b_N}-p_{a_1})/N$.

There is no universal rule for selecting $\beta$. The researcher chooses the partitioning rule, namely the boundary prices \(p_{a_1}\) and \(p_{b_N}\) with the parameter \(N\) so that the entire set of active ranges in the modeling period will be covered by $\beta$, while ensuring that the width of the bucket \(d\) is not smaller than the minimum tick size of the pool. It is essential that both the width \(d\) and the strategy parameter \(\tau\) remain fixed during model training and subsequent deployment.

\subsubsection{Strategy algorithm.}
\begin{figure*}[!t]
    \centering
    \includegraphics[width=1\textwidth]{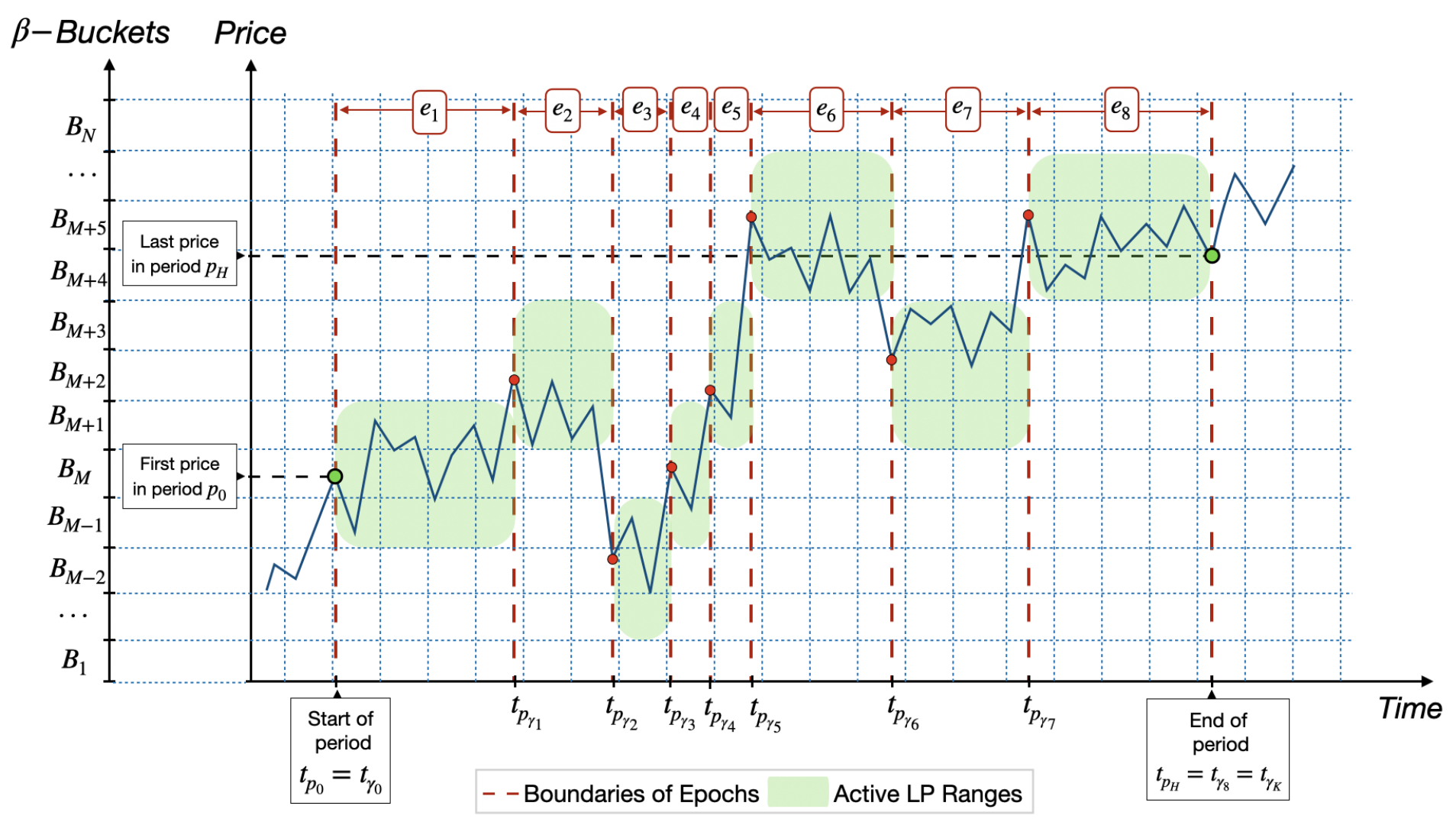}
    \smallskip
    \captionsetup{justification=centering}
    \caption{Visualization of the LP-liquidity reallocation process under the $\tau$-reset strategy ($\tau=1$) on a fixed bucket partition~$\beta$ over eight epochs.}
    \label{figure:F1}
\end{figure*}

At the moment that the LP allocates capital \(W^{\mathrm{LP}}\) to the pool as liquidity \(L^{\mathrm{LP}}\), as in \cite{ref_5} we identify \emph{reference bucket} \(B_{M}\in\beta\) with the index \( M \in [1,N] \) that contains the initial price \(p_{0}\in\bm{P}\). The parameter \(\tau\) is then fixed: It specifies how many buckets on either side of the reference bucket will also receive LP liquidity. These buckets (and their associated price ranges) are called the \emph{LP-liquid buckets} and are denoted by set $\beta^{\mathrm{LP}}_{L}= \{B_{M-\tau},\ldots,B_{M+\tau}\}$ with size $2\tau+1$. The core rebalancing rule is as follows: If the next observed price falls into a bucket with index \(M\pm\vartheta\) where \(\vartheta>\tau\), the reference index \(M\) is updated and the liquidity of the LP is reallocated to the new set of buckets, as illustrated in Fig. \ref{figure:F1}.

The instants at which liquidity is reallocated, that is, when the price exits the current set of liquid buckets from LP, are called \emph{reset moments} or \emph{epoch switches}.  
An \emph{epoch} is the time interval during which the reference bucket index \(M\) remains fixed within the modeling period.  
The total number of epochs \(K\) is fully determined by the width of the bucket \(d\) and the parameter \(\tau\) for a given price set \(\bm{P}\); in live trading, the length of an epoch is a random variable.  
The minimum value \(K=1\) on historical data arises when the initial set \(\beta^{\mathrm{LP}}_{L}\) chosen at \(p_{0}\) already covers every price in \(\bm{P}\), so no reallocation occurs except for closing the position at the end of the period.
Let the set of epochs be $\bm{E}=\{e_{1},\ldots,e_{K}\}$.
For \(K>1\), each epoch corresponds to a subset of prices from \(\bm{P}\); collectively, these form
\begin{equation}
\begin{gathered}
    \label{eqn:frml1}
    \bm{P}_{\bm{E}} = \{\bm{P}_{e_{1}},\ldots,\bm{P}_{e_{K}}\} = \\[4pt]
    = \bigl\{
          \{p_{\gamma_{0}},p_{\gamma_{0}+1},\ldots,p_{\gamma_{1}}\},
          \{p_{\gamma_{1}},\ldots,p_{\gamma_{2}}\},
          \dots,
          \{p_{\gamma_{K-1}},\ldots,p_{\gamma_{K}}\}
        \bigr\},
\end{gathered}
\end{equation}
where \(\gamma_{i}\) denotes the index of the price at which the \(i\)-th epoch ends,
\(\gamma_{0}<\gamma_{1}<\dots<\gamma_{K}\in[0,H]\) with \(\gamma_{0}=0\) and \(\gamma_{K}=H\).
The size of an epoch \(e_i\) can be quantified in two ways.  
First, as the time difference between the two reset instants that bracket the epoch, $t_{e_i}=t_{p_{\gamma_i}}-t_{p_{\gamma_{i-1}}}$,
and second, as the number of price observations it contains, $q_{e_i}=|\bm{P}_{e_i}|=\gamma_i-\gamma_{i-1}+1$.
We regard the last price of the current epoch as the first price of the next one, thereby assuming a minimum latency for allocation decision making in one block.
When the position is closed at the end of the modeling period, a corner case can occur in which an epoch contains only one price, i.e. \(p_{\gamma_{K-1}}=p_{\gamma_K}\).  
We classify such an epoch as \emph{degenerate} and at the same time the epoch size as \(q_{e_i}\ge 2\).
Denote $\bm{T}_{\bm{E}}=\bigl(t_{e_1},\ldots,t_{e_K}\bigr)$,
$\bm{Q}_{\bm{E}}=\bigl(q_{e_1},\ldots,q_{e_K}\bigr)$
as the vectors of epoch sizes expressed, respectively, in units of elapsed time and in counts of price updates; both notions of size are used in the remainder of the paper.

\subsection{Concept of Liquidity}
We have already introduced the concept of \emph{LP-liquid buckets} and spoken of deploying LP capital within price ranges "as liquidity".  
In this subsection, we formalize the term and outline how liquidity is computed.
In the Uniswap family of protocols, the constant \(L\) (historically denoted \(k\)) from the constant product function plays a central role in both the original AMM and its concentrated-liquidity extension (Uniswap v3). Algorithm for computing and using liquidity:
\begin{enumerate}
  \item Given a bucket partition \(\beta\) with fixed price boundaries, the LP allocated capital \(W^{\mathrm{LP}}_i\) (denominated in the numéraire, e.g.\ USDC) to bucket \(B_i\in\beta\).

  \item Depending on the current pool price \(p\) relative to the placement interval
        \([p_{a_i},\,p_{b_i}]\), the capital \(W^{\mathrm{LP}}_i\) is converted into the
        \emph{real reserves} \(x_i\) and \(y_i\) of bucket \(i\), corresponding to the two pool tokens
        \(A\) (e.g.\ ETH) and \(B\) (e.g.\ USDC), such that
        \[
          W^{\mathrm{LP}}_i \;=\; y_i \;+\; p\,x_i .
        \]

  \item These reserves reside inside bucket \(B_i\) as liquidity \(L_i\).
        The fundamental invariant Uniswap v3 that links \(L_i\) to the reserves \((x_i,y_i)\) within
        \([p_{a_i},p_{b_i}]\) is
        \begin{equation}
          \bigl(x_i+\tfrac{L_i}{\sqrt{p_{b_i}}}\bigr)\,
          \bigl(y_i+L_i\sqrt{p_{a_i}}\bigr)
          \;=\;
          L_i^2 ,
          \label{eq:Li-invariant}
        \end{equation}
        where \emph{real} reserves shifted by the constants fixed by the range boundaries form the so-called \emph{virtual} reserves that determine the price inside the range. Drawing on the results of~\cite{ref_7}, the paper \cite{ref_6} presents closed-form expressions for \(L_i\) in terms of the \emph{real} reserves \(x_i\) and \(y_i\) for three configurations: the price \(p\) lies inside the chosen \(i\)-th range, to its left, or to its right. Let us consider each case.
        
    \medskip
    \emph{Case 1}: \(p\le p_{a_i}\) (price at or below the lower bound),
    with \(x_i>0\) and \(y_i=0\):
    \begin{equation}
          L_i
          \;=\;
          \frac{x_i \sqrt{p_{a_i} p_{b_i}}}
               {\sqrt{p_{b_i}}-\sqrt{p_{a_i}}}.
    \end{equation}
    
    \emph{Case 2}: \(p\ge p_{b_i}\) (price at or above the upper bound),
    with \(x_i=0\) and \(y_i>0\):
    \begin{equation}
          L_i
          \;=\;
          \frac{y_i}{\sqrt{p_{b_i}}-\sqrt{p_{a_i}}}.
    \end{equation}
    
    \emph{Case 3}: \(p_{a_i}<p<p_{b_i}\) (price inside the range),
    with \(x_i>0\) and \(y_i>0\).  
    Introduce the auxiliary notation
    \[
      x_{L_i} = \frac{\sqrt{p_{b_i}p}}{\sqrt{p_{b_i}}-\sqrt{p}},
      \qquad
      y_{L_i} = \frac{1}{\sqrt{p}-\sqrt{p_{a_i}}},
    \]
    so that the \emph{real} reserves become
    \begin{equation}
        x_i = \frac{W_i^{\mathrm{LP}}\,y_{L_i}}{x_{L_i}+y_{L_i}p},
        \qquad
        y_i \;=\; W_i^{\mathrm{LP}}\,\bigl(1-
            \frac{y_{L_i}p}{x_{L_i}+y_{L_i}p}\bigl).
    \end{equation}
    The liquidity can be expressed from either reserve:
    \[
      L_i
      \;=\;
      x_i\,x_{L_i}
      \;=\;
      y_i\,y_{L_i}.
    \]
    
    In this case, the allocation of capital \(W_i^{\mathrm{LP}}\) proceeds by computing the
    reserves \emph{real} of tokens \(A\) and \(B\) that correspond to the bucket \(B_i\),
    given its price interval \([p_{a_i},p_{b_i}]\) and the price in the range \(p\).
    In contrast, in the inverse problem, when the desired \emph{real} reserves
    \(x_i>0\) and \(y_i>0\) are specified in advance but the partition \(\beta\) is not, the limits of the bucket \([p_{a_i},p_{b_i}]\) can be recovered analogously.

    \item Following the formulation of \cite{ref_6} and the
    notation of \cite{ref_5} and \cite{ref_8},
    we define the \emph{liquidity--state} function for bucket~\(i\):
    \[
      \mathcal{V}_i\!\left(L_i,p,B_i\right)
      \;=\;
      \bigl(x_i,\,y_i\bigr),
    \]
    which maps a fixed liquidity amount \(L_i\) together with a price
    \(p\) inside the bucket \(B_i=[p_{a_i},p_{b_i}]\) to the vector of
    \emph{real} token reserves. Distinct prices yield distinct reserve
    vectors: \(\mathcal{V}_i(L_i,p,B_i)\neq\mathcal{V}_i(L_i,p',B_i)\) for
    \(p\neq p'\).
    A trader’s swap that "pushes" the pool price from the range \(B_i\)
    converts the position entirely into one of the two assets:
    \[
    \mathcal{V}_i\!\bigl(L_i,p',B_i\bigr)=
    \begin{cases}
    (\tilde{x}_i,0), & p' \le p_{a_i},\\
    (0,\tilde{y}_i), & p' \ge p_{b_i}
    \end{cases}
    \]
    
    so that higher local liquidity concentration reducing price slippage. Introduce the standard notation
    \[
      \Delta^{x_i}_{p_{a_i},p_{b_i}}
      =\frac{1}{\sqrt{p_{a_i}}}-\frac{1}{\sqrt{p_{b_i}}},
      \qquad
      \Delta^{y_i}_{p_{a_i},p_{b_i}}
      =\sqrt{p_{b_i}}-\sqrt{p_{a_i}}.
    \]
    With these, the liquidity--state function admits the piece-wise form
    \begin{equation}
        \mathcal{V}_i\!\left(L_i,p,B_i\right)
        =
        \begin{cases}
        \bigl(L_i\,\Delta^{x_i}_{p_{a_i},p_{b_i}},\,0\bigr),
              & p\le p_{a_i},\\[4pt]
        \bigl(L_i\,\Delta^{x_i}_{p,p_{b_i}},\,L_i\,\Delta^{y_i}_{p_{a_i},p}\bigr),
              & p_{a_i}<p<p_{b_i},\\[4pt]
        \bigl(0,\,L_i\,\Delta^{y_i}_{p_{a_i},p_{b_i}}\bigr),
              & p\ge p_{b_i}.
        \end{cases}
        \label{eq:liquidity-state}
    \end{equation}
    
    This representation is consistent with the Uniswap v3 liquidity
    formulas derived in \cite{ref_7} and employed in
    \cite{ref_6}.

\end{enumerate}

\section{Related Work}
\label{section:relwk}
This subsection reviews existing literature on liquidity provision in CLMM markets, covering specific strategies, backtesting frameworks, and machine learning applications. Recent systematic reviews compare the architectures and performance of centralized versus decentralized exchanges (CEX and DEX)~\cite{ref_48}. Beyond core exchange mechanics, DeFi protocols are enabling novel financial interactions in adjacent sectors such as GameFi, where liquidity provision and staking mechanisms are integral to 'play-to-earn' economic models~\cite{ref_49}.

The concept of liquidity provision in concentrated liquidity markets has been studied from multiple perspectives, including impermanent loss, trading fee dynamics, DEX market design, and liquidity provision strategies. 

The option-like properties of liquidity provision in both CPMM and CLMM settings have been thoroughly analyzed, with foundational work on token reserves, swaps, fees, and liquidity calculations provided in \cite{ref_7,ref_9}. Building on this, \cite{ref_10} developed a methodology to estimate accumulated fees in liquidity pools under basic price behavior assumptions, though without explicit consideration of trading volumes.

From a trader's perspective, \cite{ref_11} compares price discovery processes in DEXs versus centralized markets and studies the behavior of informed traders. Meanwhile, \cite{ref_12} analyzes bid-ask prices in liquidity pools and derives the value of no-trade gaps. Arbitrage opportunities in DEXs have been systematically studied, including simple arbitrage strategies~\cite{ref_44} and more complex cyclic arbitrage patterns~\cite{ref_45}.

Research on impermanent loss includes \cite{ref_13}, which studies LP risks and potential returns, providing numerical estimations from real-world data. \cite{ref_14} offers the concerning estimate that approximately half of LPs lose money in their dataset, highlighting the complexity of concentrated liquidity provision. Further analysis by \cite{ref_39} examines the risk profile of liquidity providers, while \cite{ref_40} and \cite{ref_41} provide additional theoretical frameworks for understanding impermanent loss in both traditional and concentrated liquidity settings. \cite{ref_15} provides concise results for estimating impermanent loss and generalizes the concept to allow comparison with various benchmark strategies.

Hedging impermanent loss with traditional options has been explored in \cite{ref_16}, which applies arbitrage-based methods to produce both static and dynamic hedges under Black-Scholes-Merton and log-normal stochastic volatility models.

The $\tau$-reset strategy family, central to our work, builds on concepts introduced in \cite{ref_5,ref_6}. Related strategy research includes \cite{ref_17}, which considers accumulated fees, rebalancing costs, and liquidity concentration risk in continuous-time models for specific market conditions. \cite{ref_18} models liquidity dynamics in CLMM pools in continuous time with rebalancing requirements, while \cite{ref_19} approaches liquidity provision as an optimal stopping problem.

Machine learning applications in CLMM markets are gaining traction. \cite{ref_20} applies deep reinforcement learning to maximize accumulated fees while considering loss-versus-rebalancing risk, optimizing price interval parameters. Similarly, \cite{ref_21} demonstrates superior strategy performance compared to benchmarks using deep reinforcement learning on ETH/USDC and ETH/USDT pools. Beyond strategy optimization, \cite{ref_22} applies reinforcement learning to AMM design itself, aiming to reduce impermanent loss and slippage.

Market design considerations are addressed in \cite{ref_23}, which introduces generalized decentralized liquidity pools with dynamic bonding curves, and \cite{ref_24}, which extends the problem by considering the trading platform as a player competing for fees with LPs.

Competition among LPs has been modeled using mean-field games in \cite{ref_25}, analyzing resulting liquidity distributions and incorporating just-in-time liquidity. Beyond market microstructure risks, broader DeFi protocol risks including smart contract vulnerabilities, regulatory uncertainty, and systemic risks have been examined in the literature~\cite{ref_46,ref_47}. A specific focus on just-in-time liquidity appears in \cite{ref_26}.

Finally, backtesting frameworks for CLMM strategies are essential for practical validation. Our previous work \cite{ref_6} provides the foundation for the backtesting approach extended in this paper. The current research applies accumulated knowledge of concentrated liquidity mechanics to extend this framework and enable data-driven liquidity provision optimization using machine learning methods within the $\tau$-reset strategy family.

\section{Problem Statement and Proposed Solution}
\label{section:problem_solution}

This section formalizes the liquidity provision optimization problem and presents our proposed methodology for developing optimal $\tau$-reset strategies using historical data analysis.

\subsection{Problem Statement}
\label{section:Pr_St}

In this subsection, \emph{LP-balance equation} is introduced, followed by
a definition of \emph{optimal strategy} and the main research
objective.  The exposition begins with the algorithm used to compute
pool-level fee revenues and then proceeds to a detailed derivation of the
reward attributable to an individual LP.  

\subsubsection{Approach to Pool Reward Calculation.}
After partitioning the historical horizon into \(K\) epochs under the
dynamic \(\tau\)-reset rule, the liquidity corresponding to the average TVL (notation as
\(\Sigma\)) is \emph{fixed} across the partition \(\beta\) at the start of
every epoch \(e_i\).
Let vector $\bm{W}_{e_i}^{\Sigma} = \bigl(w_1^{\Sigma},w_2^{\Sigma},\ldots,w_N^{\Sigma}\bigr)$,
  $\sum_{j=1}^{N} w_j^{\Sigma} = \Sigma$,
denote the capital allocated to each bucket of $\beta$ and let $\bm{L}_{e_i}^{\Sigma} = \bigl(L_1^{\Sigma},L_2^{\Sigma},\ldots,L_N^{\Sigma}\bigr)$
be the corresponding vector of liquidity amounts (liquidity profile), determined at the
epoch-opening price \(p_{\gamma_{i-1}}\) in equation \eqref{eqn:frml1}.
During epoch \(e_i\) the pool price follows the subset
\(\bm{P}_{e_i} = \{p_{\gamma_{i-1}},\ldots,p_{\gamma_i}\}\) over the time
interval \(t_{e_i}\).
Because liquidity \(\bm{L}_{e_i}^{\Sigma}\) remains fixed in
all buckets throughout \(t_{e_i}\), each price move induced by trader swaps alters the reserves \emph{real} of tokens within the active ranges.
Changes in reserves are interpreted as trade volumes moving that determine the price path.

For every bucket \(B_n\) and every step
\(q\in\{1,\ldots,q_{e_i}-1\}\) of epoch \(e_i\), the incremental change in
token holdings produced by the historical move from
$p_{\gamma_{i-1}+q-1}$ to $p_{\gamma_{i-1}+q}$
is obtained from the liquidity-state function
\(\mathcal{V}_n\bigl(L_n^{\Sigma},p,B_n\bigr)\) as
\begin{equation}
  \Delta\mathcal{V}^{q}_{n}
  =
  \mathcal{V}_n\!\bigl(L_n^{\Sigma},p_{\gamma_{i-1}+q},B_n\bigr)
  \;-\;
  \mathcal{V}_n\!\bigl(L_n^{\Sigma},p_{\gamma_{i-1}+q-1},B_n\bigr),
  \label{eq:delta-V}
\end{equation}
where \( n \in [1,N] \). The vector \(\Delta\mathcal{V}^{q}_{n}\) thus represents the model fee-bearing
trade volume executed in bucket \(B_n\) for that price increment. 

Positive increments in the token state, corresponding to the trade volumes that traders bring into the pool, are relevant for the generation of fees under the fixed liquidity
vector \(\bm{L}_{e_i}^{\Sigma}\).
Aggregating these inflows over all buckets and all price steps of epoch
\(e_i\) yields the token-volume vector
\begin{equation}
  \bm S_{e_i}^{\Sigma}
  \;=\;
  \Gamma
  \sum_{q=1}^{q_{e_i}-1}
  \sum_{n=1}^{N}
  \bigl[\,
    \Delta\mathcal{V}^{q}_{n}
  \bigr]_{+},
  \label{eq:Sei}
\end{equation}
where \([\cdot]_{+}\) is applied to the change of reserves
\(\Delta\mathcal{V}^{q}_{n}\) and \(\Gamma\) denotes the pool’s fee tier.
After this, let
\(
  \bm{m}_{e_i}=(\,p_{\gamma_i},\,1\,)
\)
be the vector of token prices expressed in the chosen numéraire at the
epoch-closing price \(p_{\gamma_i}\).
Assuming efficient on-chain trading so that in-pool prices are close to
market prices, the market worth of the fees earned by the pool during
epoch \(e_i\) is
\begin{equation}
  F_{e_i}^{\Sigma}
  \;=\;
  \braket{\bm{m}_{e_i}, \bm S_{e_i}^{\Sigma}}.
  \label{eq:Fei}
\end{equation}

For a fixed TVL level \(\Sigma\), the cumulative pool reward across the
entire set of epochs $\bm{E}$ in the modeling period is
\begin{equation}
  F_{\bm{E}}^{\Sigma}
  \;=\;
  \sum_{i=1}^{K}
  \braket{\bm{m}_{e_i}, \bm S_{e_i}^{\Sigma}},
  \label{eq:FE}
\end{equation}
where $K$ is the total number of epochs.
The algorithm described in
Eqs.\eqref{eq:delta-V}–\eqref{eq:FE} is implemented in the backtest
tool presented in~\cite{ref_6}, which is also used as the primary
tool for the present research.

\subsubsection{Approach to LP Reward Calculation Within the Pool.}
At the beginning of epoch \(e_i\), the liquidity provider commits a capital
amount \(W_{e_i}^{\mathrm{LP}}\) to a restricted set of buckets in the
same partition \(\beta\) according to the $\tau$-reset rule.
Denote
  $\bm{W}_{e_i}^{\mathrm{LP}}
  =\bigl(w_1^{\mathrm{LP}},w_2^{\mathrm{LP}},\ldots,w_N^{\mathrm{LP}}\bigr)$,
  $\sum_{j=1}^{N} w_j^{\mathrm{LP}} = W_{e_i}^{\mathrm{LP}}$,
with
\(w_j^{\mathrm{LP}}\neq 0\) only for \(j\in[M-\tau,\,M+\tau]\), where
\(M\) is the reference-bucket index of epoch \(e_i\), hence
\begin{equation}
  \bm{W}_{e_i}^{\mathrm{LP}}
  =
  \bigl(
    0,\ldots,
    w_{M-\tau}^{\mathrm{LP}},\ldots,
    w_M^{\mathrm{LP}},\ldots,
    w_{M+\tau}^{\mathrm{LP}},
    \ldots,0
  \bigr).
\end{equation}

\noindent
Let
\(
  \bm{L}_{e_i}^{\mathrm{LP}}
  =\bigl(L_1^{\mathrm{LP}},L_2^{\mathrm{LP}},\ldots,L_N^{\mathrm{LP}}\bigr)
\)
be the corresponding liquidity vector fixed at the epoch's first price
\(p_{\gamma_{i-1}}\).
The LP’s share of liquidity in each bucket relative to the pool liquidity
\(\bm{L}_{e_i}^{\Sigma}\) is
\begin{equation}
  \bm{r}_{e_i}^{\mathrm{LP}}
  =
  \bigl(r_1^{\mathrm{LP}},r_2^{\mathrm{LP}},\ldots,r_N^{\mathrm{LP}}\bigr)
  =
  \left(
    \frac{L_1^{\mathrm{LP}}}{L_1^{\Sigma}},\,
    \frac{L_2^{\mathrm{LP}}}{L_2^{\Sigma}},\,
    \ldots,\,
    \frac{L_N^{\mathrm{LP}}}{L_N^{\Sigma}}
  \right),
  \label{eq:rLP}
\end{equation}
where $L_n^{\Sigma}>0\;\forall n$ \(\in [1,N] \).
The fee vector earned by the LP in epoch \(e_i\) is obtained by multiplying
the positive reserve increments from equation \eqref{eq:delta-V} by the liquidity shares:
\begin{equation}
  \bm S_{e_i}^{\mathrm{LP}}
  =
  \Gamma
  \sum_{q=1}^{q_{e_i}-1}
  \sum_{n=1}^{N}
  r_n^{\mathrm{LP}}\,
  \bigl[\Delta\mathcal{V}^{q}_{n}\bigr]_{+}.
\end{equation}
As in equation \eqref{eq:Fei}, the market worth in the chosen numéraire is
\begin{equation}
  F_{e_i}^{\mathrm{LP}}
  =
  \braket{\bm{m}_{e_i}, \bm S_{e_i}^{\mathrm{LP}}},
  \label{eq:_FLP}
\end{equation}
with \(\bm{m}_{e_i}=(p_{\gamma_i},\,1)\).
Aggregating over all epochs yields
\begin{equation}
  F_{\bm{E}}^{\mathrm{LP}}
  =
  \sum_{i=1}^{K}
  \braket{\bm{m}_{e_i}, \bm S_{e_i}^{\mathrm{LP}}},
\end{equation}
where \(K\) is the total number of epochs. The framework outlined above is deliberately general.  
A detailed treatment of the dynamics of the, as yet abstract, capital
variable \(W_{e_i}^{\mathrm{LP}}\) is provided in a subsequent subsection.

\subsubsection{State Assumptions and the LP-balance Equation.}
The quantities introduced above such as abstract capital allocation vector
\(\bm{W}_{e_i}^{\mathrm{LP}}\) refer to a fixed state.
This subsection formalizes how such a state is determined by its
underlying factors.

The modeling of LP remuneration proceeds in two stages.  
Stage 1 approximates the historical liquidity of the pool in the active ranges
of each epoch, producing
\(\bm{L}_{e_i}^{\Sigma}\).
Stage 2 computes the LP rewards according to
Eqs.\,\eqref{eq:rLP}–\eqref{eq:_FLP} for a given liquidity allocation vector
\(\bm{L}_{e_i}^{\mathrm{LP}}\).
Both stages are described below.

\paragraph{Stage 1: approximating the pool-liquidity profile.}
Following the parametric approach of \cite{ref_6}, a liquidity profile
($\bm{L}_{e_i}^{\Sigma}$) is fitted so that the model-implied pool fees approximate, to a prescribed
tolerance, the historical fees earned in each epoch of the modeling
period.
Previously we denoted the $\bm{V} =\{v_0,v_1,\ldots,v_H\}$ set of historical swap volumes in the chosen numéraire (e.g. USDC), now we write the aggregate volumes in epochs $\bm V_{\bm{E}}^{\text{hist}}
  =
  \bigl(v_{e_1}^{\text{hist}},\ldots,v_{e_K}^{\text{hist}}\bigr)$.
Given the pool LP fee tier \(\Gamma\), the corresponding historical fee
vector is
\[
  \bm F_{\bm{E}}^{\text{hist}}
  =
  \Gamma\,\bm V_{\bm{E}}^{\text{hist}}
  =
  \bigl(f_{e_1}^{\text{hist}},\ldots,f_{e_K}^{\text{hist}}\bigr).
\]
The approximate liquidity is then optimized so that the model fees match
\(\bm F_{\bm{E}}^{\text{hist}}\) epoch by epoch within the specified
tolerance.

Let the TVL level \(\Sigma\) be allocated across the \(N\) buckets of the
partition \(\beta\) during epoch \(e_i\) according to the weight vector $\bm{\alpha}_{e_i}^{\Sigma}
  =\bigl(\alpha_1^{\Sigma},\alpha_2^{\Sigma},\ldots,\alpha_N^{\Sigma}\bigr)$,
  $\sum_{n=1}^{N} \alpha_n^{\Sigma}=1$,
  $\alpha_n^{\Sigma}>0$.
The capital assigned to the bucket \(n\) is \(w_n^{\Sigma}=\alpha_n^{\Sigma}\Sigma\).
Following the Gaussian–curve approach in \cite{ref_6}, the weight vector
is generated by a normal density with mean \(\mu_{e_i}\) and sigma
\(\sigma_{e_i}\):
\begin{equation}
  \bm{\alpha}_{e_i}^{\Sigma}
  \;=\;
  f_{\mathcal{G}}\bigl(\mu_{e_i},\sigma_{e_i}\bigr),
\end{equation}
where \(f_{\mathcal{G}}\) maps the parameters
\((\mu_{e_i},\sigma_{e_i})\) to a discrete Gaussian weight vector over
the buckets.
These parameters are calibrated so that the model fee
\({F}_{e_i}^{\Sigma}\) matches the historical fee vector
\({f}_{e_i}^{\text{hist}}\) defined in the previous subsection.
Calibration proceeds by iterating on the Gaussian parameters
\(\mu_{e_i}\) and \(\sigma_{e_i}\) over two common grids,
  $\mu_{e_i}\in[\mu_{\min},\mu_{\max}]$,
  $\sigma_{e_i}\in[\sigma_{\min},\sigma_{\max}]$,
identical for all epochs.
For every candidate pair \((\mu_{e_i},\sigma_{e_i})\) the following
pipeline is executed:

\begin{equation}
\begin{gathered}
    \bm\alpha_{e_i}^{\Sigma}
      =f_{\mathcal{G}}(\mu_{e_i},\sigma_{e_i})
  \;\longrightarrow\;
  \bm{W}_{e_i}^{\Sigma}\bigl(\Sigma,\bm\alpha_{e_i}^{\Sigma}\bigr)
  \;\longrightarrow\;
  \bm{L}_{e_i}^{\Sigma}\bigl(\bm{W}_{e_i}^{\Sigma},p_{\gamma_{i-1}},\beta\bigr)
  \;\longrightarrow\;
    \\[4pt]
    \longrightarrow\;
  \bm S_{e_i}^{\Sigma}
  \;\longrightarrow\;
  F_{e_i}^{\Sigma}=f_{e_i}^{m\Sigma}(\mu_{e_i},\sigma_{e_i}),
  \label{eq:schPool}
\end{gathered}
\end{equation}
where \(f_{e_i}^{m\Sigma}(\mu_{e_i},\sigma_{e_i})\) denotes the model fee
(in the chosen numéraire) earned by the pool in epoch \(e_i\) when the
liquidity distribution follows the Gaussian profile with parameters
\(\mu_{e_i}\) and \(\sigma_{e_i}\). The calibration terminates at the first parameter pair
\(\bigl(\mu_{e_i}^{\ast},\sigma_{e_i}^{\ast}\bigr)\)
for which the discrepancy between the model fee and the
historical benchmark falls within the prescribed tolerance:
\begin{equation}
  \bigl(\mu_{e_i}^{\ast},\sigma_{e_i}^{\ast}\bigr)
  =
  \arg\min_{(\mu_{e_i},\sigma_{e_i})}
  \bigl|\,f_{e_i}^{m\Sigma}(\mu_{e_i},\sigma_{e_i})-f_{e_i}^{\text{hist}}\,\bigr|.
  \label{eq:optPoolT}
\end{equation}

Thus, the calibrated pair
\(\bigl(\mu_{e_i}^{\ast},\sigma_{e_i}^{\ast}\bigr)\)
achieves the closest match between the model fee and the
observed pool reward in epoch \(e_i\).
Theoretically, the first chosen pair of parameters might not be the only possible one, but
this can occur only in two cases:
\begin{enumerate}
    \item The historical prices within the active ranges of the epoch are
          nearly uniformly distributed; or
    \item The prices are concentrated in the neighborhood of the current reference bucket.
\end{enumerate}
In either situation, selecting the first admissible parameter pair does not affect the resulting shape of the optimal LP strategy.

\paragraph{Stage 2: LP balance equation.}
\label{sec:st2lpb}
The shares \(\bm{r}_{e_i}^{\mathrm{LP}}\) are derived from the liquidity
vector \(\bm{L}_{e_i}^{\mathrm{LP}}\) and an epoch-specific weight vector
  $\bm{\alpha}_{e_i}^{\mathrm{LP}}
  =
  \bigl(\alpha_1^{\mathrm{LP}},\alpha_2^{\mathrm{LP}},\ldots,
         \alpha_N^{\mathrm{LP}}\bigr)$,
  $\sum_{n=1}^{N}\alpha_n^{\mathrm{LP}}=1$,
  $\alpha_n^{\mathrm{LP}}\ge 0$,
where \(w_n^{\mathrm{LP}}=\alpha_n^{\mathrm{LP}}\,W_{e_i}^{\mathrm{LP}}\).
The $\tau$-reset strategy \(\varphi^{\tau}\) imposes sparsity on
\(\bm{\alpha}_{e_i}^{\mathrm{LP}}\): positive weights appear only in the
\((2\tau+1)\) buckets centered on the reference index \(M\). For the uniform benchmark strategy, all non-zero elements of $\bm{\alpha}_{e_i}^{\mathrm{LP}}$ are equal to $\frac{1}{2\tau + 1}$.  Hence
\begin{equation}
  \bm{\alpha}_{e_i}^{\mathrm{LP}}
  \;=\;
  \bigl(
    \underbrace{0,\ldots,0}_{M-\tau-1},
    \;\alpha_{M-\tau}^{\mathrm{LP}},\ldots,\alpha_{M+\tau}^{\mathrm{LP}},
    \underbrace{0,\ldots,0}_{N-M-\tau}
  \bigr).
  \label{eq:alpLP}
\end{equation}

Analogous to the pool–side approximation in equation \eqref{eq:schPool}, the chain of
dependencies that generates LP fees in epoch \(e_i\) under the
\(\tau\)-reset strategy \(\varphi^{\tau}\) can be summarized schematically as
\begin{equation}
\begin{gathered}
    \bm{\alpha}_{e_i}^{\mathrm{LP}}(\varphi^{\tau})
  \;\longrightarrow\;
  \bm{W}_{e_i}^{\mathrm{LP}}\!\bigl(W_{e_i}^{\mathrm{LP}},
                                   \bm{\alpha}_{e_i}^{\mathrm{LP}}\bigr)
  \;\longrightarrow\;
  \bm{L}_{e_i}^{\mathrm{LP}}\!\bigl(\bm{W}_{e_i}^{\mathrm{LP}},
                                   p_{\gamma_{i-1}},\beta\bigr)
  \;\longrightarrow\;
  \\[4pt]
    \longrightarrow\;
  \bm{r}_{e_i}^{\mathrm{LP}}\!\Bigl(
      \bm{L}_{e_i}^{\mathrm{LP}},\,
      \bm{L}_{e_i}^{\Sigma}\bigl(
          \bm{W}_{e_i}^{\Sigma}\bigl(\Sigma,
            f_{\mathcal{G}}(\mu_{e_i}^{\ast},\sigma_{e_i}^{\ast})\bigr),
          p_{\gamma_{i-1}},\beta\bigr)\Bigr)
  \;\longrightarrow\;
    \\[4pt]
    \longrightarrow\;
  \bm S_{e_i}^{\mathrm{LP}}
  \;\longrightarrow\;
  F_{e_i}^{\mathrm{LP}}
  = f_{e_i}^{\mathrm{LP}}(\varphi^{\tau}).
  \label{eq:schLPst}
\end{gathered}
\end{equation}
Here \(f_{e_i}^{\mathrm{LP}}(\varphi^{\tau})\) is the value (in the chosen
numéraire) of the fee stream paid to the LP for the supply of liquidity under strategy
\(\varphi^{\tau}\), given the approximated pool profile characterized by
\((\mu_{e_i}^{\ast},\sigma_{e_i}^{\ast})\).
Assuming the approximated profile remains fixed throughout epoch \(e_i\) and
that in-pool prices track their market counterparts, the end-of-epoch value
of the LP’s position is
\begin{equation}
    \label{eq:WB21}
  W_{\mathrm{end},e_i}^{\mathrm{LP}}
  \;=\;
  \braket{
  \bm{m}_{e_i},
  \sum_{n=1}^{N}
     \mathcal{V}_n(L_n^{\Sigma},p_{\gamma_i},B_n)
  r_n^{\mathrm{LP}}}
  \;+\;
  f_{e_i}^{\mathrm{LP}}(\varphi^{\tau}),
\end{equation}
where \(\bm{m}_{e_i}=(p_{\gamma_i},1)\) and the notation follows
\eqref{eq:Fei}–\eqref{eq:FE}. Equation \eqref{eq:WB21} expresses the
LP capital after accounting for both impermanent loss and accrued fees.
It should be noted that when \(K>1\) epochs, the researcher can do so:
\begin{enumerate}
    \item Reinvest \(W_{\mathrm{end},e_i}^{\mathrm{LP}}\) at the start of
          epoch \(e_{i+1}\), thus compounding the fee income.
    \item Refrain from reinvesting fees, leaving the working capital exposed to
          impermanent loss and subject to \(\tau\)-reset execution costs, as illustrated
          in part \,5.4 of~\cite{ref_6}; or
    \item Keep vector $\bm{W}_{e_i}^{\mathrm{LP}}$ constant across epochs, implicitly assuming an
          unlimited supply of deployable liquidity.
\end{enumerate}

\subsubsection{Optimal strategy selection.}
For each epoch \(e_i\) the search for an optimal LP strategy  
\(\varphi_{e_i}^{\text{OPT}}\) is performed over a finite family of
\(\tau\)-symmetric capital allocations.  
Let \(W_{e_i}^{\mathrm{LP}}\) be the capital to be deployed and
\(\bm{\alpha}_{e_i}^{\mathrm{LP}}\) the associated weight vector
(cf.\,\eqref{eq:alpLP}); $\tau$-symmetry requires
  $\alpha_{M-\tau+k}^{\mathrm{LP}}
  \;=\;
  \alpha_{M+\tau-k}^{\mathrm{LP}}$,
  \( k \in [0,\tau-1] \), 
where \(M\) is the reference bucket of epoch \(e_i\).
Let
  $\mathbf{S}_{e_i}^{\tau}
  = \bigl\{\varphi_{1}^{\tau},\ \varphi_{2}^{\tau},\ \ldots,\
          \varphi_{N_{\mathcal{S}}}^{\tau}\bigr\}$
be the finite family of random \(\tau\)-strategies, where
\(N_{\mathcal{S}}\) is the number of candidate strategies examined.
The optimal strategy maximizes the modeled fee income
\(f_{e_i}^{\mathrm{LP}}(\varphi^{\tau})\) defined in
\eqref{eq:schLPst}:
\begin{equation}
  \varphi_{e_i}^{\text{OPT}}
  =
  \arg\max_{\varphi^{\tau}\in\mathbf{S}_{e_i}^{\tau}}
  f_{e_i}^{\mathrm{LP}}\!\bigl(\varphi^{\tau}\bigr).
  \label{eq:optLPs}
\end{equation}

The present study focuses on LP rewards model and does not
address the complete liquidity management strategy; this larger problem is left
for our next research.  
Consequently, the buy and hold strategy is not employed as a benchmark in the current study.
The set of optimal epoch strategies $\varphi_{e_i}^{\text{OPT}}$ serves
as the target output to train the machine learning model described in the
next section.

\subsection{Machine Learning Architecture}
\label{section:ml-architecture}
The machine learning (ML) model is used to infer the optimal liquidity allocation
strategy from the prevailing market state.
Its input features are quantitative descriptors of market conditions at the
decision time; the model outputs the weight vector associated with the
optimal LP strategy.
Training is carried out on historical data that comprise, for each epoch of
the modeling period, the optimal strategies obtained by solving the
problem \eqref{eq:optLPs}; the set of features reflects the market conditions at the
beginning of the corresponding epoch.
A detailed specification of the model architecture, the feature
engineering pipeline and the hyperparameter configuration is provided in
Section \ref{section:Experiments}.

\subsubsection{ML model architecture.} 
\begin{figure*}
    \centering
    \includegraphics[width=0.5\textwidth]{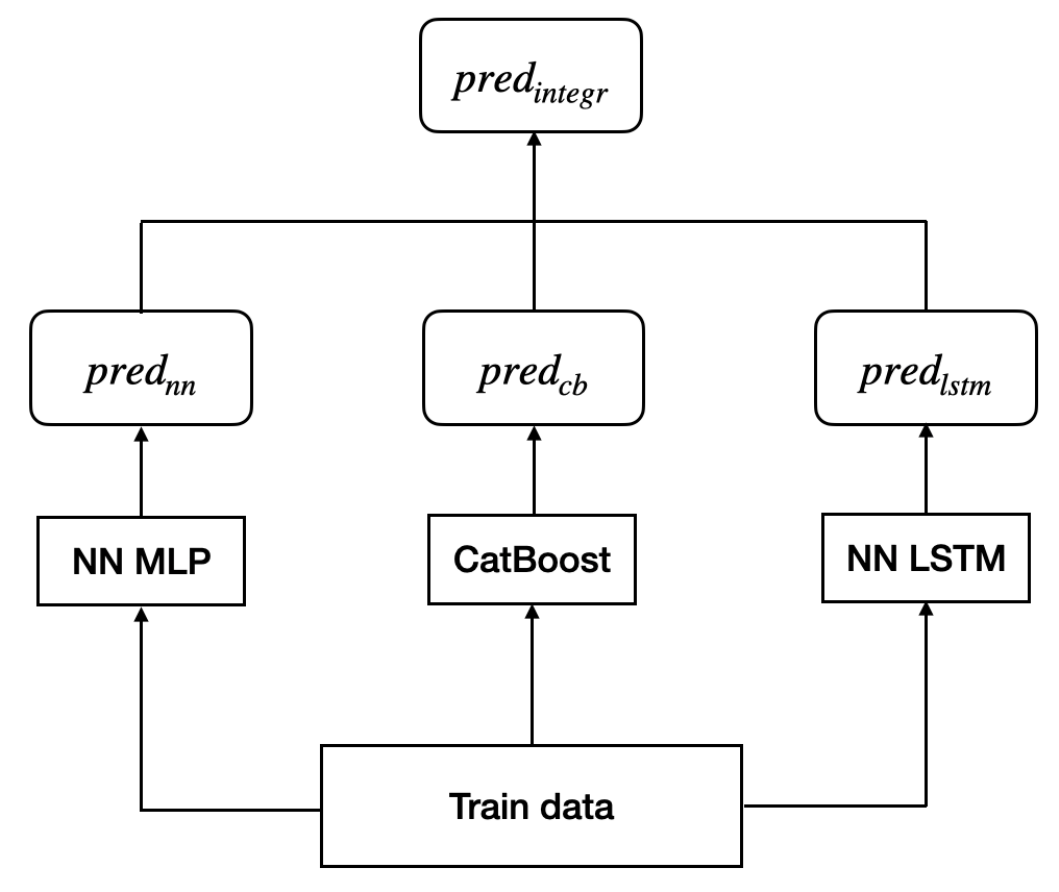}
    \smallskip
    \captionsetup{justification=centering}
    \caption{Overall architecture of the integral model.}
    \label{figure:F2}
\end{figure*}

For each epoch \(e_i\), the optimal LP strategy is represented by the weight
vector
\(
  \bm{\alpha}_{e_i}^{\mathrm{LP}}\bigl(\varphi_{e_i}^{\mathrm{OPT}}\bigr)
  \in\mathbb{R}^{N}_{+}
\),
whose non-zero entries \(2\tau+1\) occupy the buckets
\(j\in[M-\tau,\,M+\tau]\).
Collect these entries in the vector $\bm{\rho}_{e_i}^{\mathrm{LP}}
  =
  (\alpha_{M-\tau}^{\mathrm{LP}},\ldots,
    \alpha_{M}^{\mathrm{LP}},\ldots,
    \alpha_{M+\tau}^{\mathrm{LP}})
  \in\mathbb{R}^{2\tau+1}_{+}$,
and exploit the \(\tau\)-symmetry
\(\alpha_{M-\tau+k}^{\mathrm{LP}}=\alpha_{M+\tau-k}^{\mathrm{LP}}\)
for \(k\in[0,\,\tau-1]\) to define the reduced representation $\hat{\bm{\rho}}_{e_i}^{\mathrm{LP}}
  =
  (\alpha_{M}^{\mathrm{LP}},
    \,2\alpha_{M+1}^{\mathrm{LP}},
    \,\ldots,
    \,2\alpha_{M+\tau}^{\mathrm{LP}})
  \in\mathbb{R}_{+}^{\tau+1}$.
Stacking the epoch vectors yields the target vector
\[
  Y
  =
  \bigl(\hat{\bm{\rho}}_{e_1}^{\mathrm{LP}}\;
        \hat{\bm{\rho}}_{e_2}^{\mathrm{LP}}\;
        \ldots\;
        \hat{\bm{\rho}}_{e_K}^{\mathrm{LP}}\bigr)^{\!\top}
  \in \mathbb{R}^{K\times(\tau+1)},
\]
where \(K\) is the number of epochs.
Each epoch is characterized by a feature vector
\(
  \bm{\psi}_{e_i}\in\mathbb{R}^{\theta}
\),
\(\theta\) being the feature dimension, and the feature matrix is
\[
  X
  =
  \begin{pmatrix}
    \bm{\psi}_{e_1}\\
    \bm{\psi}_{e_2}\\
    \vdots\\
    \bm{\psi}_{e_K}
  \end{pmatrix}
  \in\mathbb{R}^{K\times\theta}.
\]
The learning task is to approximate the mapping
\(Y = f(X)\) using a supervised ML algorithm.
The dataset is partitioned 80/20 into training and test subsets.:
\[
  X_{\text{Train}}\in\mathbb{R}^{k_{\text{Train}}\times\theta},
  \quad
  Y_{\text{Train}}\in\mathbb{R}^{k_{\text{Train}}\times(\tau+1)},
\]
\[
  X_{\text{Test}}\in\mathbb{R}^{k_{\text{Test}}\times\theta},
  \quad
  Y_{\text{Test}}\in\mathbb{R}^{k_{\text{Test}}\times(\tau+1)},
\]
with \(k_{\text{Train}}=0.8K\) and \(k_{\text{Test}}=0.2K\).
An integrated ensemble model is used, which combines a multilayer perceptron (MLP), gradient-boosted CatBoost decision trees, and a long-short-term memory (LSTM) network. The models are trained on $(X_{\text{Train}}, Y_{\text{Train}})$, tested on $(X_{\text{Test}}, Y_{\text{Test}})$, and subsequently applied to an \textit{out-of-time} (OOT) period that simulates real-world deployment $(X_{\text{OOT}}, Y_{\text{OOT}})$. One of the simplest, yet most effective, model fusion techniques is weighted linear prediction averaging:
\[
  {pred}_{\text{integr}}
  =
  w_1\,{pred}_{\text{MLP}}
  + w_2\,{pred}_{\text{CB}}
  + w_3\,{pred}_{\text{LSTM}},
\]
where \(w_1,w_2,w_3>0\) and \(w_1+w_2+w_3=1\).
The weights are chosen to minimize the error on the train date.
The ensemble leverages the architectural heterogeneity of MLPs, CatBoost  and LSTMs to capture a broad spectrum of signal types—from static tabular relationships to non-linear interactions and latent compositional structures—within a unified learning framework.
A schematic of the ensemble is shown in Fig. \ref{figure:F2}.

Full implementation details - network hyperparameters, feature
engineering, and evaluation metrics - are provided in
Section \ref{section:Experiments}.

\subsection{Reward Modeling Approaches}
\label{sub:RApr}
Four principal approaches can be distinguished when modeling fee accrual
in DEX liquidity pools:
\begin{enumerate}
  \item Pool-level reward modeling: reproducing the aggregate fees earned by the entire pool.
  \item Single-LP reward in an isolated theoretical pool: evaluate the fees of a lone liquidity provider in a hypothetical pool that contains only the liquidity of that provider.
  \item Single-LP reward as a share of historical pool liquidity: calculate the fees of an LP relative to the actual historical liquidity present in the pool.
  \item Single-LP reward in a pool augmented beyond its historical liquidity: computing an LP’s fees after injecting additional model-generated liquidity on top of the historical pool state.
\end{enumerate}
Each approach is analyzed in detail, together with its advantages, limitations, and key modeling assumptions.

\begin{figure*}[!b]
    \centering
    \includegraphics[width=0.9\textwidth]{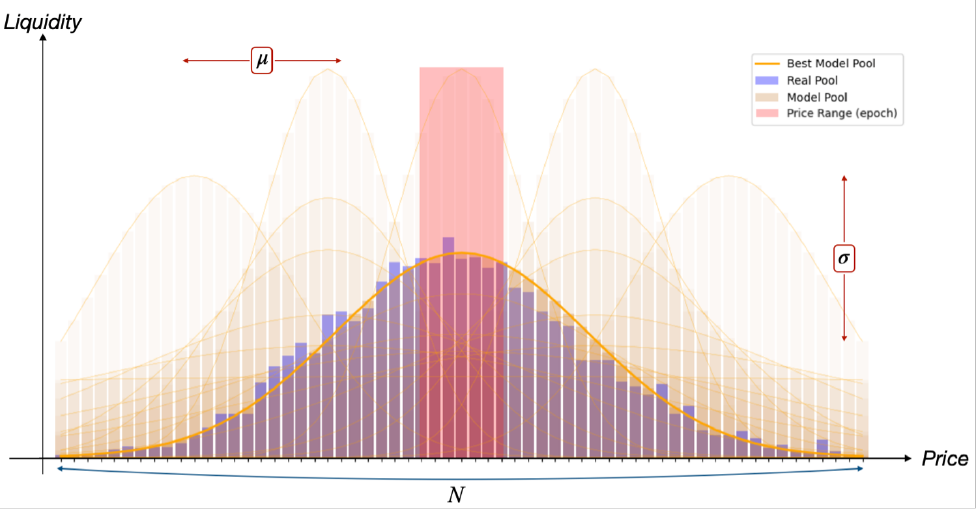}
    \smallskip
    \captionsetup{justification=centering}
    \caption{Calibration procedure for the approximating liquidity profile,
    specified by the weight vector
    \(
      \bm{\alpha}_{e_i}^{\Sigma}
      = f_{\mathcal{G}}\!\bigl(\mu_{e_i},\sigma_{e_i}\bigr)
    \)
    within epoch \(e_i\).}
    \label{figure:F3}
\end{figure*}

\subsubsection{Approach 1: pool-level reward modeling.}\label{sec:approach-1}
The first approach focuses exclusively on reproducing the historical fee income of the pool.
Given a fixed average TVL marked as \(\Sigma\), different liquidity profiles
(concentrated to varying degrees) can generate different fee levels in
the same set of active ranges, because the trade volume required to move
the price along the prescribed path \(\bm{P}\) depends on how liquidity is
distributed within the pool.

By selecting the shape parameters of the Gaussian liquidity profile according to equation \eqref{eq:optPoolT}, the modeled pool reward can be brought
within a prescribed tolerance of the historical reward realized.  
This procedure establishes a direct link to the empirical fee level and
thus supplies a natural benchmark for the calibration of the synthetic
liquidity profile.

\begin{figure*}[!t]
    \centering
    \includegraphics[width=1\textwidth]{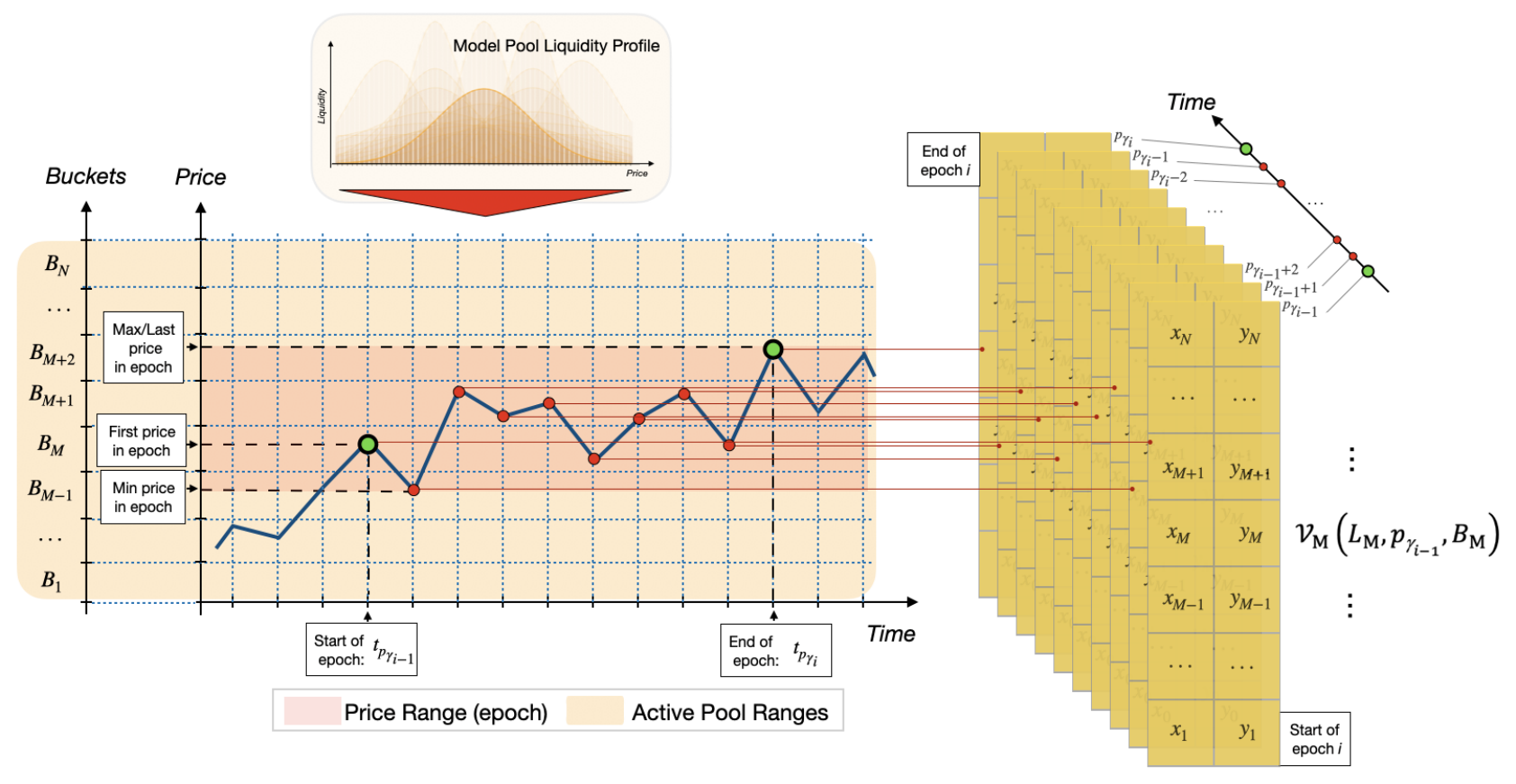}
    \smallskip
    \captionsetup{justification=centering}
    \caption{Liquidity state in each bucket under the parametric approximation \\ of the pool’s historical liquidity.}
    \label{figure:F4}
\end{figure*}

\begin{figure*}[!b]
    \centering
    \includegraphics[width=0.9\textwidth]{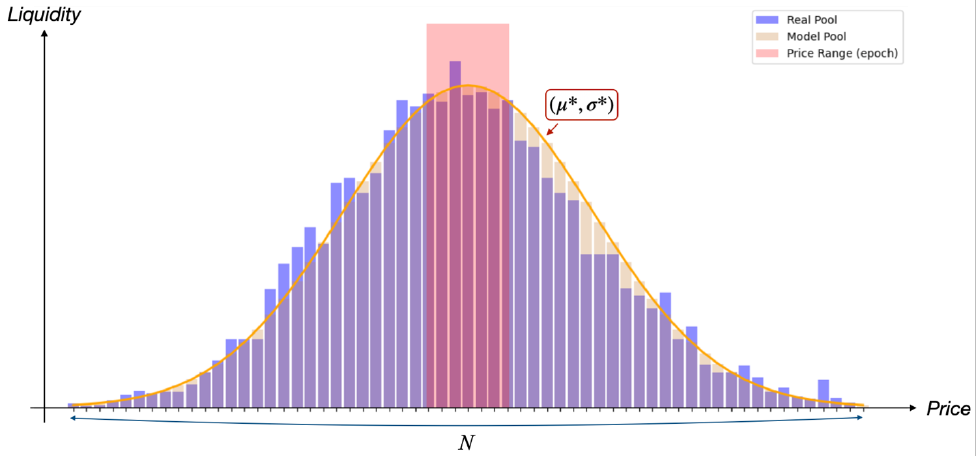}
    \smallskip
    \captionsetup{justification=centering}
    \caption{Target state for Approach 1: the parametric liquidity profile characterized by the weight vector
\(
  \bm{\alpha}_{e_i}^{\Sigma}
  = f_{\mathcal{G}}\!\bigl(\mu_{e_i}^{\ast},\sigma_{e_i}^{\ast}\bigr)
\)
within epoch \(e_i\).
}
    \label{figure:F5}
\end{figure*}

\paragraph{Methodological details.}
Following the Section \ref{section:Pr_St}, the liquidity profile of the pool is approximated within each epoch \(e_i\in\bm{E}\)
by sweeping the two Gaussian parameters
\(\mu_{e_i}\) and \(\sigma_{e_i}\) as illustrated in Figure \ref{figure:F3}.
The outer boundaries \(p_{a_1}\) and \(p_{b_N}\) of the partition
\(\beta\) are chosen to coincide with the empirical limits of liquidity
deployment observed over the historical period, while the number of
buckets \(N\) is selected so that the implied bucket width
\(d=p_{b_i}-p_{a_i}\) does not fall below the minimum tick size of the
modeled pool.
The average TVL level \(\Sigma\) is specified in accordance with
Section \ref{section:KCED}.

Once the Gaussian parameters have been swept, the liquidity profile
\(\bm{L}_{e_i}^{\Sigma}\) for each epoch \(e_i\) is fixed by solving
Eqs.\,\eqref{eq:delta-V}–\eqref{eq:Fei} over the price path
\(\bm{P}_{e_i}\).
The peak of the Gaussian curve is located at the mean price of the epoch
\(\bar{\bm{P}}_{e_i}\).
Figure \ref{figure:F4} illustrates, for a representative choice of
\(f_{\mathcal{G}}(\mu_{e_i},\sigma_{e_i})\), how the reserve state in each
bucket evolves as historical prices move.
Figure \ref{figure:F5} shows an example of the calibrated profile
\(\bm{L}_{e_i}^{\Sigma}\) obtained with the pair of parameters
\((\mu_{e_i}^{\ast},\sigma_{e_i}^{\ast})\) for which the model fee
\(f_{e_i}^{m\Sigma}(\mu_{e_i}^{\ast},\sigma_{e_i}^{\ast})\)
matches the historical fee \(f_{e_i}^{\text{hist}}\) within the researcher's tolerance.

It should be emphasized that this methodology constitutes the core of \cite{ref_6}; in the present work it is retained as an integral component of the remaining LP reward modeling approaches.

\begin{figure*}[!t]
    \centering
    \includegraphics[width=1\textwidth]{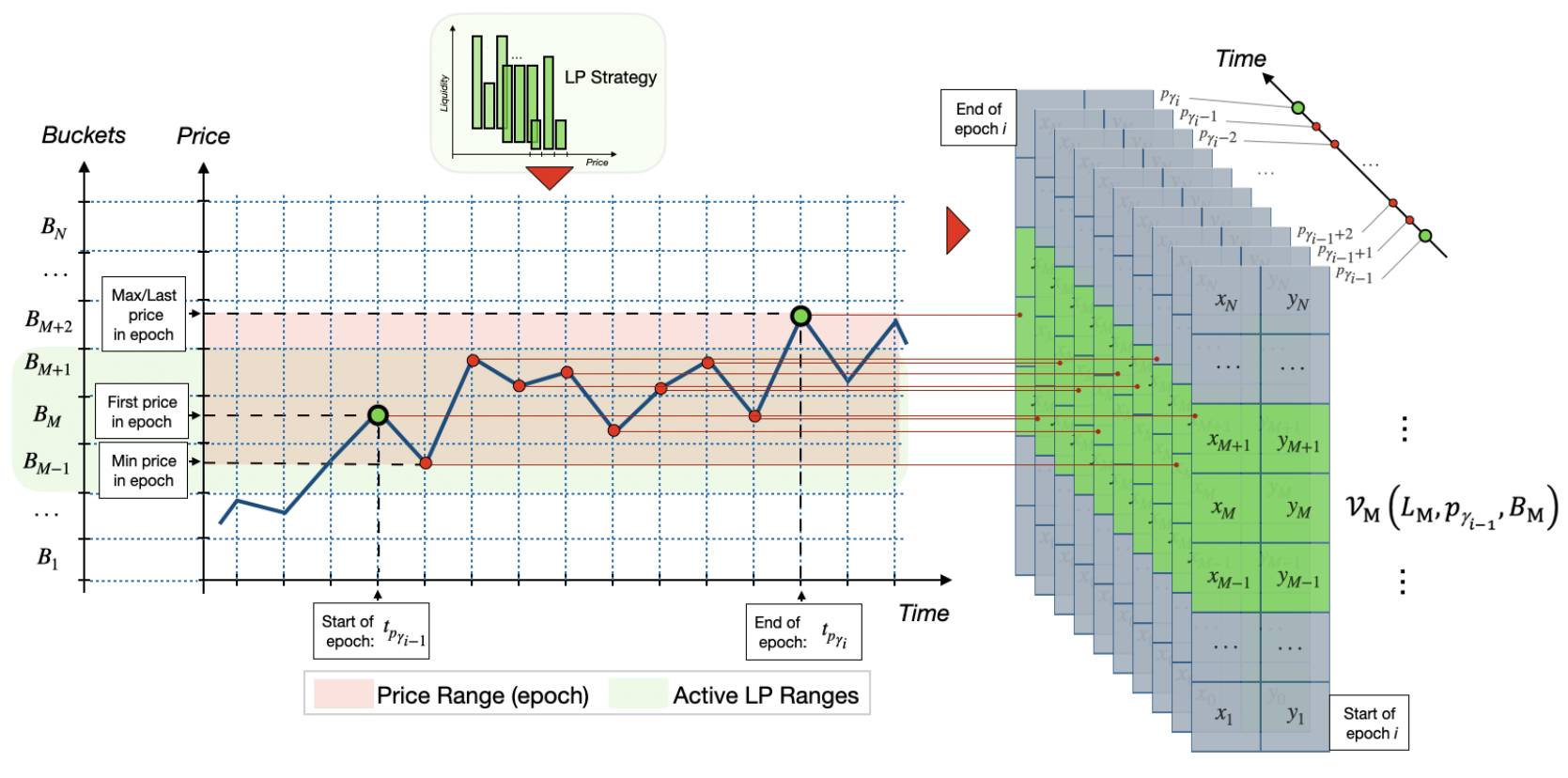}
    \smallskip
    \captionsetup{justification=centering}
    \caption{Liquidity state in each pool bucket under a random LP strategy
    \(\varphi^{\tau=1}\).
    }
    \label{figure:F6}
\end{figure*}

\subsubsection{Approach 2: single-LP rewards in an isolated, theoretical pool.}
This case, essentially a \emph{hypothetical-pool model} - considers an
individual LP as the \emph{only} liquidity provider.
Unlike Approach 1, no natural fee benchmark exists because the simulation
is decoupled from the historical trade volumes of any specific pool.  
The optimal strategy is therefore the strategy that maximizes the \emph{model}
fee level, without reference to the Gaussian parameters
\((\mu_{e_i},\sigma_{e_i})\) used to approximate historical liquidity.

Although not implemented in the empirical part of the present study, the
approach is included here for completeness.
Its optimization problem mirrors the Eqs. \eqref{eq:optPoolT}, \eqref{eq:optLPs} reads as
\begin{equation}
  \varphi_{e_i}^{\mathrm{OPT}}
  =
  \arg\max_{\varphi^{\tau}\in\mathbf{S}_{e_i}^{\tau}}
  f_{e_i}^{m\Sigma}\!\bigl(\varphi^{\tau}\bigr),
\end{equation}
where the LP strategy \(\varphi^{\tau}\) prescribes liquidity
allocation, and the pool liquidity vector depends only on the LP’s own
capital,
\(
  \bm{W}_{e_i}^{\Sigma}
  =
  \bm{W}_{e_i}^{\mathrm{LP}}\!\bigl(W_{e_i}^{\mathrm{LP}},
                                    \bm{\alpha}_{e_i}^{\mathrm{LP}}(\varphi^{\tau})\bigr).
\)

In essence, this is a special, single-provider case of Approach 1; yet it
is highly stylized, because only in a theoretical setting would the price
path mirror the historical trajectory of a real pool in the presence of an
arbitrary liquidity surface supplied by a lone LP.

\paragraph{Methodological details.}
Adopting the framework of Section \ref{section:Pr_St}, the boundaries of the outer bucket \(p_{a_1}\) and \(p_{b_N}\) are first set to span the admissible
liquidity range in the hypothetical pool.
The researcher then selects \(N\) such that the width of the bucket
$d$ does not fall below the minimum tick size of the pool.
Because the sole liquidity provider is the LP, the TVL pool equals the LP
capital \(\Sigma=W^{\mathrm{LP}}\).

For each epoch \(e_i\) and the fixed price path \(\bm{P}_{e_i}\), random
$\tau$–symmetric strategies are drawn from
\(\mathbf{S}_{e_i}^{\tau}\).
The optimal strategy \(\varphi_{e_i}^{\mathrm{OPT}}\) maximizes
the income from the LP's fee and, by construction, coincides with the pool model fee: \(\bm{L}_{e_i}^{\Sigma}\) is therefore concentrated in
\(2\tau+1\) ranges, mirroring \emph{Stage 2} from
Section \ref{section:Pr_St} but at the \emph{pool} level rather than
for an individual LP.
Figure \ref{figure:F6} illustrates how the liquidity state in each bucket changes
while historical prices evolve under one realization of the
\(\varphi^{\tau}\) strategy.

To conclude, although the isolated pool framework is conceptually valid, its practical relevance is limited; therefore, the foregoing description is sufficient, and the paper now turns to the core approach of this research.

\begin{figure*}[!t]
    \centering
    \includegraphics[width=0.9\textwidth]{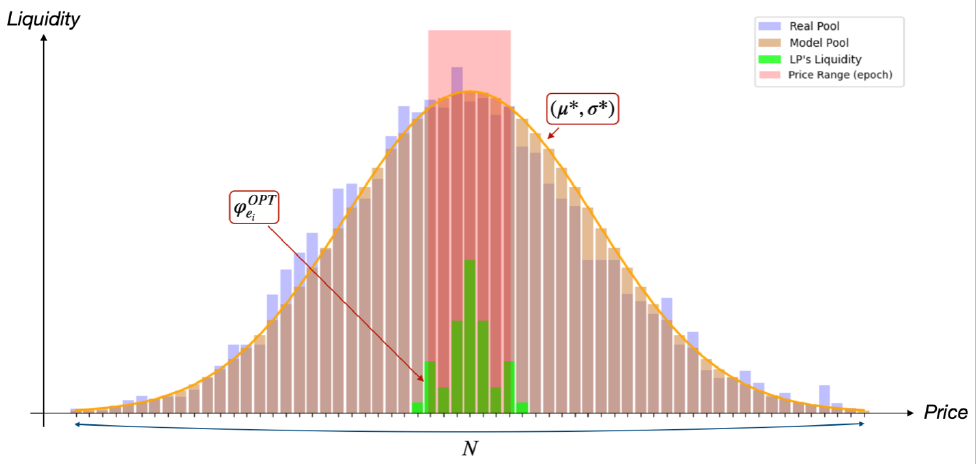}
    \smallskip
    \captionsetup{justification=centering}
    \caption{Target state for Approach 3: the calibrated liquidity profile \\
    \(
      \bm{\alpha}_{e_i}^{\Sigma}
      = f_{\mathcal{G}}\!\bigl(\mu_{e_i}^{\ast},\sigma_{e_i}^{\ast}\bigr)
    \)
    together with the corresponding optimal LP strategy
    \(\varphi_{e_i}^{\mathrm{OPT}}\) for epoch \(e_i\).
    }
    \label{figure:F7}
\end{figure*}

\subsubsection{Approach 3: LP rewards as a share of historical pool liquidity.}
\label{sec:approach-3}
This and the next approaches address the optimization problem
in \eqref{eq:optLPs}.  
The distinguishing feature of the present approach is its two-stage structure as described in Section \ref{section:Pr_St}:

\begin{enumerate}
  \item  
        \emph{Stage 1.} For each epoch \(e_i\) the liquidity profile of the pool is first calibrated by solving~\eqref{eq:optPoolT}, resulting in the pair of parameters \((\mu_{e_i}^{\ast},\sigma_{e_i}^{\ast})\) such that the model fee
        \(f_{e_i}^{m\Sigma}(\mu_{e_i}^{\ast},\sigma_{e_i}^{\ast})\)
        matches the historical fee \(f_{e_i}^{\text{hist}}\).
        This step exactly replicates Approach 1 and produces the approximate liquidity of the pool \(\bm{L}_{e_i}^{\Sigma}\).
  \item \emph{Stage 2.} 
        Conditional on \(\bm{L}_{e_i}^{\Sigma}\), the optimal LP strategy \(\varphi_{e_i}^{\mathrm{OPT}}\) is selected as the maximizer of
        \(f_{e_i}^{\mathrm{LP}}(\varphi^{\tau})\); cf.\,\eqref{eq:optLPs}.
\end{enumerate}

In this setting, the model LP reward is determined by the share vector
\(\bm{r}_{e_i}^{\mathrm{LP}}\) that quantifies the proportion of liquidity
held by the LP in each active price range of epoch \(e_i\)
(Fig. \ref{figure:F7}).
A practical implication is that the historical pool reward constitutes an
\emph{upper bound} on the attainable LP reward: the LP cannot earn more
than the fee potential embedded in the approximate pool profile specified
by \((\mu_{e_i}^{\ast},\sigma_{e_i}^{\ast})\).

\begin{figure*}[!t]
    \centering
    \includegraphics[width=1\textwidth]{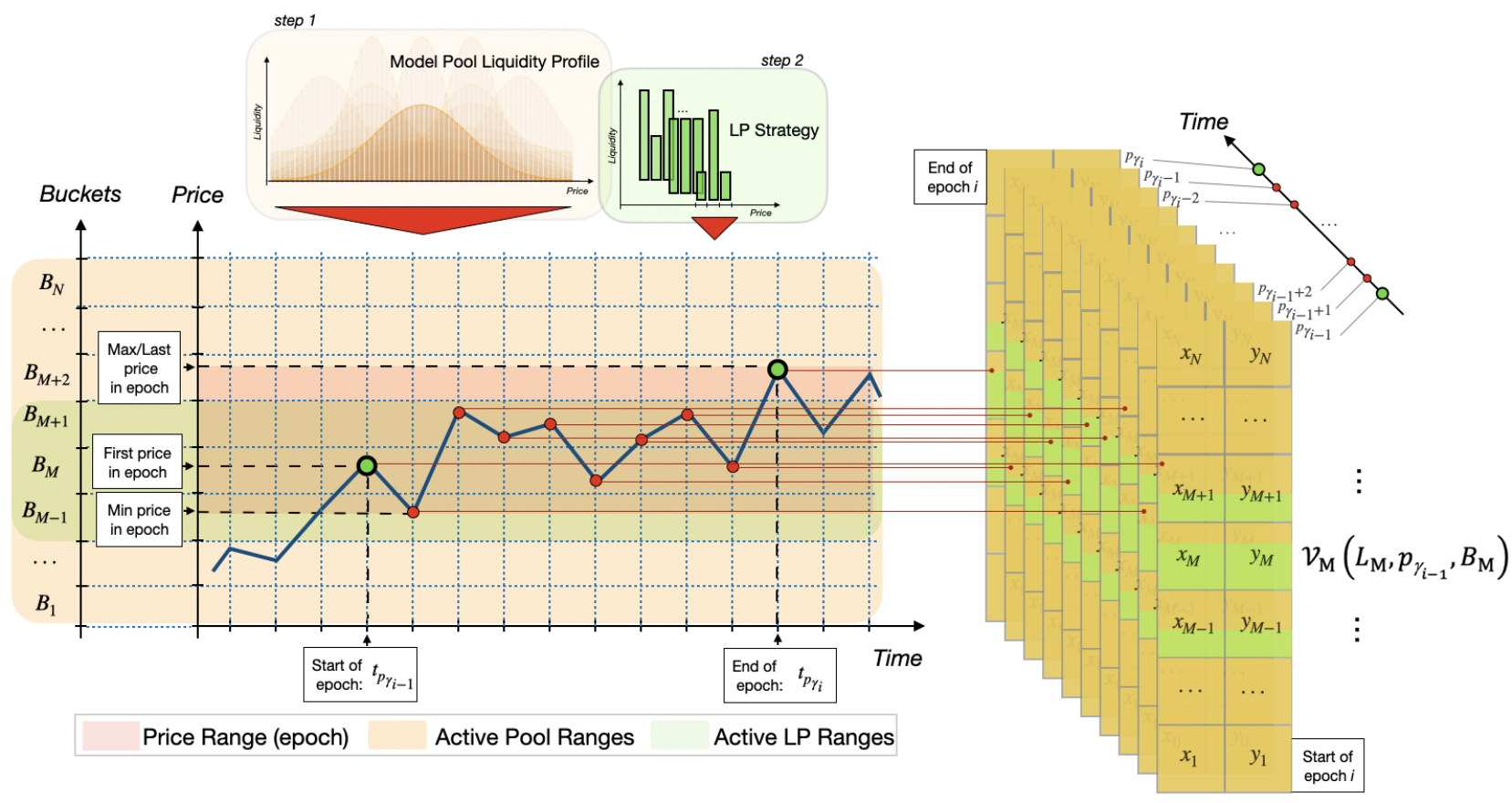}
    \smallskip
    \captionsetup{justification=centering}
    \caption{Visualization of the composite liquidity profile
    \(\bm{L}_{e_i}^{\Sigma}\), calibrated with
    \((\mu_{e_i}^{\ast},\sigma_{e_i}^{\ast})\) and augmented by the optimal
    LP allocation \(\varphi_{e_i}^{\mathrm{OPT}}\in\mathbf{S}_{e_i}^{\tau}\)
    for the case \(\tau=1\) in epoch \(e_i\).
    }
    \label{figure:F8}
\end{figure*}

\paragraph{Methodological details.}
As in Approach 1, the partition \(\beta\) and the TVL level \(\Sigma\) are fixed first.  
Given the price path \(\bm{P}_{e_i}\), the liquidity surface of the pool for epoch \(e_i\) is calibrated by sweeping the Gaussian parameters \(\mu_{e_i}\) and \(\sigma_{e_i}\) until the model fee \(f_{e_i}^{m\Sigma}(\mu_{e_i}^{\ast},\sigma_{e_i}^{\ast})\) matches the historical fee \(f_{e_i}^{\text{hist}}\) within the preset tolerance.  

With the initial LP capital \(W_{e_i}^{\mathrm{LP}}\) and the same price subset \(\bm{P}_{e_i}\), a random search of the strategy family \(\mathbf{S}_{e_i}^{\tau}\) (for the fixed parameter \(\tau\)) produces the optimal strategy \(\varphi_{e_i}^{\mathrm{OPT}}\) that maximizes the LP reward \(f_{e_i}^{\mathrm{LP}}\) within the liquidity of the approximated pool, cf.\eqref{eq:optLPs}.  
Figure \ref{figure:F8} illustrates how the liquidity state in each bucket changes while historical prices evolve, and \(\bm{L}_{e_i}^{\Sigma}\) in the \(2\tau+1\) buckets of \(\beta\) contains, among other components, the LP-position in the form of \(\bm{L}_{e_i}^{\mathrm{LP}}\) dictated by \(\varphi_{e_i}^{\mathrm{OPT}}\).

\begin{figure*}[!b]
    \centering
    \includegraphics[width=0.8\textwidth]{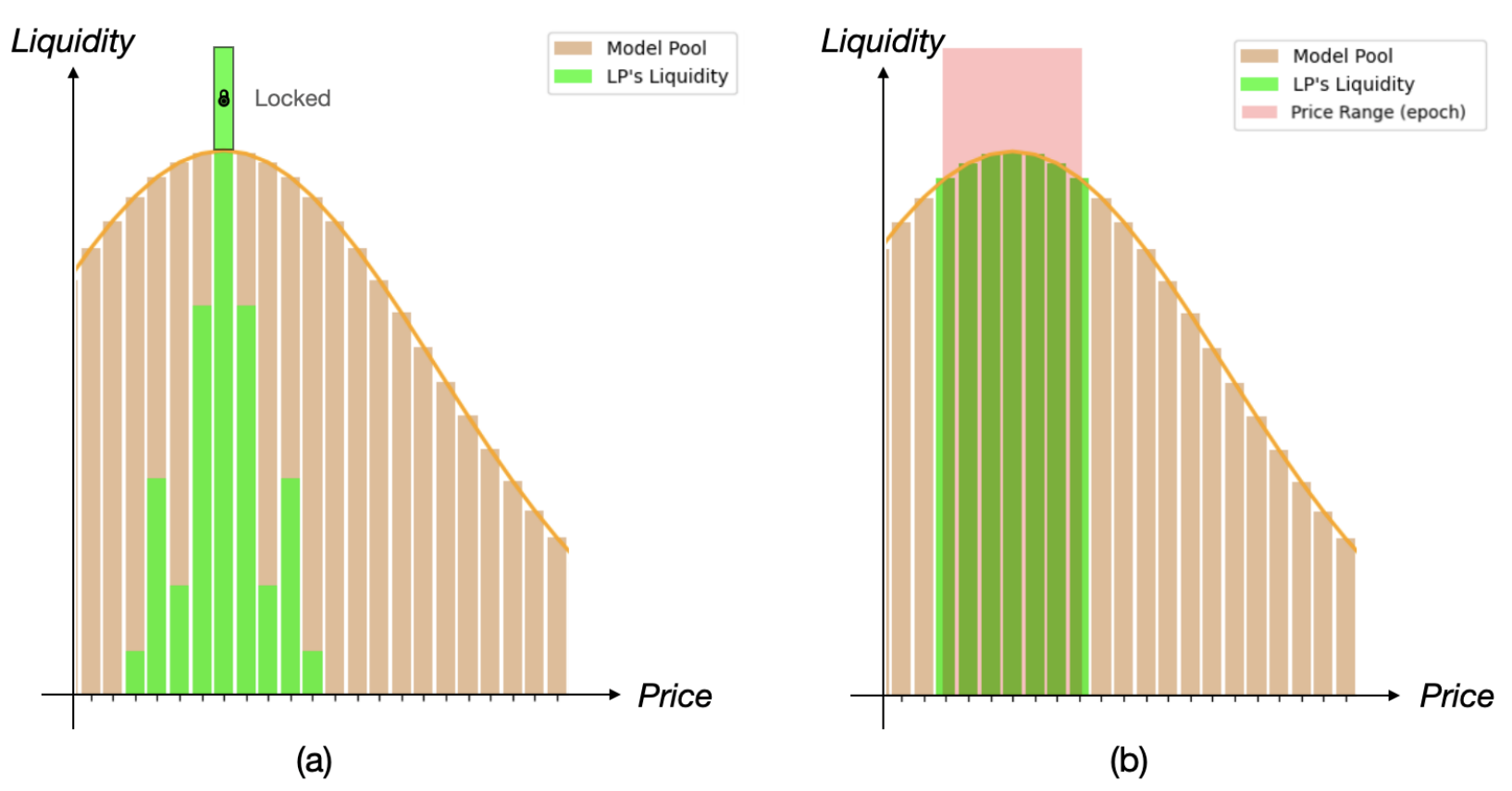}
    \smallskip
    \captionsetup{justification=centering}
    \caption{Visualization of the specific aspects of Approach 3.
    }
    \label{figure:F9}
\end{figure*}

\paragraph{Practical caveat.}
If the capital \(W_{e_i}^{\mathrm{LP}}\) allocated under a candidate
strategy is sufficiently large, it can happen that in some bucket
\(B_n\in\beta\) the liquidity of the LP exceeds the approximate liquidity of the pool
\(L_n^{\mathrm{LP}}>L_n^{\Sigma}\).  In that case, the model assumes that the
LP captures \emph{all} the historical fees generated in that bucket during
epoch \(e_i\), but never more; see Fig. \ref{figure:F9}(a).  
This situation is admittedly unrealistic because it assumes that the
LP can absorb the entire historical liquidity and collect the full fee
stream, an outcome attainable only when
\(W_{e_i}^{\mathrm{LP}}\) represents a non-negligible fraction of the TVL pool - in the extreme, matching it, Fig. \ref{figure:F9}(b).

Although this case pushes the model away from empirical plausibility,
it can be mitigated by (i) choosing
\(W_{e_i}^{\mathrm{LP}}\ll\Sigma\) in the simulations, or
(ii) imposing additional constraints on the shape of the approximating
liquidity profile to prevent a single LP from dominating any active
bucket.

\subsubsection{Approach 4: LP rewards in a pool augmented beyond its historical liquidity.}
\label{sec:approach-4}
Like Approach 3, follows a two-stage
workflow.  
First, the historical liquidity surface is recovered by Approach 1,
producing the calibrated vector
\(       \bm{\alpha}_{e_i}^{\Sigma}       = f_{\mathcal{G}}\!\bigl(\mu_{e_i}^{\ast},\sigma_{e_i}^{\ast}\bigr)     \).
Next, for a given LP capital \(W_{e_i}^{\mathrm{LP}}\) the optimal strategy
\(\varphi_{e_i}^{\mathrm{OPT}}\) is obtained from
\eqref{eq:optLPs}, maximizing the LP model fee in epoch \(e_i\).

\begin{figure*}
    \centering
    \includegraphics[width=1\textwidth]{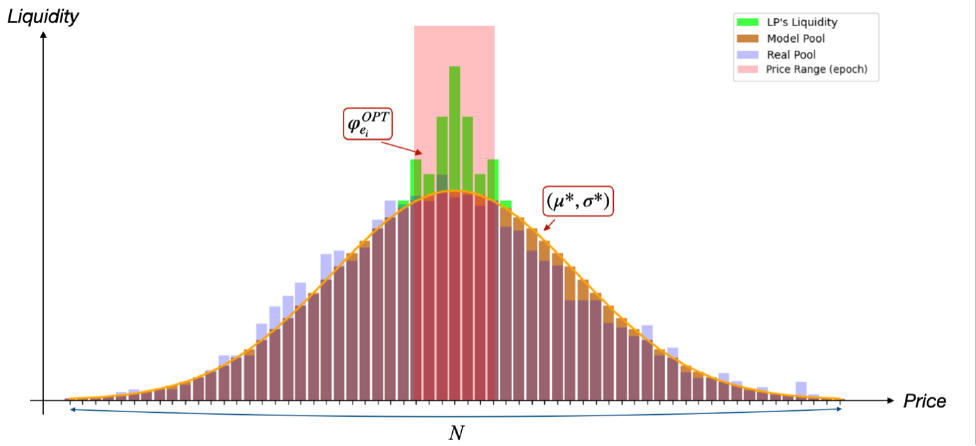}
    \smallskip
    \captionsetup{justification=centering}
    \caption{Target state for Approach 4: the calibrated liquidity profile
\(
  \bm{\alpha}_{e_i}^{\Sigma}
  = f_{\mathcal{G}}\!\bigl(\mu_{e_i}^{\ast},\sigma_{e_i}^{\ast}\bigr)
\)
together with the optimal LP strategy
\(\varphi_{e_i}^{\mathrm{OPT}}\) applied on top of the historical
liquidity in epoch \(e_i\).
    }
    \label{figure:F10}
\end{figure*}

The distinguishing element of the present approach is that LP fees are
calculated after local expansion of historical liquidity by the LP’s own liquidity
\(\bm{L}_{e_i}^{\mathrm{LP}}\)
into \(2\tau+1\) buckets centered on the reference index \(M\).
Thus, the fee share is taken relative to the \emph{augmented} liquidity
profile - historical plus the additional capital of the LP - see Fig. \ref{figure:F10}.

\paragraph{Methodological details.}
This approach follows the same two–stage workflow described in
Section~\ref{section:Pr_St}, with one modification.
After finding the Gaussian parameters
\((\mu_{e_i}^{\ast},\sigma_{e_i}^{\ast})\) so that
\(f_{e_i}^{m\Sigma}(\mu_{e_i}^{\ast},\sigma_{e_i}^{\ast})
      \approx f_{e_i}^{\text{hist}}\),
the approximated historical liquidity vector
\(\bm{L}_{e_i}^{\Sigma}\) is \emph{augmented}
by the LP’s liquidity
\(\bm{L}_{e_i}^{\mathrm{LP}}\)
generated by a candidate strategy
\(\varphi^{\tau}\in\mathbf{S}_{e_i}^{\tau}\).
The updated liquidity vector is
  $\bm{L}_{e_i}^{\Sigma+\mathrm{LP}}
  =
  \bm{L}_{e_i}^{\Sigma}\!
  \bigl(\,\bm{W}_{e_i}^{\Sigma}+\bm{W}_{e_i}^{\mathrm{LP}},
         p_{\gamma_{i-1}},\beta\bigr).$
The optimization pipeline, analogous to \eqref{eq:schLPst}, becomes
\begin{equation}
\begin{gathered}
  \bm{\alpha}_{e_i}^{\mathrm{LP}}\!\bigl(\varphi^{\tau}\bigr)
     \;\longrightarrow\;
     \bm{W}_{e_i}^{\mathrm{LP}}
       \!\bigl(W_{e_i}^{\mathrm{LP}},\bm{\alpha}_{e_i}^{\mathrm{LP}}\bigr)
     \;\longrightarrow\;
     \bm{L}_{e_i}^{\mathrm{LP}}
       \!\bigl(\bm{W}_{e_i}^{\mathrm{LP}},p_{\gamma_{i-1}},\beta\bigr)
     \;\longrightarrow\;
     \\[4pt]
  \longrightarrow\;
     \bm{r}_{e_i}^{\mathrm{LP}}
     \!\Bigl(
        \bm{L}_{e_i}^{\mathrm{LP}},\,
        \bm{L}_{e_i}^{\Sigma+\mathrm{LP}}
        \!\Bigl(
           \bm{W}_{e_i}^{\Sigma}
             \!\bigl(\Sigma,
                     f_{\mathcal{G}}(\mu_{e_i}^{\ast},\sigma_{e_i}^{\ast})\bigr)
           +\bm{W}_{e_i}^{\mathrm{LP}},
           \,p_{\gamma_{i-1}},\beta
        \Bigr)
     \Bigr)
     \;\longrightarrow\;
     \\[4pt]
  \longrightarrow\;
     \bm S_{e_i}^{\mathrm{LP}}
     \;\longrightarrow\;
     F_{e_i}^{\mathrm{LP}}
     \;=\;
     f_{e_i}^{\mathrm{LP}}\!\bigl(\varphi^{\tau}\bigr),
\end{gathered}
\end{equation}
where the share vector
\(\bm{r}_{e_i}^{\mathrm{LP}}\)
is calculated with respect to the increase in liquidity
\(\bm{L}_{e_i}^{\Sigma+\mathrm{LP}}\).
The strategy
\(\varphi_{e_i}^{\mathrm{OPT}}\)
is then selected as the maximizer of
\(f_{e_i}^{\mathrm{LP}}(\varphi^{\tau})\),
subject to the configuration of the augmented pool.

\paragraph{Practical caveat.}
Unlike Approach 3, the historical fee level no longer serves as an upper
bound for the modeled LP reward.  
If the injected capital \(W_{e_i}^{\mathrm{LP}}\) is sufficiently large,
the LP can impose an extreme liquidity concentration within the active
range, so that the historical share of liquidity held by \emph{all the other}
providers tends to zero.  In the limit, the situation reverts to the
hypothetical single-LP pool of Approach 2,
which seems unrealistic.  
This pathological result can be avoided by enforcing
\(W_{e_i}^{\mathrm{LP}}\ll\Sigma\) when simulating the strategy.

Another assumption concerns the nature of the trades that generate the
historical price set $\bm{P}$.
If it is posited that every price update was driven exclusively by
arbitrage trades executed by informed traders, then Approach 4 remains
internally consistent: the additional LP liquidity, however concentrated,
does not impede arbitrageurs from supplying the trade volume required to
reproduce the historical price path.
However, in reality, a non-arbitrage order flow is always present, and its
volume is unlikely to scale with the LP super-concentration.
Consequently, the real trade volume would fall short of that model volume, dampening the realized price volatility of the pool and
activating fewer price ranges than in the original history, which in turn
reduces the modeled LP reward.
Ideally, therefore, the practitioner should either (i) mark historical
swaps as arbitrage or non-arbitrage and retain only the former when
simulating with super-concentration, or (ii) estimate the arbitrage share
ex ante and adjust the modeled LP reward downward to reflect the
liquidity that genuine arbitrage trades can reasonably provide.

\section{Modeling on Real Data}
\label{section:Experiments}
In this section, the methodology developed above is applied to real
Uniswap~V3 data on Ethereum.  
A \emph{uniform-liquidity strategy} is used as a benchmark, while the
proposed approach is evaluated on data from USDC/ETH pools with several fee
levels, in stablecoin pools and in WBTC/ETH.  
The section ends with (i) a summary of the consolidated performance of all
pools, (ii) a comparison with an external backtesting tool,
and (iii) a high-level schematic of the application of the ML model for liquidity allocation decisions.

The step-by-step exposition in current section focuses on the USDC/ETH pool with the
0.3\%\footnote[1]{Contract 0x8ad599c3A0ff1De082011EFDDc58f1908eb6e6D8} fee tier; the accompanying Jupyter notebooks with code,
hyper-parameters, and raw data are available on GitHub \cite{BBcs}.
General details and modeling approach:
\begin{itemize}
  \item \emph{In-sample period}.  
        Swap transactions and trading volumes from 1~April~2023 to
        30~June~2024 serve as the training modeling period .
  \item \emph{Out-of-time (OOT) period.}  
        Model deployment is emulated on the perod
        1–30~September~2024.
  \item \emph{Reward approach.}  
        The third approach of Section~\ref{sec:approach-3} was adopted because it links the level of synthetic fees to the historical upper
bound, which yields conservative expectations for the OOT analysis.
\end{itemize}
Historical swap‑transaction data were retrieved using GraphQL queries to the Uniswap V3 subgraph.
\label{sec:liquidity-approx}

\subsection{Historical Liquidity Approximation and Optimal LP Strategy Search}
\begin{figure*}[!b]
    \centering
    \includegraphics[width=1\textwidth]{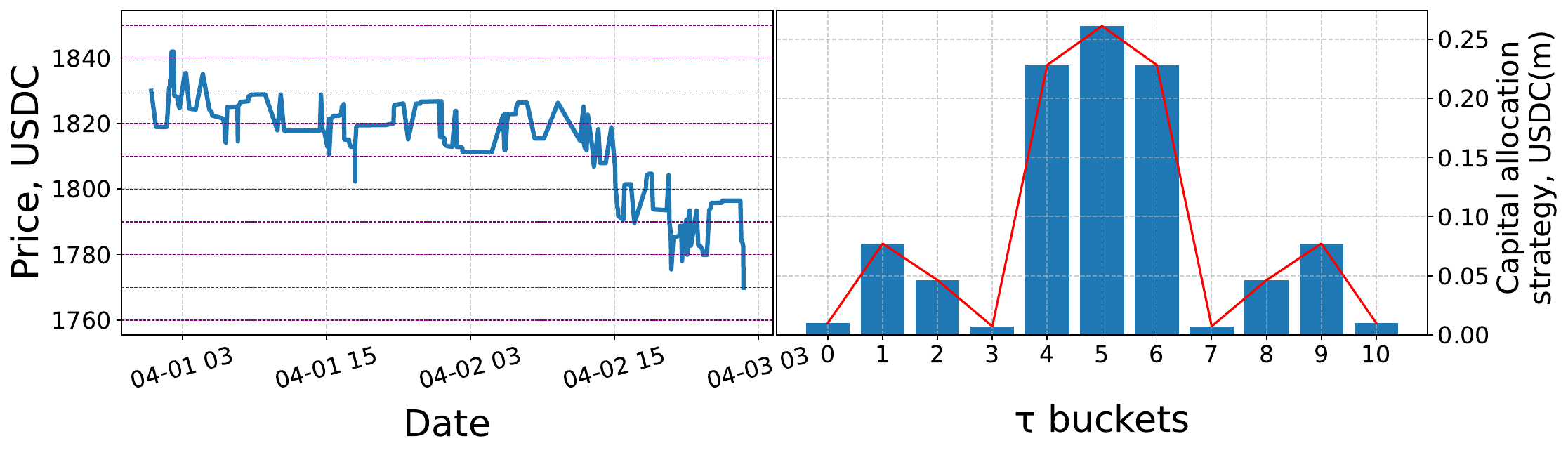}
    \smallskip
    \caption{Example of the optimal LP strategy for epoch~$e_{1}$ ($\tau=5$).}
    \label{figure:F12}
\end{figure*}
\begin{figure*}[!t]
    \centering
    \includegraphics[width=1\textwidth]{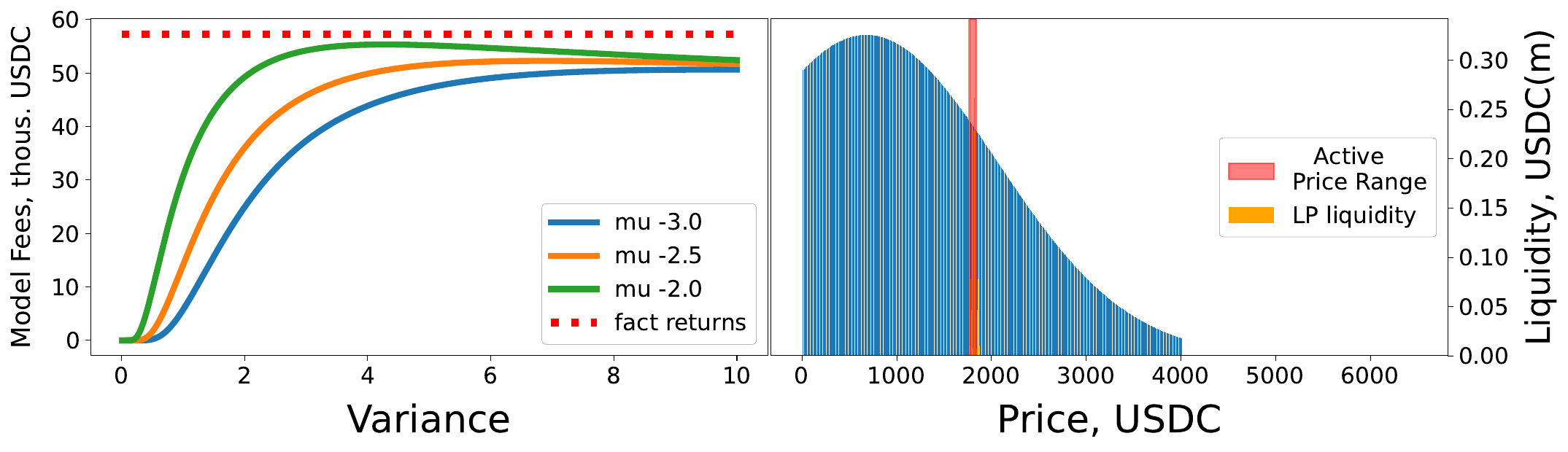}
    \smallskip
    \captionsetup{justification=centering}
    \caption{Process of calibrating the approximating liquidity profile for epoch \(e_1\).
    }
    \label{figure:F11}
\end{figure*}

According to the proposed methodology, the modeling is performed in two stages:
\paragraph{Stage 1: approximating of the historical liquidity profile.}
For every epoch~$e_i$ the pool’s liquidity shape is approximated by
sweeping the Gaussian parameters
$\mu_{e_i}\in[-3,3]$ and
$\sigma_{e_i}\in[\sqrt{0.01},\sqrt{10}]$ until the reward of the modeled pool
$f_{e_i}^{m\!\Sigma}(\mu_{e_i},\sigma_{e_i})$
deviates from the historical value $f_{e_i}^{\text{hist}}$ by no more than
$5\%$.
The average TVL is fixed at $\Sigma=80m\,$USDC, width of the bucket
$d=10$ USDC, lower boundary $p_{a_1}=0$ and $N=650$.
Applying the $\tau$-reset rule with $\tau=5$ partitions the in-sample
period into $K=820$ epochs.
The cumulative historical fee
$\sum_{i=1}^{820} f_{e_i}^{\text{hist}}\approx 24.1m$ USDC
is matched by the modeled fee
$\sum_{i=1}^{820} f_{e_i}^{m\!\Sigma}\approx 23.8m$ USDC,
that is, an approximation error of about $1.3\%$, well inside the $5\%$
tolerance.
Figure \ref{figure:F11} illustrates the calibration for the first epoch:
the optimal parameters $\mu_{e_1}^{\ast}=-2.0$ and
$\sigma_{e_1}^{\ast}=\sqrt{4.28}$ yield
$f_{e_1}^{m\!\Sigma}=55\,348$ USDC versus historical
$57\,259$ USDC (error~$3.5\%$).
The right-hand panel of Fig. \ref{figure:F11} shows the resulting liquidity profile
$\bm{L}_{e_1}^{\Sigma}$ and the active price range of epoch~$e_1$ (in red).

\paragraph{Stage 2: determination of optimal LP strategies.}
Once the historical liquidity $\bm{L}_{e_i}^{\Sigma}$ is approximated,
the optimal LP strategy is found by solving the problem~\eqref{eq:optLPs}.
At the start of each epoch the LP deploys the same initial capital
$W_{e_i}^{\mathrm{LP}} = W^{\mathrm{LP}} = 1$ million USDC;
inter-epoch capital dynamics are ignored at the training stage to isolate
each epoch.
\begin{figure*}[!b]
    \centering
    \includegraphics[width=1\textwidth]{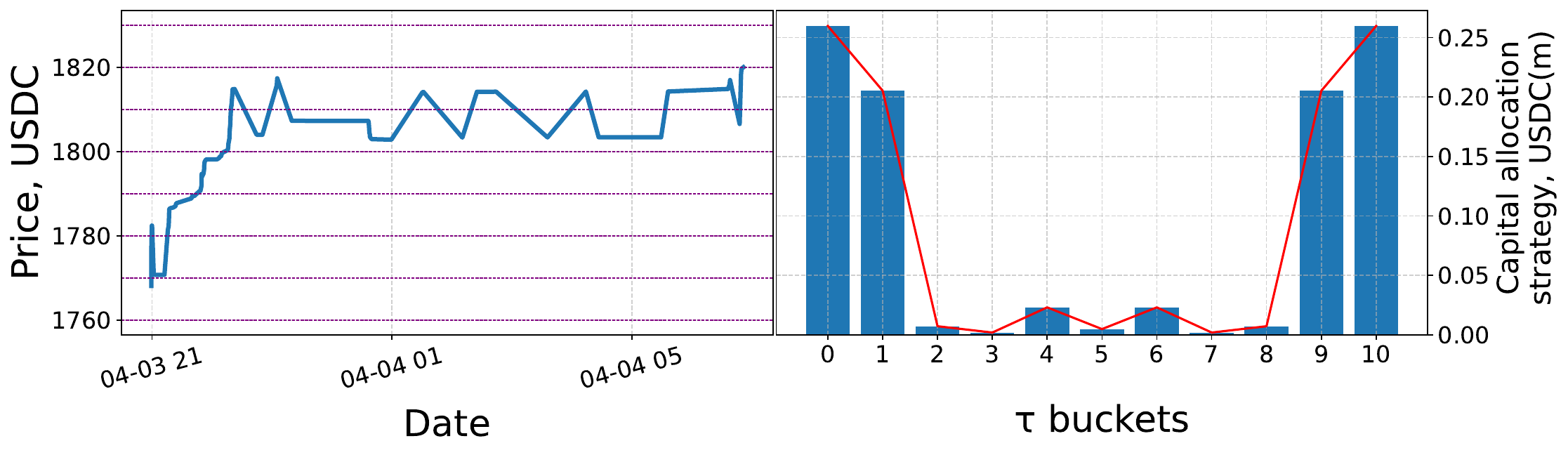}
    \smallskip
    \caption{Example of the optimal LP strategy for epoch~$e_{4}$ ($\tau=5$).}
    \label{figure:F13}
\end{figure*}
For epoch~$e_1$ the optimizer produces the vector
$\bm{W}_{e_1}^{\mathrm{LP}}$ shown in Fig. \ref{figure:F12} (right),
which aligns with the price path in the left panel and
delivers a modeled LP fee of
$f_{e_1}^{\mathrm{LP}}\bigl(\varphi_{e_1}^{\mathrm{OPT}}\bigr)
      = 39\,162$ USDC.
      
\begin{figure*}[!t]
    \centering
    \includegraphics[width=1\textwidth]{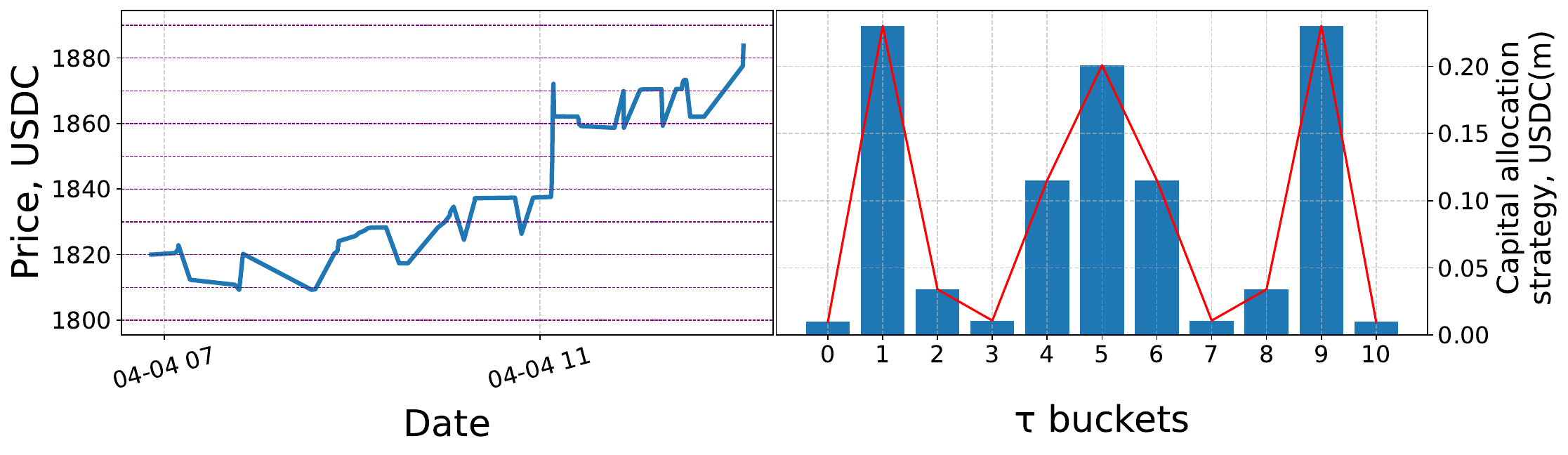}
    \smallskip
    \caption{Example of the optimal LP strategy for epoch~$e_{5}$ ($\tau=5$).}
    \label{figure:F14}
\end{figure*}

Additional examples are provided in Fig. \ref{figure:F13} (epoch $e_4$) and Fig. \ref{figure:F14}
(epoch $e_5$), where optimal allocations shift towards the buckets that
capture most of the observed price movement.
This provides an intuitive and visually compelling demonstration of the concept of optimal strategy.
Collecting optimal allocations for all $K$ epochs yields the target
vector
$Y = (\hat{\bm{\rho}}_{e_1}^{\mathrm{LP}}\;
      \hat{\bm{\rho}}_{e_2}^{\mathrm{LP}}\;
      \dots
      \hat{\bm{\rho}}_{e_K}^{\mathrm{LP}})^{\!\top}$
used to train the ML model, as defined in
Section~\ref{section:ml-architecture}.

\subsection{ML Model Training and Features Engineering}
Within the present study the architectural nuances of the machine learning
models do not constitute the main focus of the research.
Each model is deployed in a baseline configuration with no systematic
hyper-parameter tuning.
The emphasis is on demonstrating the capabilities of the proposed
methodology and the accompanying backtesting framework: in particular,
the ability (i) to determine optimal LP strategies ex post,
(ii) to record the market-state feature values at the moments of optimal
allocations, and (iii) to use the resulting patterns for future
liquidity-allocation decisions.

Feature engineering is also intentionally minimal.
All inputs are derived from only two raw numerical indicators, from which
a compact set of features is generated to capture latent patterns in
market dynamics.
The approach is generic and can be adapted to any collection of numeric
variables.
The use of external market data for feature engineering is further justified by empirical studies showing that cryptocurrency markets react distinctively to news and sentiment compared to traditional asset classes~\cite{ref_50}, and exhibit unique, stress-sensitive volatility spillovers with conventional currencies~\cite{ref_53}.
Subsequent sections confirm that even such sparse feature sets and simple
model configurations are sufficient for the goals of this work.

\subsubsection{ML Model Architecture.}
The predictions of the integral model are produced by a weighted ensemble of
three base learners: an MLP neural network, CatBoost, and an LSTM network.
The remainder of this subsection details the architectures implemented in
the experiments, including structural choices, theoretical rationale,
hyperparameter settings, regularization techniques, and evaluation metrics.
The use of an integrated ensemble model for prediction in this domain is supported by recent literature, which employs similar machine learning architectures to interpret and forecast complex, non-linear dynamics in DeFi and digital asset markets~\cite{ref_52}.

\paragraph{MLP neural network.}

Fully connected neural networks (multilayer perceptrons, or MLPs) are
universal function approximations and are widely used for regression and
classification tasks. 
The core idea is to propagate the input feature space through a sequence
of hidden layers equipped with non-linear activation functions, thereby
capturing complex relationships between inputs and targets.
Key settings:
\begin{itemize}[topsep=0pt, partopsep=0pt]
  \item Loss function: Mean Squared Error (MSE).
  \item Optimizer: Adam.
  \item Training schedule: 100 epochs with early stopping.
  \item Regularization: \texttt{EarlyStopping} callback.
  \item Validation: 5-fold cross-validation with the best model retained.
\end{itemize}
Network architecture:
\begin{itemize}[topsep=0pt, partopsep=0pt]
  \item Input layer - dimensionality equals the number of features in \(X\).
  \item Hidden layers  
        \begin{itemize}
          \item \texttt{Dense}(128, activation=\texttt{relu})  
          \item \texttt{Dense}(64,  activation=\texttt{relu})  
          \item \texttt{Dense}(32,  activation=\texttt{relu})  
          \item \texttt{Dense}(16,  activation=\texttt{relu})
        \end{itemize}
  \item Output layer: \texttt{Dense}\((\tau+1,\ \text{activation}=\texttt{softmax})\)
\end{itemize}

\paragraph{CatBoost (gradient boosting on decision trees).}
CatBoost is a gradient boost implementation optimized for categorical
features and known for its robustness to overfitting \cite{ref_28}.  
It is used here as a high-accuracy regression learner on tabular data. Key settings:
\begin{itemize}[topsep=0pt, partopsep=0pt]
  \item Loss function: \texttt{MultiRMSE}.
  \item Tree depth: 6.
  \item Number of iterations: 1000.
  \item Boosting type: \texttt{Plain}.
  \item Early stopping: enabled.
  \item Validation: 5-fold cross-validation with model selection.
\end{itemize}

\paragraph{LSTM (Long Short-Term Memory network).} LSTM networks are a class of recurrent neural networks (RNNs) designed to
handle sequential data and long-term dependencies. 
Although each training instance in this work represents an aggregated
snapshot of features at a single point in time, an LSTM is included as an
alternative to the MLP, capitalizing on its ability to model internal
patterns within the aggregated feature structure.
Key settings:
\begin{itemize}[topsep=0pt, partopsep=0pt]
  \item Loss function: Mean Squared Error (MSE).
  \item Optimizer: Adam.
  \item Training schedule: 100 epochs with \emph{early stopping}.
  \item Regularization: 20\,\% dropout after the LSTM layer.
  \item Validation: 5-fold cross-validation.
\end{itemize}
Network architecture:
\begin{itemize}[topsep=0pt, partopsep=0pt]
  \item Input - vector of one-dimensional temporal features \(X\).
  \item Hidden layer: \texttt{LSTM}(128, activation=\texttt{relu}).
  \item Dropout layer — rate 0.2.
  \item Output layer: \texttt{Dense}\((\tau+1,\ \text{activation}=\texttt{softmax})\)
\end{itemize}
\vspace{\baselineskip}
Although every training instance in the data set is an \emph{aggregated
snapshot} of market features taken at a single time stamp, an LSTM network
is nonetheless included as an architectural alternative to the MLP.
The objective is \emph{not} to extract temporal autocorrelations across
successive observations - by construction, such correlations are absent - but
rather to exploit the LSTM gating mechanism as a powerful
non-linear\slash combinatorial feature extractor.
Thanks to its cell state and input/forget gates, an LSTM can learn
structured interactions and hierarchical patterns that arise inside the
high-dimensional vector of engineered features, even when those features
are computed from historical time series and then collapsed into a single
observation.

A further consideration is that LP reallocation events (i.e., the epoch
boundaries) may be widely spaced and occur at irregular intervals, making
classical sequence-modeling assumptions difficult to satisfy.
By embedding an LSTM block in the ensemble, the model gains a neural
component whose internal dynamics differ markedly from those of an MLP;
this architectural diversity is expected to enhance the ensemble’s ability
to generalize outside of the sample.

\paragraph{Integral model
(ensemble by weighted averaging).}
The final prediction is obtained by a weighted sum of the three base
models.  
The optimal weight vector
\(\bm{w}=\{w_{1},w_{2},w_{3}\}\)
is found by minimizing the mean square error with regularization \(L_{2}\) (\(\alpha=0.001\)) under the constraints
\(w_{j}>0\) and \(\sum_{j=1}^{3} w_{j}=1\).
Despite its architectural simplicity, the ensemble introduces methodological novelty by embedding structurally diverse models into a domain-specific backtesting framework for liquidity provision

Optimization is carried out with the
\textit{Sequential Least Squares Programming} (SLSQP) algorithm, an
efficient method for constrained nonlinear optimization. 
SLSQP is particularly well suited here because
(i)~it directly handles the simplex constraints on the weights,
(ii)~the ensemble layer involves only three parameters, so the
computational burden per iteration is negligible, and
(iii)~the problem is convex, ensuring a globally consistent solution.

\medskip
\subsection{Feature Engineering}
Centralized exchange (CEX) data serve as the information set to characterize the market state in each allocation decision.
Historical market data are sourced through the
\texttt{live\_trading\_indicators}\footnote[2]{https://pypi.org/project/live-trading-indicators/} library, which offers unified access to
multiple cryptocurrency exchanges.
For the USDC/ETH pool, the study employs one-minute Open-High-Low-Close-Volume (OHLCV) bars for the
USDC/USDT and BTC/USDT pairs on Binance, covering
1~March~2023 to 30~September~2024; the inclusion of BTC captures potential
cross-asset comovement with ETH.
For other pools, the same procedure is followed: features are generated for both the target pair and the BTC.

\begin{figure*}[!b]
    \centering
    \includegraphics[width=1\textwidth]{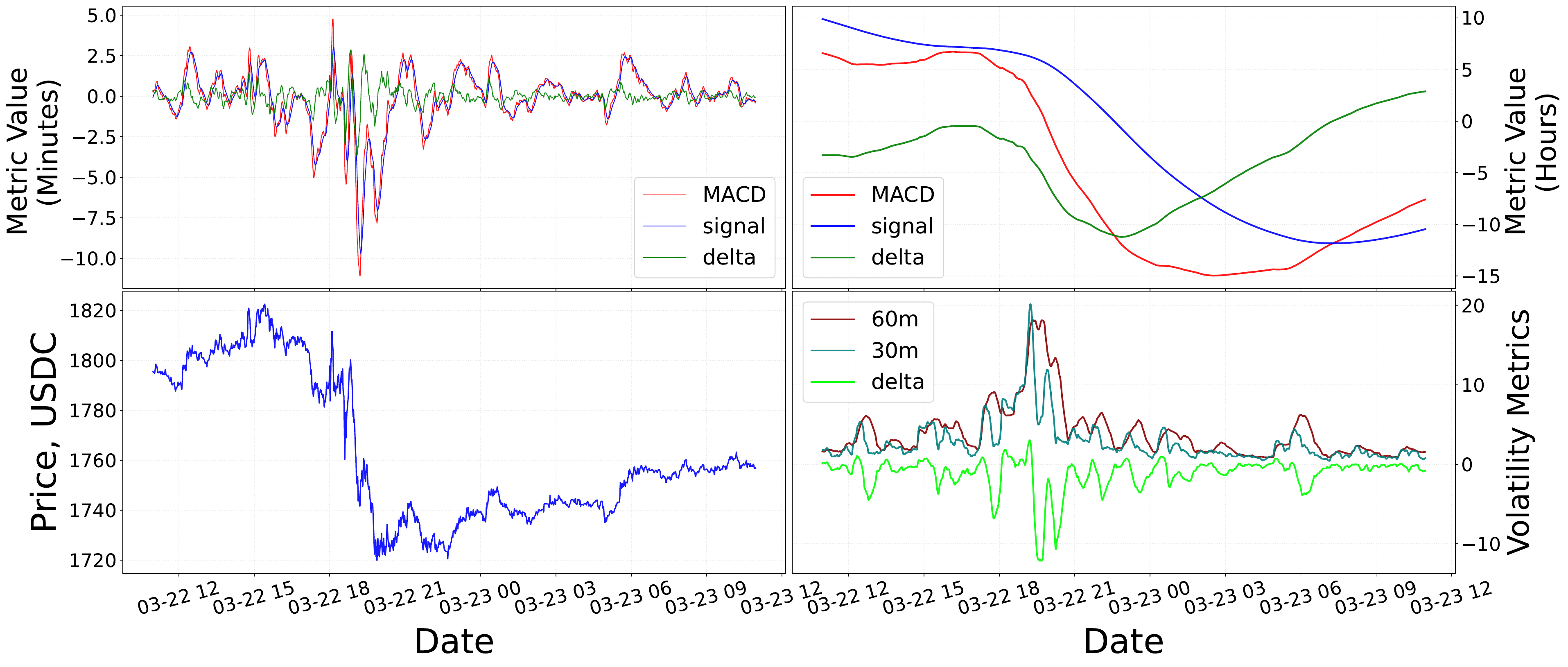}
    \smallskip
    \caption{Examples of features used in ML models.}
    \label{figure:Z1}
\end{figure*}

Each bar provides the closing price and traded volume; the timestamp is
used as the index.
For every epoch, the feature vector is generated at the
beginning of the epoch (index \(\gamma_{i-1}\) in Section~\ref{sec:bst}) by
selecting the nearest CEX timestamp.
The other indicators are the exponential moving average (EMA) and
the moving average convergence divergence (MACD) — fundamental tools of
technical analysis that highlight trend and momentum
behavior. 


The market features for the ML model are derived from two pairs of cryptocurrency: USDC/ETH and BTC/USDT according to the following algorithm:
\begin{enumerate}
  \item \textit{Data retrieval.}
        One-minute OHLCV bars are downloaded from Binance for the symbols
        \texttt{ethusdt} and \texttt{btcusdt}.
        The closing price (\textit{feature 1}) and trade volume
        (\textit{feature 2}) are retained.

  \item \textit{Volume clearing.}
        Zero volumes, if any, are replaced with a small constant ($10^{-8}$) to avoid numerical errors in subsequent calculations.

  \item \textit{Sliding-window indicators (minute scale).}
        \begin{enumerate}
          \item Two exponential moving averages (EMAs) with periods
                12~min and 26~min are computed
                (\textit{features 3 and 4}).
          \item The MACD line is defined as the difference between the two
                EMAs (\textit{feature 5});
                the signal line (\textit{feature 6}) is the 9-period EMA
                of the MACD, and their difference
                (\textit{feature 7}) completes the trio.
        \end{enumerate}

  \item \textit{Sliding-window indicators (hour scale).}
        Steps in 3 are repeated in the hourly aggregation (12 h and 26 h windows), producing \textit{features 8-12}.

  \item \textit{Volume-volatility measures.}
        Rolling standard deviations of the trade volume are calculated
        over 30- and 60-minute windows
        (\textit{features 13 and 14}); their difference provides
        \textit{feature 15}, quantifying the change in volume turbulence.
\end{enumerate}

For each pair, the above procedure outputs 15 features (Fig. \ref{figure:Z1}); the same
calculations for the second pair and the pairwise ratios between the two
sets enlarge the feature vector to 45 dimensions.  
Thus, at the start of each epoch~$e_i$ the market state is represented by
\(\bm{\psi}_{e_i}\in\mathbb{R}^{45}\) and the complete
design matrix is
\(X\in\mathbb{R}^{820\times45}\) (Section~2.5).

\textit{Look-back horizon.}
At every liquidity reallocation (or initial provisioning), the feature
algorithm consumes a one week history
(\(1440\times7\) minutes) preceding the decision time.
This window can be shortened or extended for specific indicators if
required, but a uniform horizon is adopted here for the sake of clarity.

\textit{Remarks on feature selection.}
Standard feature selection techniques could be applied to reduce
multicollinearity; however, the complete set is deliberately retained to
emphasize the generality of the methodology rather than to maximize
predictive accuracy through aggressive pruning.
In addition, all three base learners are trained in the same feature
matrix, which simplifies comparison and integration of the ensemble, albeit at
the cost of avoiding model-specific feature customization - an acceptable
trade-off given that architecture optimization is not the main objective of
the study.

\subsection{Application of ML Models to the OOT Sample}
\label{section:ExperimentsOOT}
The out–of–time (OOT) period comprises all swap transactions
and volumes recorded in the pools between 1~and~30~September~2024.
With the exception of the locked total value $\Sigma$, all modeling
hyper-parameters are retained from the training part.

In live operation, the model requires only the current feature vector to output an allocation decision.
However, for retrospective fee backtesting, the full workflow of
Section~\ref{sec:liquidity-approx} must be repeated, still relying on the
third reward modeling approach, but \emph{without} enumerating random
LP strategies.
The optimal strategy for epoch~$e_i$ is supplied directly by the ML model
and is denoted
\(\varphi_{e_i}^{\text{mOPT}}\).

The bucket grid~$\beta$ is fixed (its boundaries may differ from the
training grid, but the tick size~$d$ must be unchanged), and the historical
liquidity in the active ranges now approximating with an updated average TVL of
$\Sigma = 40m$ USDC.
Apart from this change, the algorithm mirrors the training procedure,
subject to the additional considerations described next.

\subsubsection{Specific Assumptions for OOT.}
A more refined liquidity approximation scheme is employed for the OOT
period in order to produce a tighter fit to the observed swap flow.
\begin{figure*}[!b]
    \centering
    \includegraphics[width=0.65\textwidth]{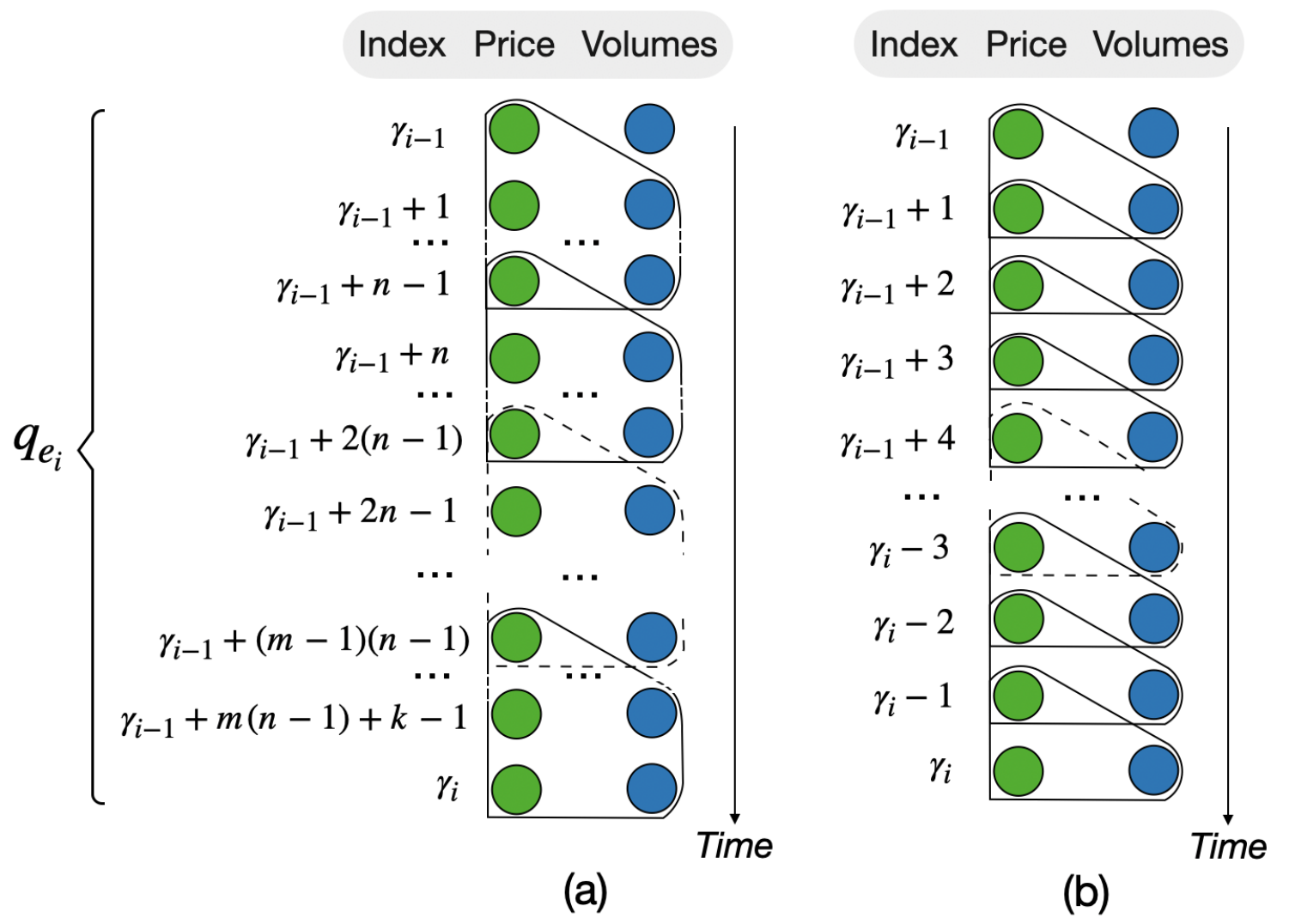}
    \smallskip
    \captionsetup{justification=centering}
    \caption{Schematic illustration of the sub-epoch segmentation algorithm within
    epoch~$e_i$ \\ for the cases \(n>2\) (a) and \(n=2\) (b).
    }
    \label{figure:W1}
\end{figure*}

In the advanced approach, the approximate liquidity profile is \emph{not}
kept fixed over the whole epoch~$e_i$.
Instead, a sliding window of length $n$ trades is fitted successively
$m$ times within the epoch.
The historical pool fee is therefore decomposed as
\(
  f_{e_i}^{\text{hist}}
  = \sum_{j=1}^{m} f_{e_{i_j}}^{\text{hist}},
\)
so that $e_i$ is represented by $m$ \emph{sub-epochs}, each endowed with
its own calibrated parameters
\(\bigl(\mu_{e_{i_j}}^{\ast},\sigma_{e_{i_j}}^{\ast}\bigr)\) for
\( j \in [1,m] \).
The number of sub-epochs is

\begin{equation}
  m \;=\;
  \Bigl\lfloor\frac{q_{e_i}-1}{\,n-1}\Bigr\rfloor,
  \label{eq:m-subepochs}
\end{equation}

\noindent
where \( q_{e_i} \in \bm{Q}_{\bm{E}} \) and can be represented as \(q_{e_i}=n+(m-1)(n-1)+k\), where \(k\) is the remainder of trades that
is insufficient to form another full window and is therefore appended to
the last sub-epoch.
This granular calibration reproduces the historical fee flow almost down
to the individual-swap level, delivering a more faithful representation of
pool dynamics in the OOT period (Fig. \ref{figure:W1}).

The window length~$n$ is user-defined and controls the degree of
\emph{flexibility} of the model liquidity:
LP capital as vector \(\bm{L}_{e_i}^{\mathrm{LP}}\) is still fixed at the start of
$e_i$, but the provider’s share evolves with each sub-epoch,
producing distinct ratio vectors $m$ instead of a single
\(\bm{r}_{e_i}^{\mathrm{LP}}\).
Flexibility in epoch is quantified by

\begin{equation}
  F_{e_i} \;=\;
  \frac{m}{q_{e_i}-1},
  \label{eq:plasticity}
\end{equation}

\noindent
so that \(F_{e_i}=1\) for \(n=2\) (maximal dynamics,
one approximate per price change) and \(F_{e_i}\approx 0\) for \(n=q_{e_i}\) when \(q_{e_i}\gg2\)
(a single static approximate, as in the training part).

To prevent unrealistically large LP shares in narrow ranges
(cf. Fig. \ref{figure:F9}), a conservative rule is imposed:
whenever the found approximated parameter
\(\mu_{e_{i_j}}^{\ast} = \mu_{\min}\),
the LP fee contribution for that sub-epoch is set to zero.
Denoting this \emph{liquidity-rule form} as \emph{LRF}, curtails excessive fee projections
that could arise from an over-concentrated placement.

\begin{figure*}[!b]
    \centering
    \includegraphics[width=1\textwidth]{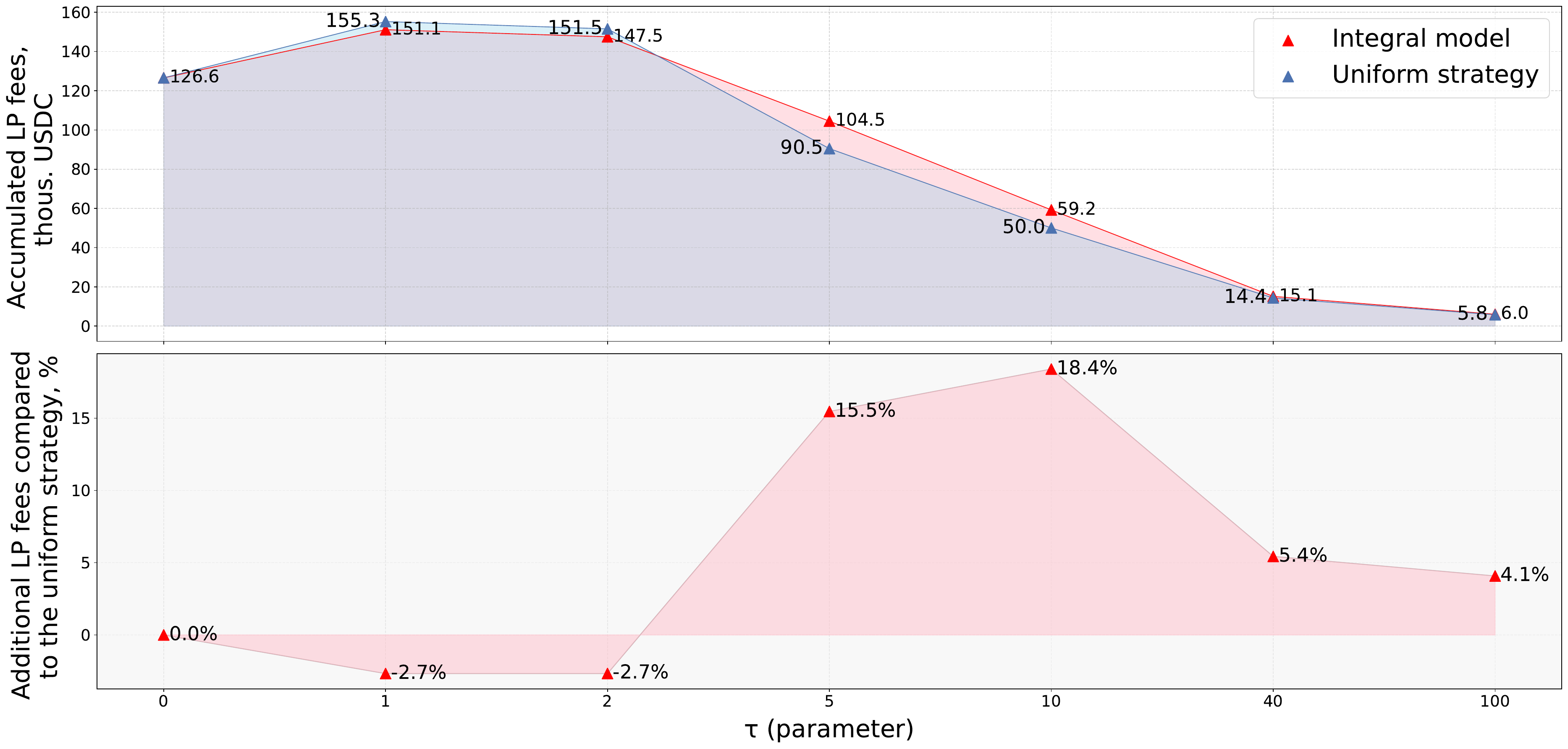}
    \smallskip
    \captionsetup{justification=centering}
    \caption{Accumulated LP fees depending on $\tau$ \\ on the out-of-time period (September 2024) for USDC/ETH 0.3\% pool.}
    \label{figure:JJ1}
\end{figure*}

\subsubsection{Results on the OOT period.}
Using the same procedure as in Section~\ref{sec:liquidity-approx},
the OOT price set $\bm{P}$, the bucket grid~\(\beta\)
and the \(\tau\)-reset rule partition the period
(1–30~September~2024) into \(K=45\) epochs with \(\tau=5\).
Applying the algorithm of Section~\ref{sec:approach-3} together with the
flexibility of
the model liquidity
(\(n=2, F_{e_i}=1\)) and the \emph{LRF} constraint,
problem~\eqref{eq:optPoolT} is solved for every sub-epoch.

The aggregated historical fee for the OOT period is
$\sum_{i=1}^{45} f_{e_i}^{\text{hist}} \approx 0.311m$ USDC;
the model reproduces this value with
$\sum_{i=1}^{45} f_{e_i}^{m\!\Sigma} \approx 0.306m$ USDC,
which corresponds to an approximation error of approximately \(1.6\%\).

The pool is then analyzed with respect to the impact of the allocation
parameter~\(\tau\) on modeled LP rewards.
Figure \ref{figure:JJ1} and Table \ref{tab:t1} compare the results of the optimal ML model
strategy with a uniform-liquidity baseline. The relationship between $\tau$ values and the corresponding metrics is demonstrated, including epochs, average time, number of elements, fee values, and percentages of ML model efficiency.
The case \(\tau=0\) represents a single-bucket placement in the reference
bucket \(B_M\).

It should be emphasized that the initial capital 
$W^{\mathrm{LP}} = 1$m USDC; however, unlike in the training
part, the working capital evolves over time
\(\bigl(W_{e_i}^{\mathrm{LP}}\neq W^{\mathrm{LP}}\) for \(i> 1\bigr)\).
Now, two additional components are taken into account:
\begin{itemize}
  \item \emph{Gas expenditure.}
        Deploying liquidity in a single range is charged
        \(430\,000\)~gas units, whereas burning liquidity costs
        \(215\,000\)~gas units.
        If the amount of LP liquidity in a given bucket remains unchanged
after reallocation, the associated gas cost is ignored, in line with the convention of~\cite{ref_6}. In this research, we do not consider the costs associated with token swaps necessary to convert into the desired asset for liquidity allocation.
        A constant gas price of \(20\,\text{Gwei}\) is assumed.
  \item \emph{Impermanent loss and downside (market) risk.}
        Beyond accounting for impermanent loss during epoch transitions, should the pool price fall below the lower boundary
        \(p_{a_{M-\tau}}\) of the outermost LP-liquid bucket
        \(B_{M-\tau}\) in epoch~\(e_i\),
        the mark-to-market losses are fixed and decrease the LP’s capital.
\end{itemize}
Throughout the OOT backtesting, all fees are \emph{not} reinvested between
epochs; this separation isolates the modeled reward stream from the
evolving principal \(W^{\mathrm{LP}}\). 
It is worth adding that impermanent loss in AMMs also can be interpreted as an AMM-specific manifestation of adverse selection known from traditional trading, where market makers systematically trade against better-informed (arbitrage) order flow.

For the USDC/ETH pool with 0.3\% fee level, the experimental
framework reveals a clear relationship between liquidity concentration
(depending on $\tau$) and expected reward, with a seemingly counterintuitive
exception for \(\tau=0\) and \(\tau=1\) (Fig. \ref{figure:JJ1}).
The left-hand skew arises from two opposing factors:

\begin{enumerate}
  \item Heightened negative impact of impermanent loss and more frequent
        realization of market risk when the liquidity is so concentrated.
  \item The upper bound imposed by the historical fee volume in each
        active range (Fig. \ref{figure:F9}).  
        In live trading a tighter concentration could in principle yield higher fees, but in backtesting the payoff is capped by historical swap volumes.
\end{enumerate}

\vspace*{-3em} 
\begin{table}
\captionsetup{justification=centering}
\caption{Analysis of strategy metrics depending on the $\tau$ parameter \\ on the out-of-time period (September 2024) for USDC/ETH 0.3\% pool}\label{tab1}
\centering
\setlength{\tabcolsep}{6pt}
\begin{tabular}{@{}c
                S[table-format=4.0]
                S[table-format=3.1]
                S[table-format=4.0]
                C{1.8cm}
                C{1.8cm}
                c@{}}
\toprule
\multirow{2}*{$\tau$} & 
\multicolumn{1}{c}{\multirow{2}*{\shortstack{Number \\ of Epochs $K$}}} &
\multicolumn{1}{c}{\multirow{2}*{\shortstack{Avg.  ${t}_{e_{i}}$ \\ (hour)}}} & 
\multicolumn{1}{c}{\multirow{2}*{\shortstack{Avg. ${q}_{e_{i}}$ \\ (numb.)}}} & 
\multicolumn{2}{c}{LP Fees, thous. USDC} & 
\multicolumn{1}{c}{\multirow{2}*{\shortstack{ML eff., \\ \%}}} \\
\cmidrule(lr){5-6}
 & & & & \multicolumn{1}{c}{ML model} & \multicolumn{1}{c}{uniform} & \\
\midrule

0    & 1946   & 0.3    & 4     & 127.6   & 127.6    & 0.0 \\
1    & 658    & 1      & 13    & 151.1   & 155.3    & -2.7 \\
2    & 213    & 2      & 39    & 147.5   & 151.5    & -2.7 \\
5    & 45     & 15     & 187   & 104.5   & 90.5     & +15.5 \\
10   & 16     & 45     & 525   & 59.2    & 50.0     & +18.4 \\
40   & 1      & 719    & 8396  & 15.1    & 14.4     & +5.4 \\
100  & 1      & 719    & 8396  & 6.01    & 5.8      & +4.1 \\

\bottomrule
\end{tabular}
\label{tab:t1}
\end{table}

\begin{figure*}
    \centering
    \includegraphics[width=1\textwidth]{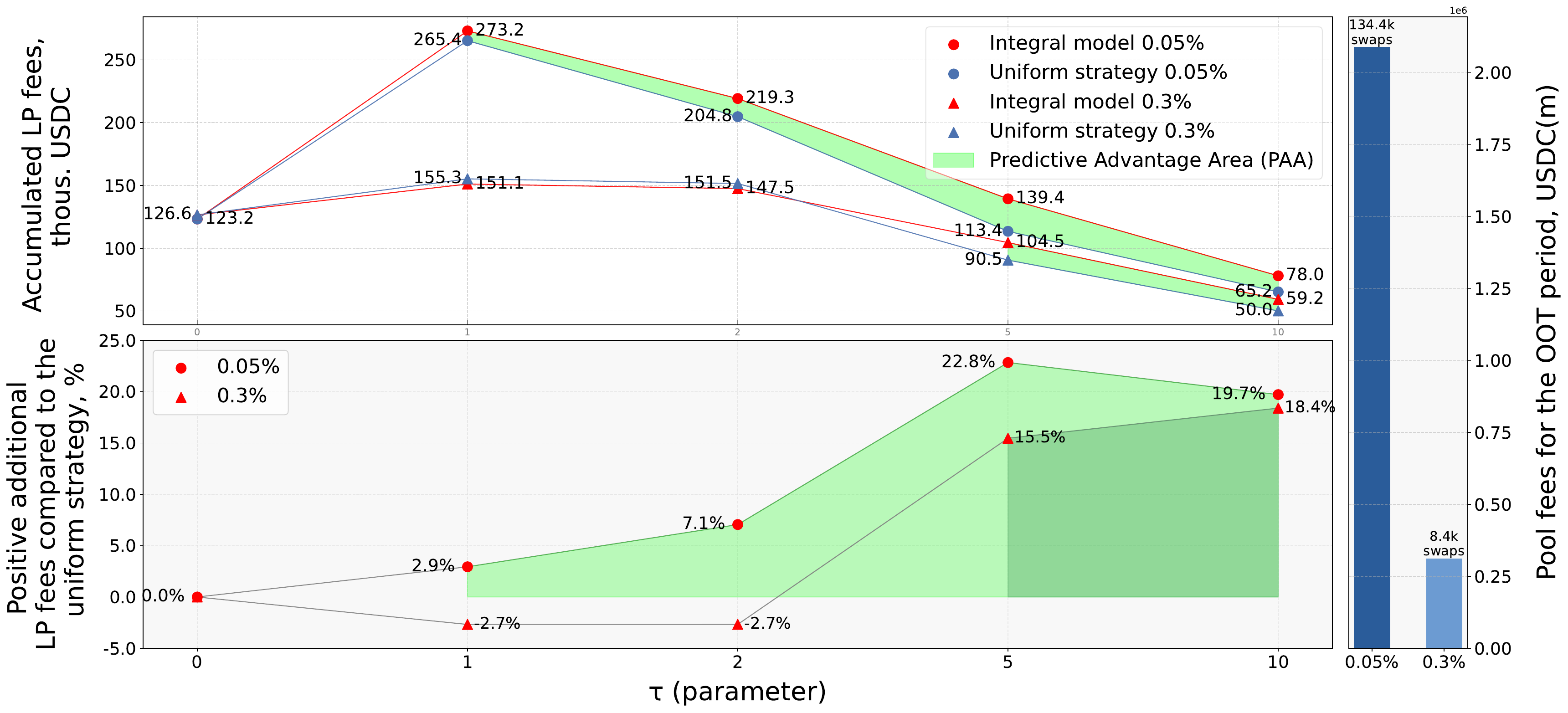}
    \smallskip
    \captionsetup{justification=centering}
    \caption{Accumulated LP fees as a function of~$\tau$ on the
           out-of-time period (September 2024) for the USDC/ETH pools with
           fee tiers of 0.05\,\% and 0.3\,\%.}
    \label{figure:JJ2}
\end{figure*}

Together, these effects explain the drop in modeled rewards for
the narrowest allocations despite the intuitive appeal of a tighter
concentration.
Analyzing the experimental results, we derive the following key conclusions:

\begin{enumerate}
\item \emph{Concentration versus profitability.}
      A more concentrated liquidity placement generally yields
      higher fees, yet a clear profitability frontier exists: beyond a
      certain point negative drivers—impermanent loss (IL), downside
      price moves, and gas expenditure—can erode the initial capital
      \(W^{\mathrm{LP}}\) so severely that the earned fees fail to offset
      the losses (\ref{app:App11}).

\item \emph{Reward ceiling under the third modeling approach.}
      When modeling inside the historical-liquidity envelope the attainable fee level is capped by
      the historical pool rewards.
      With a large model capital \(W^{\mathrm{LP}}\) it might appear always beneficial to
      spread liquidity across a greater number of ranges
      (\emph{i.e.}\ larger~\(\tau\)); however, this impression is
      misleading.
      A high capital level merely causes the LP to occupy 100\,\% of
      several buckets while suffering less from negative drivers—a modeling artifact.
      The solution is straightforward: select \(W^{\mathrm{LP}}\ll\Sigma\).
      The present study intentionally chooses a high \(W^{\mathrm{LP}}\)
      to highlight this issue.

\item \emph{When is an ML strategy worthwhile?}
      \begin{enumerate}
        \item Increasing \(\tau\) (with fixed bucket width~\(d\)) lowers
              modeled fees for \emph{both} the ML and the uniform
              strategies, bringing their results closer together ($\tau$=40 or 100).
              Furthermore, a broader allocation demands that the ML model cover a
              larger volatility band over a longer horizon, forcing the
              optimal strategy towards a uniform placement.
        \item Conversely, a very narrow placement
              (\(\tau = 1\) or \(2\) for the present pool) can under-perform
              the uniform benchmark; the latter’s universality wins out
              on short prediction horizons.
              The case \(\tau=0\) is identical for both approaches.
        \item We formalize the concept of \emph{predictive-advantage area} (\emph{PAA}) can be identified
              in which the ML ensemble clearly outperforms the uniform
              strategy; for the USDC/ETH~0.3\,\% pool this occurs at
              \(\tau = 5\) and~\(10\), where the absolute fee difference
              is maximal.
      \end{enumerate}

\item \emph{Impact of pool liquidity.}
      Figure \ref{figure:JJ2} contrasts the current pool (USDC/ETH,
      0.3\,\%) with a second pool (USDC/ETH 0.05\%\footnote[3]{Contract 0x88e6A0c2dDD26FEEb64F039a2c41296FcB3f5640}) for
      \(\tau\le 10\).
      The second pool exhibits an order-of-magnitude larger trade count
      (134\,k swaps vs 8.4\,k) and a higher fee volume
      ($\sim$2.1m USDC vs $\sim$0.3m USDC) over the same OOT period.
      The price dynamics for the selected period in both pools are presented in~\ref{app:App12}.
      Despite the different fee tiers, the qualitative patterns are
      similar.
      Owing to the higher turnover, the identical ML architecture has
      greater scope to surpass the uniform baseline in the more liquid
      pool; its \emph{PAA} even extends to \(\tau = 1\). This also can be explained by the fact that higher TVL in more liquid pools reduces the price impact of non-arbitrage trades (non-market factor), thereby enhancing the predictive power of market-feature based ML models in such pools.
\end{enumerate}

\subsubsection{Results Across All Pools in the Study.}
The proposed methodology and modeling framework were applied to a set
of Uniswap V3 pools, including stable-coin pairs, WBTC/ETH and USDC/ETH
with different fee tiers.  
For every pool the initial LP capital was fixed at
\(W^{\mathrm{LP}} = 1m\ \text{USDC}\) over the same out-of-time period
(1 to 30 September 2024) with the same specific assumptions.  
The summary statistics are reported in Table~\ref{tab:t2}.

The results demonstrate the versatility of the approach across diverse
trading pairs and liquidity profiles.  
In all cases, the working capital dynamics was tracked without reinvesting the accrued fees between epochs, and the average approximation error
for historical pool volumes did not exceed 2\%.  
Stable \emph{PAA} area emerges at
\(\tau = 5\) and \(\tau = 10\), where the ML model strategy
outperforms the uniform baseline — a steady observation that holds across all
examined pools and highlights promising directions for future research on
optimal liquidity management in decentralized markets.

\begin{table}[!t]
\centering
\renewcommand{\thetable}{2} 
\caption{Comparative Analysis of Uniswap V3 Pools Metrics}
\label{tab:uniswap_analysis}
\setlength{\tabcolsep}{3pt}
\small
\begin{tabular}{@{}>{\raggedright\arraybackslash}p{2.2cm}*{4}{S[table-format=4.2]}c@{}}
\toprule[1.5pt]
\multirow{3}{*}{Metric} & \multicolumn{4}{c}{\footnotesize Trading Pools} & \\
\cmidrule(lr){2-5}
     & \multicolumn{1}{c}{\footnotesize\makecell{USDC/ETH\\0.3\%}}
     & \multicolumn{1}{c}{\footnotesize\makecell{USDC/ETH\\0.05\%}}
     & \multicolumn{1}{c}{\footnotesize\makecell{WBTC/ETH\\0.3\% \tablefootnote[4]{Contract 0xCBCdF9626bC03E24f779434178A73a0B4bad62eD}}}
     & \multicolumn{1}{c}{\footnotesize\makecell{USDC/USDT\\0.01\% \tablefootnote[5]{Contract 0x3416cF6C708Da44DB2624D63ea0AAef7113527C6}}} \\
\midrule[0.8pt]

Avg. TVL, \$M & 40 & 140 & 65 & 35 & \\
Swap Count, k & 8.4 & 134.4 & 3.2 & 26.7 & \\
Range Size ($d$) & {10 USDC} & {10 USDC} & {1.23e-4 WBTC } & {1.57e-4 USDC} & \\

\midrule[0.5pt]
\textbf{Trading Vol.} & \multicolumn{4}{c}{} & \multirow{4}{*}{\rottext{\scriptsize USD M}} \\
\indentitem Actual & 103.9 & 4179.1 & 124.1 & 595.1 & \\
\indentitem Model & 102.3 & 4073.1 & 121.9 & 583.8 & \\
\indentitem error & 1.6\% & 2.5\% & 1.7\% & 1.9\% & \\

\midrule[0.5pt]
\textbf{Pool Fees} & \multicolumn{4}{c}{} & \multirow{14}{*}{\rottext{\scriptsize thousand USD}} \\
\indentitem Actual & 311.8 & 2089.6 & 372.1 & 59.5 & \\
\indentitem Model & 306.9 & 2036.5 & 365.8 & 58.4 & \\
\addlinespace[0.1em]
\textbf{LP Fees*} & \multicolumn{4}{c}{} & \\
\indentitem $\tau=2$ 
    & 151.5/ 
    & 204.8/
    & 125.8/ 
    & 3.5/
    & \\
\indentitem 
    & \ \ \ \ \textbf{147.5$\scriptscriptstyle-3\%$}
    & \ \ \ \ \textbf{219.3$\scriptscriptstyle+7\%$}
    & \ \ \ \ \textbf{132.8$\scriptscriptstyle+6\%$}
    & \ \ \ \ \ \ \ \textbf{3.7$\scriptscriptstyle+5\%$}
    & \\
\indentitem $\tau=5$ 
    & \bfseries 90.5/
    & \bfseries 113.4/
    & \bfseries 64.6/
    & \bfseries 1.6/
    & \\
\indentitem
    & \ \ \ \ \ \textbf{104.5$\scriptscriptstyle+15\%$}
    & \ \ \ \ \ \textbf{139.4$\scriptscriptstyle+23\%$}
    & \ \ \ \ \ \ \ \textbf{79.5$\scriptscriptstyle+23\%$}
    & \ \ \ \ \ \ \ \ \ \ \textbf{1.9$\scriptscriptstyle+16\%$ }
    & \\
\indentitem $\tau=10$ 
    & \bfseries 50.1/
    & \bfseries 65.2/
    & \bfseries 37.1/
    & \bfseries 0.9/
    & \\
\indentitem 
    & \ \ \ \ \ \ \ \textbf{59.2$\scriptscriptstyle+18\%$}
    & \ \ \ \ \ \ \ \textbf{78.0$\scriptscriptstyle+20\%$}
    & \ \ \ \ \ \ \ \textbf{42.1$\scriptscriptstyle+13\%$}
    & \ \ \ \ \ \ \ \ \ \textbf{1.1$\scriptscriptstyle+17\%$}
    & \\
\addlinespace[0.1em]
\textbf{LP Costs} & \multicolumn{4}{c}{} & \\
\indentitem $\tau=2$ & 33.4 & 37.4 & 18.5 & 5.9 & \\
\indentitem $\tau=5$ & 15.4 & 18.7 & 9.0 & 1.7 & \\
\indentitem $\tau=10$ & 10.5 & 10.5 & 5.2 & 0.7 & \\

\midrule[0.5pt]
\textbf{Epochs} & \multicolumn{4}{c}{} & \\
\indentitem $\tau=2$ & 213 & 241 & 123 & 38 & \\
\indentitem $\tau=5$ & 45 & 55 & 27 & 5 & \\
\indentitem $\tau=10$ & 16 & 16 & 8 & 1 & \\
\bottomrule[1.5pt]
\end{tabular}

\begin{center}
\begin{minipage}{11.7cm}
\footnotesize
\noindent
*For LP Fees, first value shows results of uniform strategy, second shows \textbf{ML model} with \% additional LP fees compared to the uniform strategy. WBTC-denominated pools show values in USD, converted at the last market price in the OOT period.
\end{minipage}
\end{center}
\label{tab:t2}
\end{table}

\subsection{Liquidity Management Example: "As-Is" and Modified $\tau$-reset Strategies}
The primary focus of this research is on the quantitative assessment of LP rewards obtained through optimal liquidity placement based on historical data using ML models. However, as previously noted, this alone is insufficient for effective liquidity management as a whole. This is due, among other things, to a number of adverse factors discussed in Section~\ref{section:ExperimentsOOT}, which are characteristic of both concentrated liquidity market makers (e.g., IL) and reset strategies in particular (e.g., adverse capital revaluation when the pool price crosses the lower bound of an LP’s active range - downside risk). We will refer to these as "natural" adverse factors.
Furthermore, in order to obtain realistic out-of-time (OOT) backtesting results, it is necessary to account for the influence of other participants, rather than modeling all other LPs as a single static liquidity provider. Key examples of such influences include the following:

\begin{itemize}
    \item \textbf{Just-in-Time (JIT) liquidity} — a strategy in which other LPs add liquidity to the pool immediately prior to a swap execution to capture profit with minimal risk and then promptly remove it~\cite{ref_35}.
    
    \item \textbf{Maximal Extractable Value (MEV)} — a practice used by block producers (validators or miners) to increase their revenue by selectively including, excluding, or reordering transactions when forming a block, includes frontrunning and so-called sandwich attacks~\cite{ref_36}. It should be noted that JIT-strategy is part of MEV.
\end{itemize}

Although the examples listed above are full-fledged research topics in their own right, in the context of this work they serve merely as justification for the need to introduce additional assumptions in modeling. Ignoring interactions with other pool participants leads to an overestimation of strategy effectiveness during backtesting.

\begin{figure}[!t]
  \centering
  \includegraphics[width=0.9\linewidth]{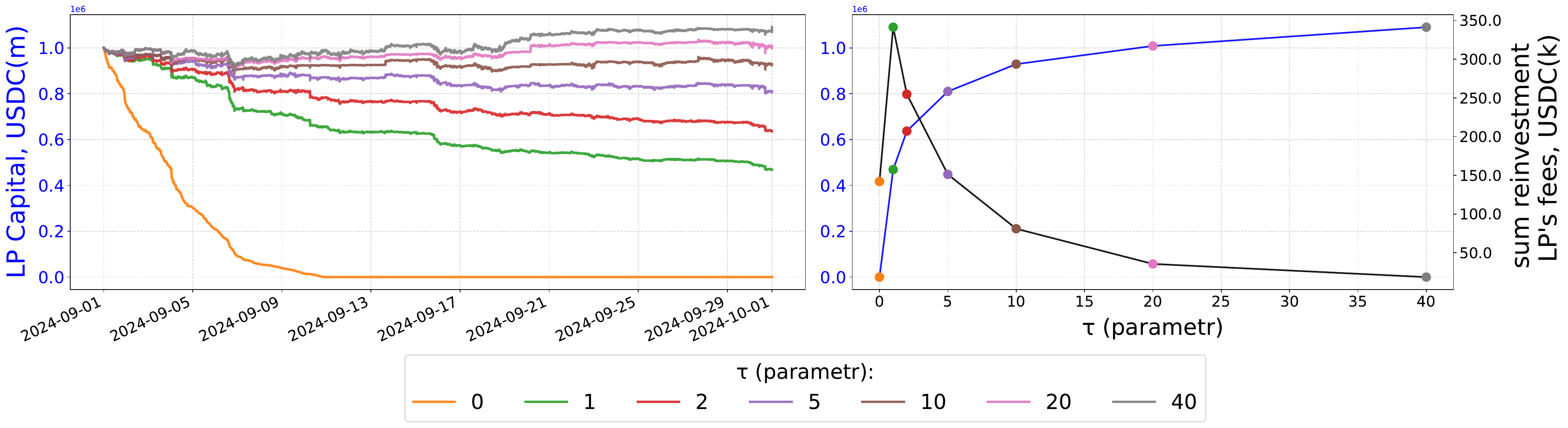}
  \captionsetup{justification=centering}
  \caption{The state of total liquidity positions in USDC depending on the $\tau$-parameter. \\ Application of "As-Is" $\tau$-reset strategies (integral models) on the USDC/ETH 0.05\% pool \\ with width of the bucket
$d=10$ USDC and \(W^{\mathrm{LP}} = 1m\ \text{USDC}\).}%
  \label{fig:A100}
\end{figure}

In this research, we take into account only the "natural" adverse factors. The left panel of Figure \ref{fig:A100} shows the dynamics of active LP capital in USDC when applying the "As-Is" strategies developed in the USDC / ETH 0.05\% pool, as a function of the parameter~$\tau$. Modeling here is similar to the Section~\ref{section:ExperimentsOOT}: the initial capital for all strategies is \(W^{\mathrm{LP}} = 1m\ \text{USDC}\), but with reinvestment of the rewards after each liquidity relocation (epoch). A clear inverse relationship is observed between capital preservation and the number of relocations, which is governed by the parameter~$\tau$. Choosing $\tau$, the LP determines the capital concentration and thereby sets the price curve with specific levels of convexity, as described in~\cite{ref_36_2}.

The right panel of Figure~\ref{fig:A100} also demonstrates an inverse relationship between the cumulative reinvested model LP rewards and the final level of LP capital in USDC at the end of the OOT period. Table~\ref{tab:tau_metrics2} presents comparative results for the application of the optimal "As-Is" strategies and the buy and hold (B\&H) strategy. Here, unlike the modeling of LP's rewards alone, the buy and hold strategy serves as a natural benchmark.
Additionally, we use classical performance metrics such as \textit{Max Drawdown} and the \textit{Sharpe Ratio} to evaluate and compare the effectiveness of the strategies.

\vspace*{-1.5em}
\begin{table}[htbp]
\centering
\renewcommand{\thetable}{3}
\captionsetup{justification=centering}
\caption{Analysis of strategies metrics depending on the $\tau$ parameter ("As-Is" integral models). \\ On the OOT period (September 2024)  for the pool USDC/ETH 0.05 $\%$}
\label{tab:tau_metrics2}
\setlength{\tabcolsep}{4pt}
\begin{tabular}{
    @{}l
    *{6}{S[table-format=5.1]}
    S[table-format=4.1]
    @{\hspace{4pt}\vrule\hspace{4pt}} 
    S[table-format=4.1]
    @{}c@{}
}
\toprule[1.5pt]
$\tau$-parameter & 0 & 1 & 2 & 5 & 10 & 20 & 40 & {B\&H} & \\
\midrule[0.8pt]
Epochs (reallocations) & 6646* & 641 & 241 & 55 & 16 & 4 & 1 & \textemdash & \multirow{5}{*}{\hspace{4pt}\rotatebox[origin=c]{90}{\scriptsize th. USD}} \\
Costs & 143.4 & 58.9 & 37.4 & 18.7 & 10.5 & 5.3 & 2.7 & \textemdash & \\
Sum reinvest. fees & 142.1 & 341.3 & 254.7 & 151.2 & 81.1 & 35.6 & 18.6 & \textemdash & \\
LP capital & 0.0 & 469.0 & 636.8 & 809.8 & 929.0 & 1008.4 & 1090.2 & \bfseries 1035.5 & \\
\midrule[0.1pt]
Compound annual returns & \scriptsize -100.0\% & \scriptsize -100.0\% & \scriptsize -99.6\% & \scriptsize -92.0\% & \scriptsize -58.7\% & \scriptsize 10.5\% & \scriptsize 181.9\% & \scriptsize 51.9\% & \\
Max Drawdown (MDD) & \scriptsize -100.0\% & \scriptsize -53.3\% & \scriptsize -36.5\% & \scriptsize -19.9\% & \scriptsize -10.9\% & \scriptsize -10.3\% & \scriptsize -9.9\% & \scriptsize -16.2\% & \\
Sharpe Ratio (annual)** & \scriptsize \textemdash & \scriptsize \textemdash & \scriptsize -13.5 & \scriptsize -7.7 & \scriptsize -3.1 & \scriptsize 0.4 & \scriptsize 3.4 & \scriptsize 0.9 & \\
\bottomrule[1.5pt]
\end{tabular}

\vspace{0.4em}
\begin{minipage}{14.5cm}
\footnotesize
\noindent
* {} 16667 epochs without complete depletion of capital. \\
** with risk-free rate 0\% and 365 crypto trading days per year.
\end{minipage}
\end{table}

The analysis conducted demonstrates the destructive impact of ill-considered liquidity management: excessive relocation frequency is a key driver of accelerated depletion of active LP capital—even when liquidity is placed optimally in theory and expected LP's rewards are high, the strategy quickly loses its effectiveness.

Let us now consider in more detail the sensitivity of capital preservation to relocation frequency in the case of $\tau = 5$, for which, according to the results of Table~\ref{tab:uniswap_analysis}, the use of advanced ML models is more justified compared to a uniform strategy. To this end, we introduce several modifications of the $\tau$-reset strategy aimed at reducing sensitivity to frequent relocations while preserving the generalization ability of the ML model:

\begin{itemize}
    \item $\tau+\eta$ - reset strategy: in addition to the $2\tau+1$ \emph{LP-liquid buckets} defined by the methodology in Section~\ref{sec:bst}, $2\eta$ empty "side" ranges are introduced. Crossing into these ranges does not trigger a liquidity relocation. In this case, the set of \emph{LP-liquid buckets} is $\beta_L^{\mathrm{LP}} = \{B_{M-\tau-\eta}, \dots, B_{M+\tau+\eta}\}$, where the allocation weights $\alpha_{M-\tau-\eta+k}^{\mathrm{LP}} = \alpha_{M+\tau+\eta-k}^{\mathrm{LP}} = 0$ for $k \in [0, \eta-1]$, and $M$ denotes the index of the reference bucket. The total coverage thus becomes $2(\tau + \eta) + 1$ buckets. This symmetric strategy reduces excessive relocations during sharp price swings; however, capital remains subject to the "natural" adverse factors, although to a lesser extent.

    \item $\tau+\eta^{\mathrm{up}}$ and $\tau+\eta^{\mathrm{down}}$ - reset strategies: in these asymmetric variants, the empty protective buckets are placed only in one direction: either toward the price increase or decrease, relative to the reference bucket. For instance, in the case of $\tau+\eta^{\mathrm{down}}$, we have $\beta_L^{\mathrm{LP}} = \{B_{M-\tau-\eta^{\mathrm{down}}}, \dots, B_{M+\tau}\}$, with $\alpha_{M-\tau-\eta^{\mathrm{down}}+k}^{\mathrm{LP}} = 0$ for $k \in [0, \eta^{\mathrm{down}} - 1]$. The total coverage thus becomes in this case is $2\tau + \eta^{\mathrm{up/down}} + 1$ buckets. When $\eta^{\mathrm{down}} > 0$, the strategy reduces relocations during sharp price declines and utilizes capital more efficiently during price increases.
\end{itemize}

Table~\ref{tab:eta_metrics} presents a comparison of key metrics for the integral model using the $\tau+\eta^{\mathrm{down}}$-reset strategy with $\tau = 5$, across different values of the $\eta^{\mathrm{down}}$ parameter:

\vspace*{-1.5em}
\begin{table}[htbp]
\centering
\renewcommand{\thetable}{4}
\captionsetup{justification=centering}
\caption{Analysis of strategies metrics depending on the $\eta^{down}$ parameter (integral model). \\ On the OOT period (September 2024) for the pool USDC/ETH 0.05 $\%$, $\tau=5$ and $W^{\mathrm{LP}} = 1$\,m USDC}
\label{tab:eta_metrics}
\setlength{\tabcolsep}{4pt}
\begin{tabular}{
    @{}l
    *{9}{S[table-format=4.1]}
    @{\hspace{4pt}\vrule\hspace{4pt}} 
    S[table-format=4.1]
    @{}c@{}
}
\toprule[1.5pt]
$\eta^{down}$-parameter & 0 & 5 & 10 & 15 & 20 & 25 & 30 & 35 & 40 & {B\&H} & \\
\midrule[0.8pt]
Epochs (reallocations) & 55 & 31 & 21 & 13 & 9 & 11 & 11 & 4 & 4 & \textemdash & \multirow{7}{*}{\hspace{4pt}\rotatebox[origin=c]{90}{\scriptsize th. USD}} \\
Positive realloc. & \scriptsize 51.9\% & \scriptsize 66.7\% & \scriptsize 75.0\% & \scriptsize 83.3\% & \scriptsize 87.5\% & \scriptsize 90.0\% & \scriptsize 90.0\% & \scriptsize 100.0\% & \scriptsize 100.0\% & \textemdash \\
Costs & 18.7 & 10.7 & 7.2 & 4.5 & 3.2 & 3.8 & 3.8 & 1.5 & 1.5 & \textemdash \\
Sum reinvest. fees & 151.2 & 119.7 & 95.4 & 72.7 & 86.6 & 85.6 & 65.6 & 35.9 & 35.9 & \textemdash \\
LP capital & 809.8 & 868.5 & 917.0 & 989.4 & 1054.4 & 1009.7 & 999.1 & 1093.4 & 1093.4 & \bfseries 1035.5 & \\
\midrule[0.1pt]
Compound annual returns & \scriptsize -92.0\% & \scriptsize -81.6\% & \scriptsize -64.7\% & \scriptsize -12.0\% & \scriptsize 88.9\% & \scriptsize 12.2\% & \scriptsize -1.1\% & \scriptsize 192.1\% & \scriptsize 192.1\% & \scriptsize 51.9\%  & \\
Max Drawdown (MDD) & \scriptsize -19.9\% & \scriptsize -17.9\% & \scriptsize -13.9\% & \scriptsize -16.3\% & \scriptsize -11.8\% & \scriptsize -13.5\% & \scriptsize -12.6\% & \scriptsize -12.9\% & \scriptsize -12.9\% & \scriptsize -16.2\% & \\
Sharpe Ratio (annual)* & \scriptsize -7.7 & \scriptsize -3.8 & \scriptsize -2.3 & \scriptsize -0.2 & \scriptsize 1.7 & \scriptsize 0.3 & \scriptsize 0.05 & \scriptsize 2.5 & \scriptsize 2.5 & \scriptsize 0.9 & \\
\bottomrule[1.5pt]
\end{tabular}

\vspace{0.4em}
\begin{minipage}{16.5cm}
\footnotesize
\noindent
* with risk-free rate 0\% and 365 crypto trading days per year.
\end{minipage}
\end{table}

The results show that deliberately reducing sensitivity to sharp but temporary price drops leads to a decreased number of relocations (epochs), while the final capital at the end of the OOT period exhibits an upward trend. Moreover, as $\eta^{\mathrm{down}}$ increases, the number of relocations triggered by price increases surpasses those triggered by price declines. This further contributes to capital preservation, since relocations during price drops lead to the conversion of liquidity into a depreciating risky asset, thereby locking in losses—a manifestation of the second "natural" adverse factor. We observe that the application of the asymmetric strategy outperforms both the benchmark and the classical $\tau$-reset strategy, delivering more robust results when accounting for reallocation fees and "natural" adverse factors.

\begin{figure}
  \centering
  \includegraphics[width=0.9\linewidth]{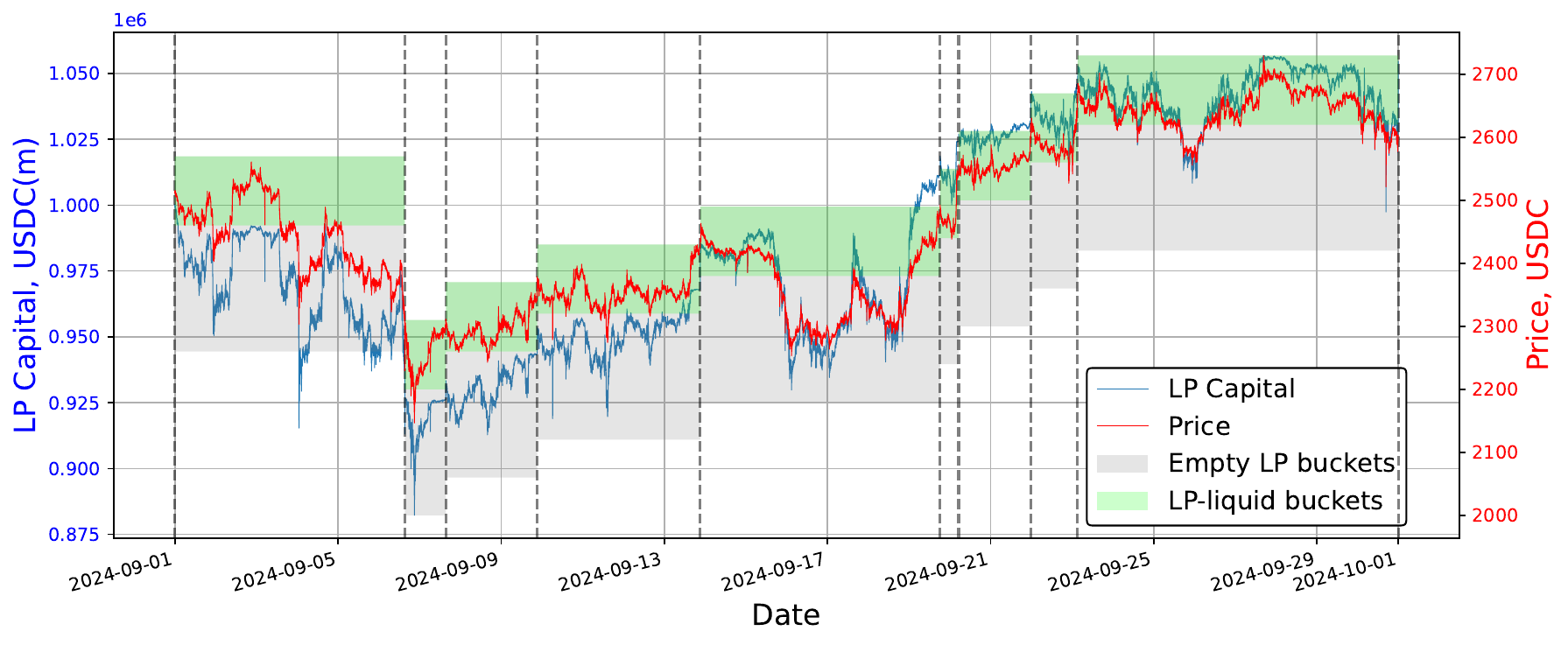}
  \captionsetup{justification=centering}
  \caption{The state of total liquidity positions in USDC. Application of $\tau+\eta^{\mathrm{down}}$ - reset strategies (integral model) on the USDC/ETH 0.05\% pool with $\tau = 5$, $\eta^{\mathrm{down}} = 20$ and \(W^{\mathrm{LP}} = 1m\ \text{USDC}\).}%
  \label{fig:A200}
\end{figure}

This approach maintains the predictive power of the ML model, unlike the naïve increase in $\tau$, which effectively causes the performance of the integral model to converge toward that of a uniform strategy (see Section~\ref{section:ExperimentsOOT}). Figure~\ref{fig:A200} shows the dynamics of active LP capital under the $\tau+\eta^{\mathrm{down}}$ - reset strategy with $\eta^{\mathrm{down}} = 20$.

As seen in the figure, the modified strategy utilizes capital efficiently: for a significant portion of time, the price remains within the \emph{LP-liquid buckets}, while the empty buckets act as a protective buffer. This modification of the baseline strategy delivers better performance metrics compared to the buy and hold strategy, and unlike configurations with $\eta^{\mathrm{down}} = 35$ or $40$, it remains more dynamic. It is important to note that the $\tau+\eta$, $\tau+\eta^{\mathrm{up}}$ and $\tau+\eta^{\mathrm{down}}$-reset strategies are merely extensions of the baseline $\tau$-reset strategy and thus do not require retraining of the ML models developed earlier. 
As a result of the analysis, the following conclusions can be formulated:
\begin{itemize}
    \item Reducing the frequency of relocations and prioritizing relocations during upward price movements are effective capital protection measures. However, these measures inherently favor static liquidity provision strategies, thereby diminishing the added value of using advanced ML models.
    \item In the context of dynamic liquidity management, applying the methodology proposed in this research—alongside hedging mechanisms—appears to be a necessary condition for achieving sustainable returns and outperforming passive strategies. Otherwise, observed results may merely reflect coincidental market conditions and exhibit high volatility, similar to the behavior of the buy and hold strategy.
\end{itemize}

This extended problem formulation—constructing an integrated active liquidity placement strategy with hedging—will be the subject of our follow-up research.

\subsection{Comparison with an Alternative Tool}

\begin{figure}[!b]
  \centering
  \includegraphics[width=0.9\linewidth]{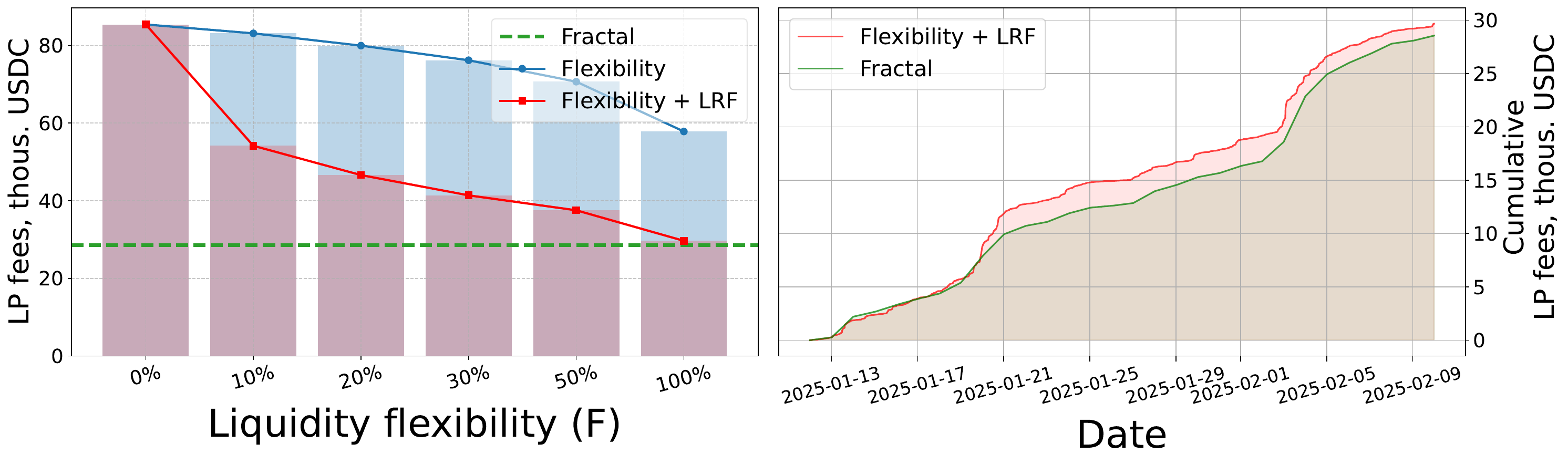}
  \captionsetup{justification=centering}
  \caption{Backtested LP fees produced by the present framework and by
           \texttt{Fractal} under varying modeling assumptions
           (USDC/ETH 0.3\,\%, 12~Jan–10~Feb 2025).}%
  \label{fig:JJ3}
\end{figure}

To assess general validity, the proposed methodology was compared with
$\texttt{Fractal}$ - a modular backtesting library for DeFi strategies
released by Logarithm Labs \cite{ref_34}.
The key difference between the two tool chains is that
$\texttt{Fractal}$ directly ingests historical per-range liquidity,
whereas the present framework requires only trade prices and volumes,
recovering the liquidity surface via the approximation approach of
Section~\ref{sub:RApr}.

\texttt{Fractal} employs a forward-simulation approach to estimate the LP rewards based on historical Uniswap V3 pool activity from The Graph. The framework iteratively ingests historical data (observations in terms of Fractal): including swap volumes, tick prices, and liquidity snapshots, and then updates the internal state of Uniswap V3 entity, producing backtested performance metrics over time.

The core of fee modeling is aligned with the methodology outlined in the Uniswap V3 whitepaper \cite{ref_2}. To model liquidity placement, the deposit amount is decomposed into token \(A\) and token \(B\) allocations, and the strategy's liquidity contribution is calculated using the canonical Uniswap V3 liquidity formulae. 
The potential fees earned from the strategy are calculated accounting for the added LP liquidity.
This value is then integrated into the Uniswap V3 fee model to estimate the fee income attributable to the LP's position for a given fee tier and price range. By iterating this process over the full historical time series, \texttt{Fractal} produces backtested strategy results that reflect dynamic fee accruals under realistic market conditions.

The USDC/ETH 0.3$\%$ pool examined earlier was re-evaluated on the
period 12~January to 10~February 2025 under identical conditions:
\(W^{\mathrm{LP}} = 1m\ \text{USDC}\) and a single epoch
(\(K=1\)), with one uniform bucket spanning the full price range.
Figure \ref{fig:JJ3} contrasts the backtested fee income obtained from
the two instruments.
The left panel shows the sensitivity of the modeled LP fees to the
liquidity–flexibility parameter~\(F_{e_1}\) and \emph{LRF} constraint; the
right panel shows the cumulative fees trajectories over time.
With maximum flexibility (\(F_{e_1}=1\)) the two approaches differ by less
than 4\% (28\,562.3 USDC vs 29\,674.2 USDC).

Close agreement with a tool that relies on actual historical
liquidity provides additional evidence for the validity of the
approximation-based methodology developed in this study.

\subsection{Results and Analysis}
\label{subsection:results_analysis}

Our experimental evaluation demonstrates the effectiveness of the proposed methodology across multiple dimensions:

\textbf{Liquidity Approximation Accuracy.} Across all tested pools, our parametric reconstruction method achieved historical fee approximation errors of, on average, about 2\%, validating the reliability of our approach even without access to historical liquidity data. This represents a significant practical advantage, as complete historical liquidity states are often unavailable or costly to obtain.

\textbf{Machine Learning Strategy Performance.} The integrated ML ensemble consistently outperformed uniform benchmark strategies in the predictive-advantage area (PAA), typically corresponding to $\tau$ values between 5 and 10. For the USDC/ETH 0.3\% pool at $\tau=5$, the ML strategy achieved 15.5\% higher fees than the uniform benchmark, while at $\tau=10$ the improvement reached 18.4\% on the OOT period with given range size parameter
$d$ from Table~\ref{tab:t2}. Similar patterns emerged across other pools, with improvements ranging from 13\% to 23\% depending on the pool characteristics and $\tau$ parameter.

\textbf{Parameter Sensitivity Analysis.} We identified a clear trade-off between liquidity concentration and profitability. While tighter concentration ($\tau=1,2$) often underperformed due to impermanent loss and downside risk, and broader allocation ($\tau\geq40$) converged toward uniform strategy performance, intermediate values ($\tau=5,10$) provided the optimal balance, maximizing ML model advantage.

\textbf{Pool Characteristics Impact.} More liquid pools (e.g., USDC/ETH 0.05\%) showed greater absolute fee improvements and extended PAA ranges, suggesting that our methodology scales effectively with pool activity. The framework maintained consistent performance across diverse asset pairs including stablecoins (USDC/USDT) and volatile pairs (WBTC/ETH).

\textbf{Modified Strategy Enhancements.} By introducing asymmetric $\tau+\eta^{\mathrm{down}}$-reset strategies, we demonstrated further performance improvements. For the pool USDC/ETH 0.05\% with $d=10$ USDC, $\tau=5$ and $\eta^{\mathrm{down}}=20$, the modified strategy achieved 88.9\% annualized returns compared to 51.9\% for buy-and-hold, while reducing maximum drawdown from 16.2\% to 11.8\%, and significantly improving over the “As-Is” strategy for $\tau=5$ with annualized returns of $-92.0\%$.

\textbf{Validation Against Alternative Tools.} Comparison with the Fractal backtesting library showed close agreement (error within 4\%), providing external validation of our liquidity approximation methodology despite its independence from historical liquidity data.

These results collectively demonstrate that our framework provides a robust, practical solution for optimizing liquidity provision in CLMMs, with consistent outperformance across diverse market conditions and pool characteristics.

\subsection{Overall Architecture of the Decision-Making System}
\begin{figure}
  \centering
  \includegraphics[width=0.7\linewidth]{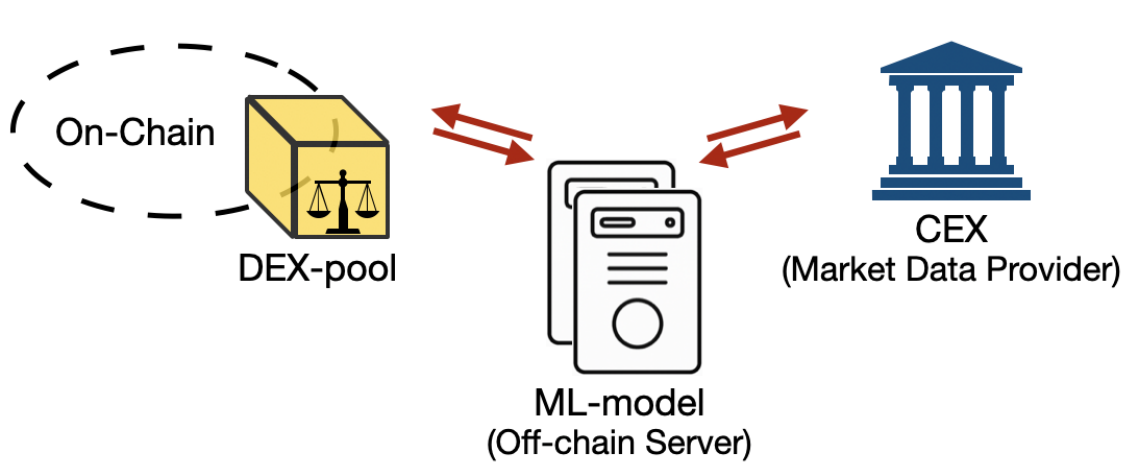}
  \captionsetup{justification=centering}
  \caption{Simplified interaction diagram for the ML-driven liquidity
           management system.}
  \label{fig:JJ4}
\end{figure}

In general terms, Figure \ref{fig:JJ4} outlines how the proposed ML model is deployed within a DEX pool and consists of the following components:
\begin{enumerate}
  \item \emph{DEX pool (on-chain).}  
        Holds the liquidity position and the current LP strategy.
        Whenever the pool price exits the \emph{LP-liquid buckets} and triggers the \(\tau\)-reset rule, an event is emitted to an off-chain server.

  \item \emph{ML engine (off-chain).}  
        Upon receiving the on-chain event, the server queries an external market data source (eg, a CEX) to build the feature vector for the current decision point.

  \item \emph{Strategy update.}  
        The ML model infers the optimal LP allocation and submits the corresponding transaction to the DEX, thereby rebalancing the liquidity ranges.
\end{enumerate}

From a technical standpoint, the interaction between the components can be organized as follows:
\begin{itemize}
  \item Communication between the DEX pool and the off-chain server can rely on standard Web3 / JSON-RPC interfaces
(e.g. an Ethereum node\footnote[6]{https://www.alchemy.com/overviews/rpc-node}).
  \item Market data is retrieved from CEX through REST or WebSocket endpoints (OHLCV streams, order-book snapshots, etc.).
  \item Liquidity reallocation can be executed through a dedicated smart contract wrapper or by directly interacting with the pool position manager contract.
\end{itemize}
  The external data feed does not need to be a centralized exchange; decentralized oracles or DEX aggregators can serve the same purpose while reducing dependence on a single data provider.

\section{Discussion and Implications}
\label{section:discussion}

\subsection{Theoretical Implications for Market Microstructure}
Our findings have several important implications for market microstructure theory in decentralized exchanges.

The consistent outperformance of moderate $\tau$ values over both narrower and broader allocations demonstrates the existence of an optimal concentration level that balances fee capture against impermanent loss risk. This finding parallels inventory management decisions by traditional market makers, who must balance profit from bid-ask spreads against adverse price movement risk. The similarity suggests that despite different institutional arrangements, fundamental economic principles govern liquidity provision decisions across market structures.

The success of our machine learning approach--which uses centralized exchange data to inform decentralized market strategies--highlights the interconnectedness of these markets. This finding contributes to ongoing debates about price discovery in decentralized versus centralized exchanges by demonstrating that informational efficiency in one venue can be leveraged for strategic advantage in another.

The sensitivity of strategy performance to $\tau$ parameters has implications for market stability. If many LPs adopt similar optimal parameters, liquidity could become concentrated in similar price ranges, potentially decreasing the price impact for small or medium sized trades but increasing price impact for significantly large trades. This concentration effect warrants further study regarding market resilience during periods of high volatility. The effect should be taken into account in further studies considering the mutual influence of liquidity providers on each others' strategies. 

\subsection{Practical Implications}
Our framework provides LPs with actionable tools for strategy development. The identification of predictive-advantage areas shows that the $\tau$-reset strategy can be tuned for specific market conditions, while asymmetric strategy benefits ($\tau$+$\eta$ modifications) provide mechanisms for managing downside risk. Practically, LPs can implement these insights through automated strategy execution, potentially improving risk-adjusted returns.

AMM protocol designers can use our findings to inform future developments. The demonstrated importance of balancing concentration benefits against transaction costs and impermanent loss suggests protocols might benefit from mechanisms that help LPs optimize this balance automatically. Additionally, our liquidity reconstruction methodology could be integrated into protocol analytics, improving transparency.

As institutional participation in DeFi grows, our framework offers a rigorous approach to evaluating liquidity provision as an asset class. The ability to quantify historical performance accurately, account for relevant costs, and optimize strategy parameters provides institutional investors with analytical tools needed for informed allocation decisions.

\subsection{Limitations and Future Research}
While our framework advances the analysis of liquidity provision in CLMMs, several limitations warrant acknowledgment.

Our current model simplifies interactions among multiple LPs. Future work could incorporate game-theoretic elements to analyze strategic interactions more completely, potentially using mean-field approaches or agent-based modeling.

Our analysis focuses on single-protocol strategies. In practice, LPs often deploy capital across multiple protocols simultaneously. Developing frameworks for cross-protocol optimization represents an important extension.

As the boundary between DeFi and traditional finance continues to blur, research exploring integrated strategies--such as using traditional derivatives to hedge impermanent loss--will become increasingly valuable.

Our findings regarding optimal concentration levels could inform the design of next-generation AMM protocols. Research exploring how protocol mechanisms might automatically optimize liquidity distribution based on market conditions could significantly improve capital efficiency and market quality.

\section{Conclusion}
\label{section:Conclusions}

The transformation of liquidity provision from passive activity to active portfolio management in concentrated liquidity AMMs represents a significant development in market microstructure. This evolution creates both challenges and opportunities for market participants, requiring new analytical tools for performance evaluation and strategy optimization.

This paper has developed and validated a comprehensive framework for addressing these challenges. Our parametric liquidity reconstruction methodology enables accurate historical performance analysis even when complete liquidity data is unavailable, addressing a critical practical limitation in DeFi research. Applying this framework to $\tau$-reset strategies across multiple market conditions, we identify consistent patterns in optimal strategy parameters and demonstrate the potential for machine learning to enhance strategy performance.

Beyond methodological contributions, our findings offer important insights into the economics of automated market making. The existence of predictive-advantage areas in strategy parameter space reveals how liquidity providers can balance competing risks and returns. The effectiveness of asymmetric strategy modifications demonstrates how targeted design choices can mitigate specific risks. And the interconnectedness of centralized and decentralized markets, evidenced by our machine learning approach, highlights the complex informational relationships in modern financial ecosystems.

As decentralized finance continues to evolve and integrate with traditional financial systems, frameworks like the one developed here will become increasingly important for participants, researchers, and regulators alike. By bridging methodological innovation with financial economic insights, this work contributes to our understanding of how market making adapts to new technological paradigms while remaining grounded in fundamental economic principles.

\newpage
\appendix                 
\setcounter{section}{0}   
\renewcommand\thesection{Appendix~\Alph{section}} 

\section{The state of total liquidity positions versus accumulated fees during the out‑of‑time (OOT) simulation period}\label{app:App11}

\begin{figure}[!ht]
  \centering
  \includegraphics[width=.9\textwidth]{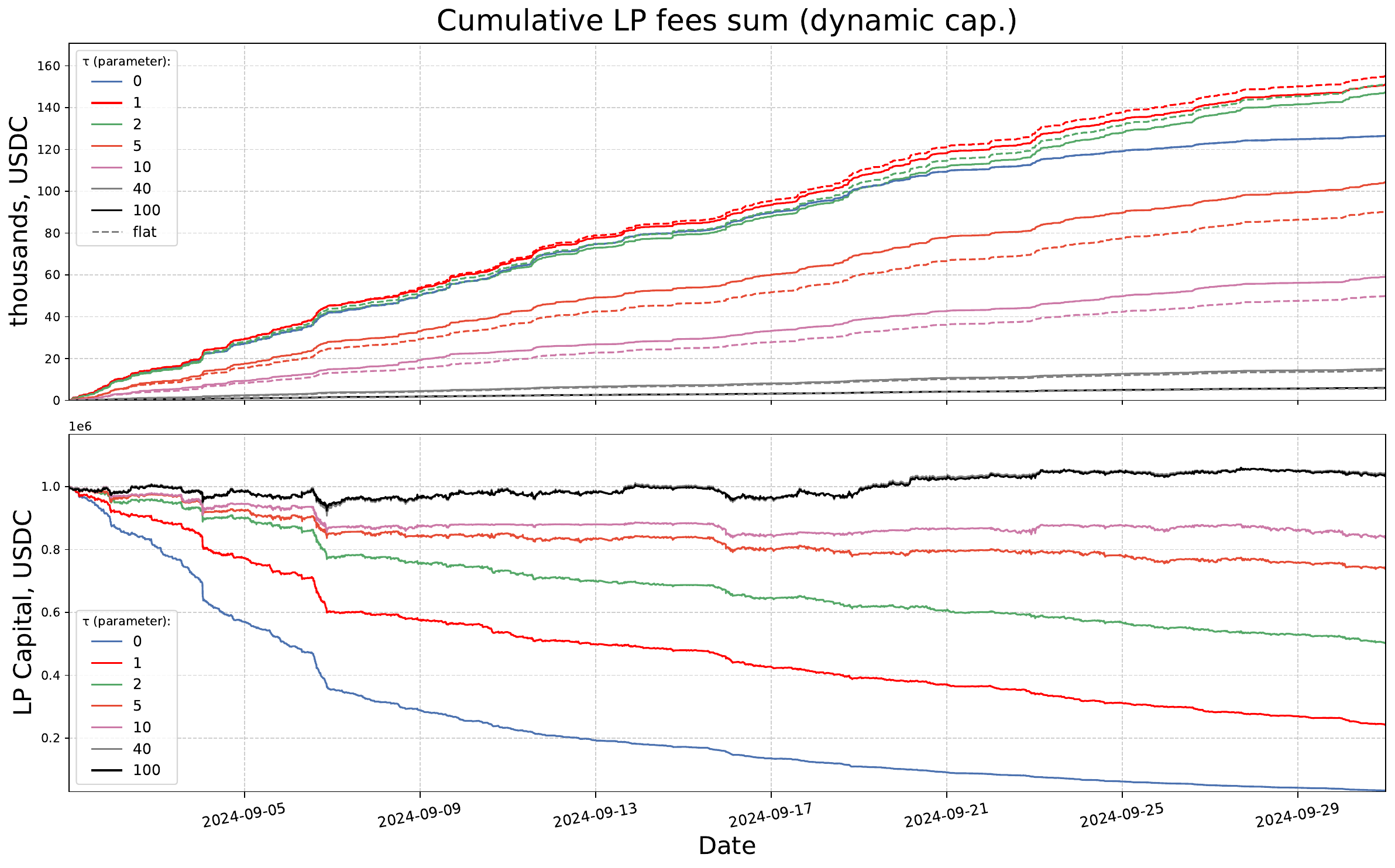}
\end{figure}

\section{Comparative price analysis between the USDC/ETH 0.05\,\% and 0.3\,\% pools during the OOT period}\label{app:App12}

\begin{figure}[!ht]
  \centering
  \includegraphics[width=.9\textwidth]{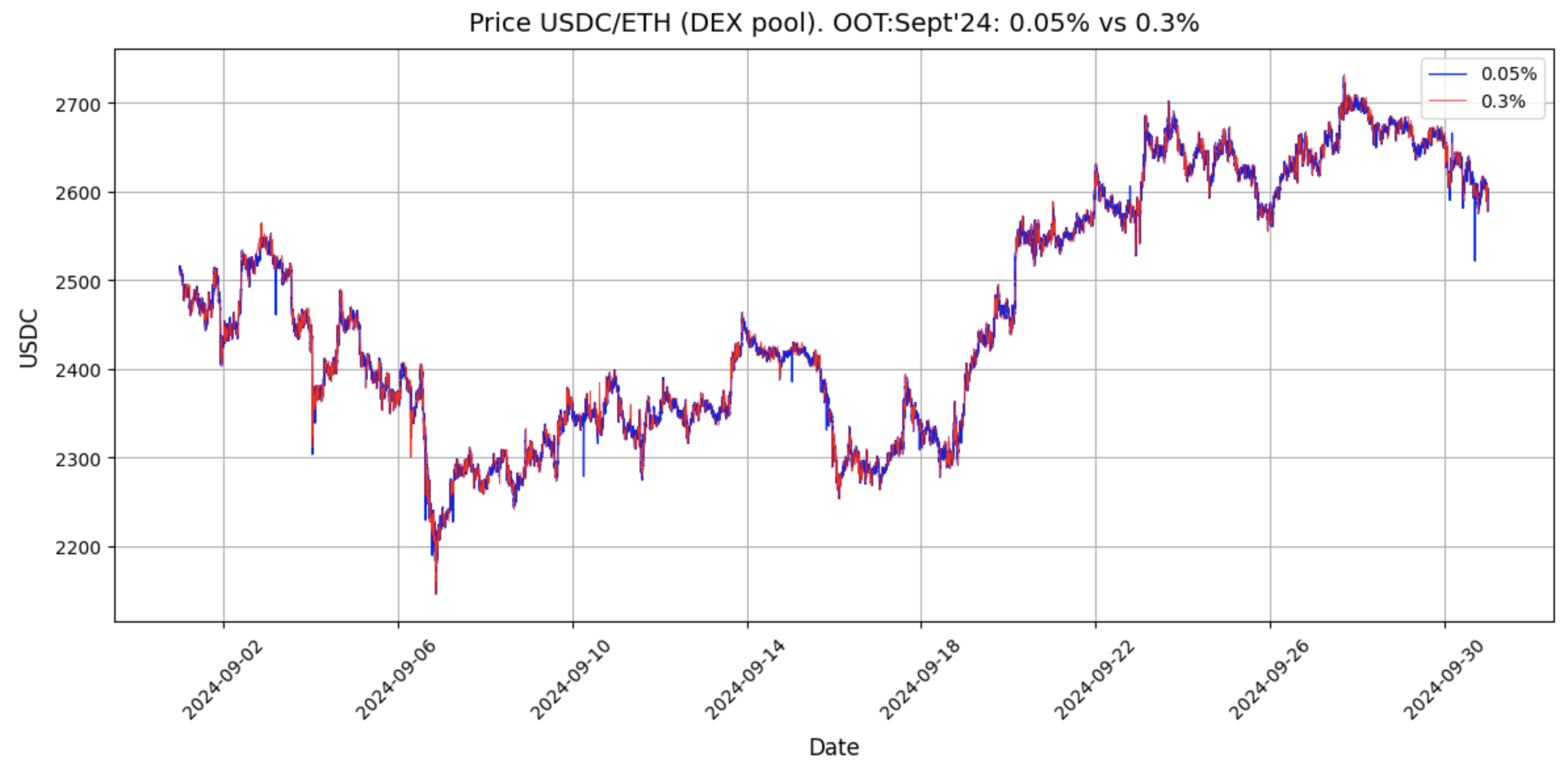}
\end{figure}

\section*{}
\addcontentsline{toc}{section}{Notation}
\renewcommand{\arraystretch}{1.1}
\setlength{\tabcolsep}{6pt}

\begin{longtable}{@{}p{4.2cm}p{9.8cm}@{}}
\toprule
\textbf{Symbol} & \textbf{Description} \\ \midrule
\endfirsthead
\multicolumn{2}{@{}l}{\textit{(continued)}}\\ \toprule
\textbf{Symbol} & \textbf{Description}\\ \midrule
\endhead
\bottomrule
\endfoot

$\Gamma$ & Fee tier of the DEX pool (\textit{in contract terms})\\
$\bm P=\{p_0,\dots,p_H\}$ & Set of historical pool prices over the modeling period\\
$\bm V=\{v_0,\dots,v_H\}$ & Set of historical trade volumes aligned with~$\bm P$\\
$\Sigma$ & Average TVL of the pool during the modeling period\\
$\beta=\{B_1,\dots,B_N\}$ & Set of partition of the price axis into $N$ buckets\\
$p_{a_i},p_{b_i}$ & Lower/upper bounds of the $i$‑th bucket $B_i\in\beta$\\
$d$ & Constant width of each bucket in~$\beta$\\
$W^{\mathrm{LP}}$ & Liquidity provider’s capital (in the numéraire)\\
$L^{\mathrm{LP}}$ & Liquidity provided by the LP\\
$B_M$ & Reference bucket\\
$\tau$ & General parameter of the $\tau$‑reset strategy (number of ranges)\\
$\beta^{\mathrm{LP}}=\{B_{M-\tau},\ldots,B_{M+\tau}\}$ & Set of LP-liquid buckets\\
$K$ & Number of epochs in the modeling period\\
$\gamma_i$ & Index of the last price at $i$-th epoch\\
$\bm E=\{e_1,\dots,e_K\}$ & Ordered list of epochs\\
$\bm T_{\bm E}=(t_{e_1},\dots,t_{e_K})$ & Vector of epoch durations\\
$\bm Q_{\bm E}=(q_{e_1},\dots,q_{e_K})$ & Vector of epoch lengths (in number of trades)\\
$W^{\mathrm{LP}}_i$ & LP capital allocated to bucket $B_i\in\beta$\\
$x_i,y_i$ & \emph{Real} reserves of tokens A and B in bucket $B_i\in\beta$\\
$L_i$ & Liquidity allocated to bucket $B_i\in\beta$\\
$p$ & Current pool price\\
$\mathcal V_i(L_i,p,B_i)$ & Liquidity–state function returning $(x_i,y_i)$ for given~$p$\\
$\bm W_{e_i}^{\Sigma}=(w_1^{\Sigma},\dots,w_N^{\Sigma})$ & Vector of $\Sigma$ across buckets during epoch $e_i$\\
$\bm L_{e_i}^{\Sigma}=(L_1^{\Sigma},\dots,L_N^{\Sigma})$ & Liquidity vector corresponding to $\bm W_{e_i}^{\Sigma}$\\
$\Delta \mathcal{V}_{n}^{q}$ & Increment of \emph{real} reserves in bucket $B_n$ at price change\\
$\bm S_{e_i}^{\Sigma}$ & Vector of token‑wise pool fees in epoch $e_i$\\
$\bm{m}_{e_i}$ & Vector of token prices expressed in
the chosen numéraire \\
$F_{e_i}^{\Sigma}$ & Value of $\bm S_{e_i}^{\Sigma}$ in the numéraire\\
$F_{\bm E}^{\Sigma}$ & Aggregate pool fees over modeling period\\
$W_{e_i}^{\mathrm{LP}}$ & LP capital at the start of epoch $e_i$\\
$\bm W_{e_i}^{\mathrm{LP}}=(w_1^{LP},\dots,w_N^{LP})$ & Bucket‑wise vector of $W_{e_i}^{\mathrm{LP}}$\\
$\mathbf r_{e_i}^{\mathrm{LP}}=(r_1^{LP},\dots,r_N^{LP})$ & Vector of LP share in total pool liquidity per bucket\\
$\bm S_{e_i}^{\mathrm{LP}}$ & Vector of token‑wise LP fees for epoch $e_i$\\
$F_{e_i}^{\mathrm{LP}}$ & Value of $\bm S_{e_i}^{\mathrm{LP}}$ in the numéraire\\
$F_{\bm E}^{\mathrm{LP}}$ & Aggregate LP fees over the modeling period\\
$\bm V_{\bm E}^{\text{hist}}=(v_{e_1}^{\text{hist}},\dots,v_{e_K}^{\text{hist}})$ & Vector of historical trade volumes per epoch\\
$\bm F_{\bm E}^{\text{hist}}=(f_{e_1}^{\text{hist}},\dots,f_{e_K}^{\text{hist}})$ & Vector of historical pool fees per epoch\\
$\bm\alpha_{e_i}^{\Sigma}$ & Weight vector of buckets (entire pool) in epoch $e_i$\\
$f_{\mathcal G}(\mu_{e_i},\sigma_{e_i})$ & Gaussian mapping that produces $\bm \alpha_{e_i}^{\Sigma}$\\
$f_{e_i}^{m\Sigma}$ & Model fees of pool for epoch $e_i$\\
$(\mu_{e_i}^\ast,\sigma_{e_i}^\ast)$ & Approximated parameters under which $|f_{e_i}^{m\Sigma}-f_{e_i}^{\text{hist}}|$\\
$\bm\alpha_{e_i}^{\mathrm{LP}}$ & LP weight vector over buckets in epoch $e_i$\\
$f_{e_i}^{\mathrm{LP}}(\varphi^\tau)$ & LP fees under strategy $\varphi^\tau$ in epoch $e_i$\\
$W^{\mathrm{LP}}_{\text{end},e_i}$ & LP capital at the end of epoch $e_i$ \\
$\mathbf{S}_{e_i}^\tau$ & Family of random $\tau$‑reset strategies\\
$\varphi_{e_i}^{\mathrm{OPT}}$ & Optimal LP strategy in epoch $e_i$\\
$\bm{\rho}_{e_i}^{\mathrm{LP}}$ & $(2\tau{+}1)-$size vector of non‑zero elements of $\bm\alpha_{e_i}^{\mathrm{LP}}$\\
$\bm{\hat\rho}_{e_i}^{\mathrm{LP}}$ & $(\tau{+}1)-$size vector of symmetry‑reduced $\bm{\rho}_{e_i}^{\mathrm{LP}}$\\
$Y$ & Vector of optimal strategies over modeling period\\
$\bm \psi_{e_i}$ & Feature vector describing market state at start of epoch $e_i$\\
$X$ & Feature matrix for the ML model\\
$\varphi_{e_i}^{\text{mOPT}}$ & ML‑predicted optimal strategy for epoch $e_i$\\
$m$ & Number of sub‑epochs in epoch $e_i$\\
$F_{e_i}$ & Liquidity flexibility metric of epoch $e_i$\\
$\eta$ & Parameter of the symmetric strategy (number of empty ranges)\\
$\eta^{\mathrm{up}}, \eta^{\mathrm{down}}$ & Parameters of the asymmetric strategy (number of empty ranges)\\
\emph{LRF} & Liquidity‑rule form constraint\\
\emph{PAA} & Predictive‑advantage area\\
\bottomrule
\end{longtable}

\end{document}